%% file: main.tex
\documentclass[%
 reprint,
aps,prl,
superscriptaddress,
floatfix,
nofootinbib,
]{revtex4-2}
\input{template}
\usepackage{graphicx}
\usepackage{dcolumn}
\usepackage{bm}
\usepackage{fix-cm}
\usepackage{mathrsfs}
\usepackage{hyperref}

\begin{document}

\crefname{section}{Supplemental Material}{Supplemental Material}
\Crefname{section}{Supplemental Material}{Supplemental Material}
\pagenumbering{arabic}

\title{Self-testing in Fermionic Networks}

\author{Fatemeh Moradi Kalarde}
\email{fatemeh.moradi-kalarde@inria.fr}
\affiliation{
 \mbox{Inria, CPHT, LIX, CNRS, Ecole polytechnique, Institut Polytechnique de Paris, Palaiseau, France}
}

\author{Luis Muñecas Tomás}
\email{luis.munecas\_tomas@nokia.com}
\affiliation{
 \mbox{Inria, CPHT, LIX, CNRS, Ecole polytechnique, Institut Polytechnique de Paris, Palaiseau, France}
}
\affiliation{
 Nokia Bell Labs, 12 Rue Jean Bart, 91300 Massy, France
}
\author{Marc-Olivier Renou}

\affiliation{
 \mbox{Inria, CPHT, LIX, CNRS, Ecole polytechnique, Institut Polytechnique de Paris, Palaiseau, France}
}

\begin{abstract}
Self-testing certifies quantum states and measurements from observed correlations alone, without assumptions on the inner workings of the devices. Its standard formulation, however, models the devices as manipulating qudits, of arbitrary or even infinite dimension. Yet febits, information encoded in the presence or absence of a fermion in a mode, can produce correlations in quantum networks that no qudit model reproduces under the same locality constraints. We first show that the modeling choice in terms of qudits or febits matters for self-testing: some correlations that self-test a unique qudit strategy admit two fermionic realizations inequivalent under local parity-preserving isometries. We then develop a fermionic self-testing framework and show that the correlations of a three-party rewiring network single out febit states and measurements up to such isometries, providing the first genuine fermionic self-test.
\end{abstract}

\maketitle

\newpage

\paragraph*{Introduction---}\label{sec:introduction}
Bell's theorem established a fundamental distinction between quantum correlations and those compatible with local classical models \cite{bell1964einstein}. Beyond this separation, understanding the structure of quantum correlations raises two questions: which observed correlations uniquely determine the physical realization underlying an experiment, and, given such correlations, what is that realization? These questions are explored in self-testing \cite{vsupic2020self}. By identifying the underlying states and measurements from observed correlations alone, self-testing provides a device-independent certification of quantum systems, i.e. without assumptions on the inner workings of the devices which produced them. For instance, any strategy reaching the maximal quantum value $2\sqrt{2}$ of the CHSH expression is equivalent, up to local isometries, to Pauli measurements on a maximally entangled two-qubit state~\cite{popescu1992which,mckague2012robust}.

The standard formulation of self-testing seeks to prove the unique realization of observed correlations \emph{assuming they were obtained within standard quantum information theory}, hereafter called qubit quantum information theory (Q-QIT). Self-testing within Q-QIT has been developed primarily for \emph{centralized} Bell scenarios, in which a single source distributes a joint state to all parties, both in the bipartite Alice and Bob scenario~\cite{mayers2004self,mckague2012robust,coladangelo2017all} and in its multipartite generalizations~\cite{mckague2014self,wu2014robust,pal2014device,vsupic2018self,baccari2020scalable}. It has since been extended to \emph{network} scenarios, where several independent sources each distribute systems to subsets of the parties~\cite{branciard2010characterizing,fritz2012beyond,tavakoli2022bell}. Examples include self-tests of the Bell-state measurement in entanglement swapping~\cite{renou2018self,bancal2018noise,vsupic2022genuine}, of all pure multipartite entangled states up to complex conjugation~\cite{vsupic2023quantum}, and partial self-tests in the triangle network~\cite{sekatski2023partial}.
In all these examples, \emph{devices are modeled within Q-QIT}, that is as manipulating qudits of arbitrary (possibly infinite) dimension, and composite systems are described by tensor products of local Hilbert spaces. This is the appropriate description of information carried by distinguishable particles or by bosonic modes~\cite{friis2016reasonable}.

\begin{figure*}[t]
\centering
\resizebox{0.95\textwidth}{!}{\input{Figure1TikZ.tex}}
\caption{Centralized and network self-testing in Q-QIT and in F-QIT.
(a) In centralized scenarios, with a single source distributing a joint state, both theories admit the same set of correlations and the same self-tests.
(b) Independent sources reveal a distinction: the correlations of a fixed three-party loop, previously shown to self-test a unique Q-QIT strategy, admit two inequivalent F-QIT strategies, as we prove in Theorem~\ref{thm:three-party-loop-informal}, while the correlations obtained across a hidden network containing all possible wirings, which no Q-QIT strategy reproduces, self-test a unique F-QIT strategy (our second result, \cref{thm:fermionic-self-testing}).}
\label{fig:self-testing-overview}
\end{figure*}
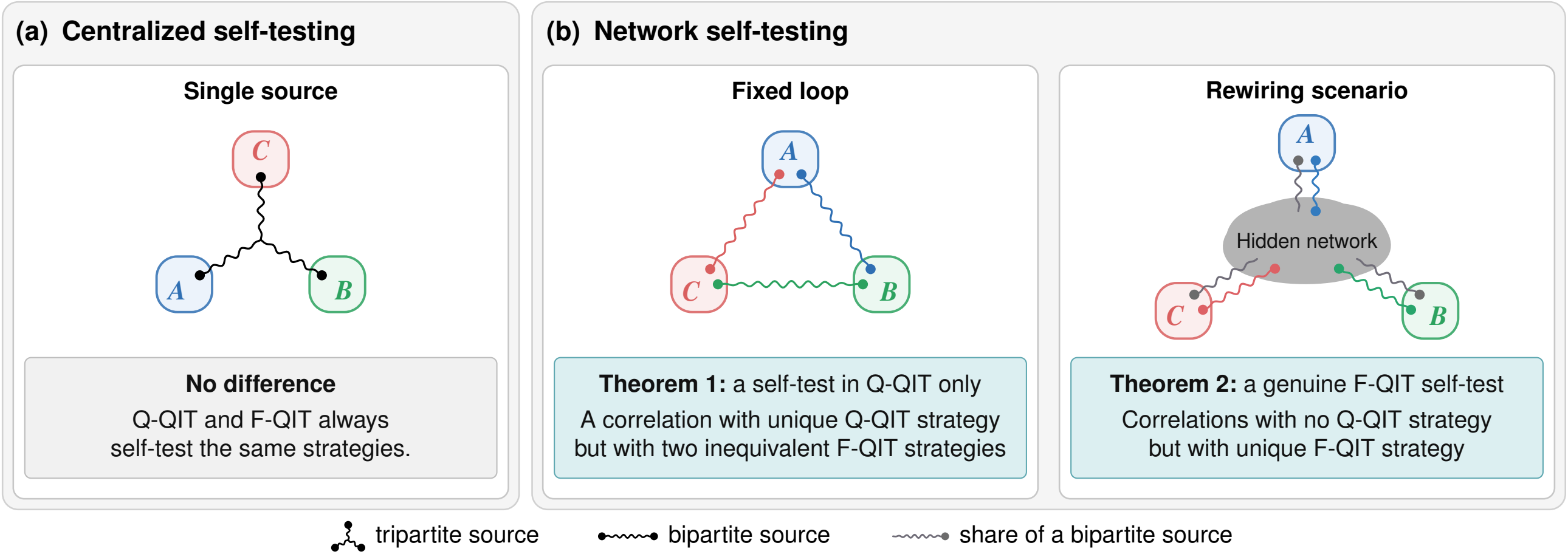

However, this tensor-product description of composite systems is a postulate, and Ref.~\cite{kalarde2026fermions} shows that it does not capture all physical systems: in a network experiment, indistinguishable fermions can generate correlations that no qudit strategy reproduces under the same locality constraints. This establishes an operational distinction between Q-QIT and fermionic quantum information theory (F-QIT), and raises two natural questions. Can correlations self-test a unique qudit strategy, yet admit several fermionic realizations that are inequivalent (under local parity-preserving isometries)? And can correlations that no qudit strategy reproduces self-test a fermionic realization? Since centralized scenarios admit no correlations achievable only by fermionic strategies~\cite{kalarde2026fermions}, answering the second requires a network scenario (see \cref{fig:self-testing-overview}).

\begin{figure*}[!t]
    \centering
\resizebox{0.95\textwidth}{!}{\input{Figure2TikZ.tex}}
    \caption{Self-testing scenario. (a) Each party independently prepares a bipartite state, retains one subsystem, and transmits the other through the network. After redistribution according to an undisclosed wiring, each party performs an input-dependent local measurement and produces an output. (b) The five wirings: two three-party loops, $ABC$ and $ACB$, and three two-party loops, $AB$, $AC$, and $BC$, with the excluded party receiving its own subsystem back.}
    \label{fig:scenario}
    \label{fig:your_label}
\end{figure*}
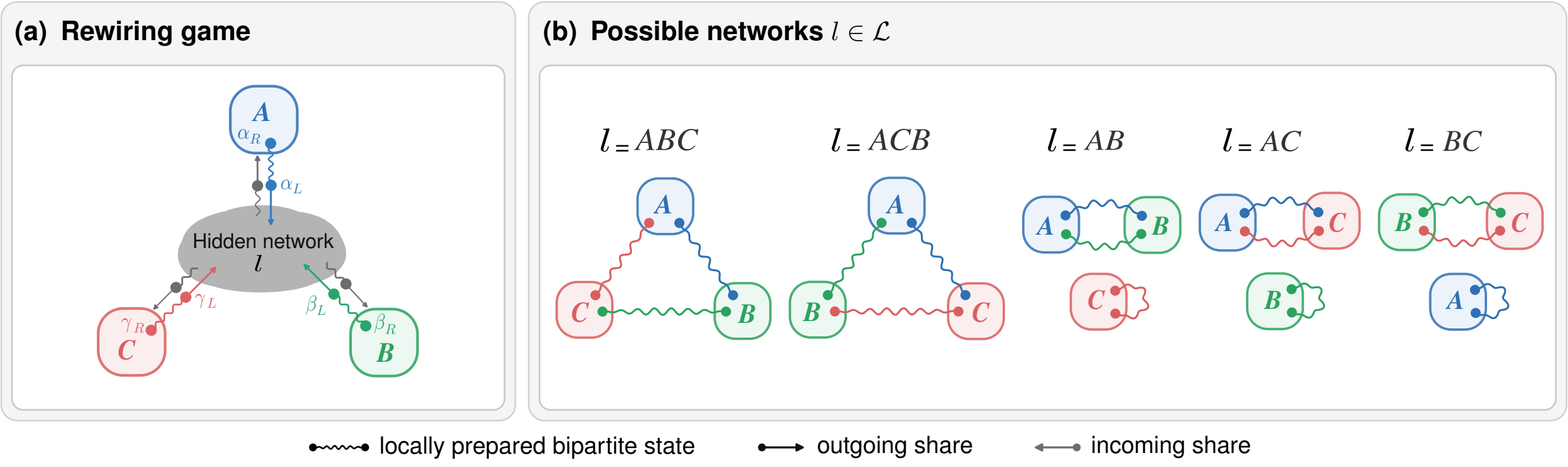

In this work, we introduce a self-testing framework for F-QIT strategies and use it to answer both questions in the rewiring experiment described below. The same sources and measurement devices are connected by a hidden wiring, forming either a three-party loop or a two-party loop in which the third party receives its own subsystem back (see Fig.\ref{fig:scenario}). We first show that the correlations of a single three-party loop, previously shown to self-test a unique qudit strategy, admit two fermionic realizations that are inequivalent under local parity-preserving isometries. We then show that the full family of correlations, obtained over all wirings, self-tests the fermionic states and measurements up to local parity-preserving isometries, yielding the first genuine instance of fermionic self-testing.

Before presenting our main results, we briefly review the F-QIT formalism, recall self-testing in the bipartite scenario within Q-QIT, and discuss its extension to F-QIT.

\paragraph*{Fermionic quantum information theory---}
\label{sec:F-QIT}
We start with a brief review of F-QIT~\cite{vidal2021quantum}, in which information is encoded in the occupation of fermionic modes. A single mode, which we call a \emph{febit}, has a two-dimensional space spanned by $\ket{0}$ (empty) and $\ket{1}$ (occupied). For mode $ i \in \{1,\ldots,M \}$, we denote its creation and annihilation operators by $f_i^\dagger$ and $f_i$ which satisfy the canonical anticommutation relations.

The anticommutation relations make febits differ from qubits, as the ordering of fermionic modes matters. For instance, writing $\ket{n_1,n_2}_{1,2}:=(f_1^\dagger)^{n_1}(f_2^\dagger)^{n_2}\ket{\Omega}$, with $\ket{\Omega}$ the vacuum, we have
\begin{equation}
    \ket{1,1}_{2,1}
    =
    f_2^\dagger f_1^\dagger\ket{\Omega}
    =
    -f_1^\dagger f_2^\dagger\ket{\Omega}
    =
    -\ket{1,1}_{1,2}.
\end{equation}
This sign may become operationally relevant when occupied modes are exchanged between parties, as in Fig.~\ref{fig:scenario}. It is convenient to use the wedge product notation~\cite{vidal2021quantum}, which indicates that modes should first be reordered in the canonical increasing ordering to keep track of such signs when composing systems. We write the occupation-number basis of the $2^M$-dimensional fermionic Fock space as
\begin{equation}
    \label{eq:WedgeDefiniton}
    \ket{n_1}_1\wedge\cdots\wedge\ket{n_M}_M
    :=
    (f_1^\dagger)^{n_1}\cdots(f_M^\dagger)^{n_M}\ket{\Omega},
\end{equation}
and extend this definition by linearity in each factor to states on disjoint sets of modes. When no confusion arises and the modes are in the canonical ordering, we omit the wedge symbol and the mode labels, writing $\ket{n_1,\ldots,n_M}:=\ket{n_1}_1\wedge\cdots\wedge\ket{n_M}_M$, sometimes inserting commas only to separate the modes of different parties.

Unlike in Q-QIT, physical states and operations in F-QIT are constrained by the \emph{parity superselection rule}~\cite{friis2016reasonable,johansson2016comment,vidal2021quantum}. The rule forbids coherent superpositions of even and odd parity states, while allowing incoherent mixtures of the two. Moreover, for spatially separated parties holding disjoint sets of modes, physical local observables are even polynomials in the local creation and annihilation operators, as required for consistency with the no-signaling principle~\cite{johansson2016comment}. Such observables preserve the local fermion-number parity, and observables associated with different parties commute. Physical Hamiltonians and the unitaries they generate must also preserve parity.

Finally, the wedge product extends to parity-preserving operators. For two disjoint sets of modes $A, B$ and states $\ket{\psi_A},\ket{\psi'_A}$ with the same definite parity, and likewise $\ket{\phi_B},\ket{\phi'_B}$, we define
\begin{equation}
   \ketbra{\psi_A}{\psi'_A}\wedge\ketbra{\phi_B}{\phi'_B}
:=\big[\ket{\psi_A}\wedge\ket{\phi_B}\big]
\big[\ket{\psi'_A}\wedge\ket{\phi'_B}\big]^\dagger,
\end{equation}
and extend this definition bilinearly to all parity-preserving local operators.

\paragraph*{Self-testing in the Bell scenario: Q-QIT and F-QIT---}
\label{sec:Bell-Selftesting}
We first recall self-testing in the bipartite Bell scenario within Q-QIT (see Fig.~\ref{fig:self-testing-overview}(a)). Two spacelike-separated parties $A$ and $B$ share a state $\ket{\psi}\in\mathcal H_A\otimes\mathcal H_B$ and, upon receiving inputs $x,y\in\{0,1\}$, measure local observables $A_x$ and $B_y$ with outcomes $a,b\in\{\pm1\}$. Since local changes of basis and the addition of degrees of freedom on which the measurements act trivially leave the correlations $p(a,b\mid x,y)$ unchanged, self-testing asks whether these correlations determine the state and measurements \emph{up to local isometries}~\cite{mayers2004self}; see \cref{sec:TheoreticalDefs} for a precise definition. The paradigmatic example is the CHSH expression $\beta=\langle A_0(B_0+B_1)+A_1(B_0-B_1)\rangle$~\cite{chshref}, bounded by $2$ for local hidden-variable models and by $2\sqrt2$ for quantum strategies~\cite{Cirelson1980QuantumGO}. The maximal quantum value is attained by the reference Q-QIT strategy
\begin{equation}\label{eq:CHSHstrategy}
\begin{aligned}
  &\ket{\phi_{\mathrm q}^{+}}=\tfrac1{\sqrt2}\big(\ket{0,0}+\ket{1,1}\big),\\
  &A_0=\sigma_X,\quad A_1=\sigma_Z,\qquad
  B_{0,1}=\frac{\sigma_X\pm\sigma_Z}{\sqrt2},
\end{aligned}
\end{equation}
and, conversely, any strategy attaining it is equivalent to this reference strategy up to local isometries~\cite{mckague2012robust}.

In F-QIT, a single febit per party cannot violate the CHSH inequality: the parity superselection rule restricts local observables to be diagonal in the occupation-number basis, i.e., to lie in the span of $\id$ and $\sigma_Z$, so that they all commute. With two febits per party, one can instead encode a qubit in the even-parity subspace, replacing $\ket0$ and $\ket1$ by $\ket{00}$ and $\ket{11}$, respectively. The parity-preserving operators $\sigma_X^f=\ket{00}\!\bra{11}+\ket{11}\!\bra{00}$ and $\sigma_Z^f=\ket{00}\!\bra{00}-\ket{11}\!\bra{11}$ then play the role of the Pauli operators, and applying this encoding to \cref{eq:CHSHstrategy} yields a fermionic strategy, with state $\ket{\phi_{\mathrm f}^{+}}=\tfrac1{\sqrt2}(\ket{00,00}+\ket{11,11})$, that attains $2\sqrt2$. To self-test this strategy, we must adapt the notion of local isometry to F-QIT: subsystems compose through the wedge product rather than the tensor product, and local isometries must preserve parity. 
With these adaptations, detailed in \cref{sec:TheoreticalDefs}, the Q-QIT self-test carries over: any fermionic strategy attaining $2\sqrt2$ is equivalent to this one up to local parity-preserving isometries.

The same encoding, replacing each qubit by two febits, carries the self-testing argument over to other centralized scenarios, such as those involving GHZ states (see \cref{lem:fermionic_Bell_relations}), as swapping $\ket{n_1n_1}$ with $\ket{n_2n_2}$ never introduces a sign. For these correlations, the answer to both questions raised in the introduction is negative: they self-test a unique strategy in both theories, and, like all centralized correlations, they can be reproduced by qudit strategies. We therefore turn to the network scenario described below, where we answer both questions affirmatively.

\paragraph*{The fermionic self-testing scenario---}
\label{sec:fermionic-blind-loop}

We recall the three-party rewiring scenario introduced in
\cite{kalarde2026fermions}, see also Fig.~\ref{fig:scenario}. It involves three spacelike-separated parties, \(A\), \(B\), and \(C\), who are allowed to communicate in a single round through a communication network. Each party locally and independently prepares a bipartite fermionic system, labelled
\(\alpha\), \(\beta\), and \(\gamma\) for \(A\), \(B\), and \(C\), respectively. The two subsystems of a system \(\theta\in\{\alpha,\beta,\gamma\}\) are denoted by \(\theta_R\) and \(\theta_L\). The party retains \(\theta_R\) and transmits \(\theta_L\) through the communication network.

In a single communication round, the network only redistributes the transmitted subsystems according to one of five wirings $l\in \mathcal{L}:=\{ABC,ACB,AB,AC,BC\}$.

The wirings \(ABC\) and \(ACB\) are the two directed three-party loops. For instance, in the wiring \(ABC\), the subsystem sent by \(A\) is delivered to \(B\), the subsystem sent by \(B\) to \(C\), and the subsystem sent by \(C\) to \(A\). The wirings \(AB\), \(AC\), and \(BC\) are the three possible two-party loops. In each case, the party not named in the loop label receives its own transmitted subsystem back. The protocol and the five possible hidden wirings are illustrated in \cref{fig:scenario}.

The wiring \(l\) is not disclosed to the parties, and the source preparations are fixed independently of it. After the redistribution, each party \(i\) receives a randomly chosen input \(x_i\in\{0,1\}\) and produces an output \(a_i\in\{0,1_-,1_+,2\}\) by a local measurement on the two subsystems \(i_R\) and \(i_L\) it holds. The correspondence between these party-based labels and the source-based labels \(\theta_R,\theta_L\) depends on \(l\).

The measurement may depend only on locally available information: the party's name, its input, and the identity of the party that sent the received subsystem. It may not depend on \(l\) itself, nor on which party received the subsystem it sent. Consequently, for the same input, two wirings that induce the same local subsystem configuration at a party must correspond to the same local measurement.

\emph{Trusted fermionic strategy.—}
For each \(\theta\in\{\alpha,\beta,\gamma\}\), the corresponding party prepares the single-fermion state \begin{equation}
\label{eq:trusted-source-state}
\begin{aligned}
\ket{\psi_\theta^+}_{\theta_R\theta_L}
&=
\frac{1}{\sqrt{2}}
\left(
\ket{0,1}_{\theta_R\theta_L}
+
\ket{1,0}_{\theta_R\theta_L}
\right)\\
&=
\frac{1}{\sqrt{2}}
\left(
f_{\theta_L}^{\dagger}
+
f_{\theta_R}^{\dagger}
\right)
\ket{\Omega},
\end{aligned}
\end{equation}
where \(\ket{\Omega}\) denotes the global vacuum and \(f_{\theta_R}^{\dagger}\) and \(f_{\theta_L}^{\dagger}\) are the fermionic creation operators of modes \(\theta_R\) and \(\theta_L\) respectively. The initial global state is therefore
\begin{equation}
\label{eq:trusted-global-state}
\ket{\psi}
=
\ket{\psi_\alpha^+}
\wedge
\ket{\psi_\beta^+}
\wedge
\ket{\psi_\gamma^+}.
\end{equation}

After the redistribution, party \(i\) performs an input-dependent, parity-preserving four-outcome PVM on the modes \(i_R,i_L\). The PVM elements are denoted by
\(\{\Pi_i^0,\Pi_i^2,\Pi_{i,x_i}^{1_+},\Pi_{i,x_i}^{1_-}\}\).
The outcomes \(0\) and \(2\) correspond to input-independent projectors onto the zero-particle state \(\ket{00}_{i_Ri_L}\) and the two-particle state \(\ket{11}_{i_Ri_L}\), respectively. The outcomes \(1_+\) and \(1_-\) correspond to projectors onto
\(u_{i,x_i}^{+}\ket{01}_{i_Ri_L}
+
v_{i,x_i}^{+}\ket{10}_{i_Ri_L}\) and
\(u_{i,x_i}^{-}\ket{01}_{i_Ri_L}
+
v_{i,x_i}^{-}\ket{10}_{i_Ri_L}\), respectively. These states form an input-dependent orthonormal basis of the one-particle sector. The coefficients are specified in \hyperref[sec:DetailedScenario]{the End Matter}.  All these PVM elements preserve the local fermion-number parity and are therefore compatible with the parity superselection rule.

The same preparations and measurements are used for every hidden wiring. They generate the family of fermionic distributions
\begin{equation}
\label{eq:trusted-correlations}
\left\{
P_f^l(a_A,a_B,a_C\mid x_A,x_B,x_C)
\right\}_{l\in \mathcal{L}},
\end{equation}
which we use below as the reference correlations.

We first consider the correlations of the fixed three-party loop \(ABC\).
They self-test a unique strategy within Q-QIT, but not within F-QIT, where
two inequivalent classes of fermionic strategies remain.\footnote{An analogous phenomenon occurs for token-counting correlations in loop networks: the classical realization is essentially unique, while the same correlations admit quantum realizations that are nonequivalent under local isometries~\cite{PhysRevA.105.022408,sekatski2023partial}.}

\begin{theorem}[Fixed-loop self-testing]
\label{thm:three-party-loop-informal} 
Consider the correlations \(P_f^{ABC}\) of the trusted fermionic strategy on the fixed loop \(ABC\).
\begin{enumerate}
\item[(i)] \emph{Q-QIT:} \(P_f^{ABC}\) self-tests a unique qubit strategy up to local isometries, obtained by reading the trusted strategy in the tensor-product formalism, with \(\wedge\) replaced by \(\otimes\).
\item[(ii)] \emph{F-QIT:} \(P_f^{ABC}\) admits exactly two nonequivalent realisations up to local parity-preserving isometries: the trusted single-mode strategy, and its double-mode encoding obtained by replacing \(\ket0\) and \(\ket1\) with \(\ket{00}\) and \(\ket{11}\) respectively.
\end{enumerate}
\end{theorem}

\begin{proof}[Proof's intuition]
\emph{(i) Q-QIT.} A direct computation shows that the trusted strategy with \(\wedge\) replaced by \(\otimes\) reproduces \(P_f^{ABC}\). Then, coarse-graining the outcomes \(1_+\) and \(1_-\) into a single outcome \(1\) gives a \emph{token-counting strategy}: each source distributes one token between its two endpoints with equal probabilities, and each party counts the tokens it receives (the projectors for \(1_+\) and \(1_-\) sum to \(\ketbra{01}{01}+\ketbra{10}{10}\)). In the loop \(ABC\), the outcome \((1,1,1)\) occurs only when either all sources retain their tokens or all transmit them around the loop. Representing \(\ket{01}\) and \(\ket{10}\) by \(\ket{\tilde0}\) and \(\ket{\tilde1}\), respectively, we obtain the conditional state for the outcome \((1,1,1)\):
\[
\ket{\mathrm{GHZ_3}^{+}}=\tfrac1{\sqrt2}\bigl(\ket{\tilde{0}\tilde{0}\tilde{0}}+\ket{\tilde{1}\tilde{1}\tilde{1}}\bigr).
\]
The proof of uniqueness combines two self-testing results. First, the coarse-grained token-counting correlations partially self-test the strategy~\cite{PhysRevA.105.022408,sekatski2023partial}: each source must distribute a single token, and each coarse-grained measurement must count the received tokens. Second, the statistics of the \(1_\pm\) outcomes conditioned on \((1,1,1)\) self-test \(\ket{\mathrm{GHZ_3}^{+}}\) and the measurements on the one-token sector, which completes the characterization (see Theorem~B.2 and its proof in the Supplemental Material of Ref.~\cite{kalarde2026fermions}).
\\
\emph{(ii) F-QIT.} Encoding the qubit strategy in case (i) into febits by replacing \(\ket0\) and \(\ket1\) with \(\ket{00}\) and \(\ket{11}\), respectively, preserves the correlation, as in the Bell scenario; hence, both the single-mode and double-mode fermionic realizations reproduce \(P_f^{ABC}\). The double-mode realization is equivalent to the qubit strategy. Note that the single-mode realization (the trusted fermionic strategy) also yields \(\ket{\mathrm{GHZ_3}^{+}}\) when conditioned on the coarse-grained outcome \((1,1,1)\), since the cyclic exchange in a three-party loop is an even permutation of single fermions.
Both self-testing ingredients extend to F-QIT (\cref{thm:exact-fermionic-TC} for token counting and \cref{lem:fermionic_Bell_relations} for GHZ self-testing), yielding a characterization of the loop, up to one ambiguity left by the parity superselection rule: the two realizations are related by the encoding \(V:\ket0\mapsto\ket{00},\ \ket1\mapsto\ket{11}\) on each source subsystem, which is a legitimate local isometry in Q-QIT but not in F-QIT, since it changes the relative parity of the branches \(\ket0\) and \(\ket1\). As we show in \cref{sec:fixed-loop-self-testing}, no local parity-preserving isometry can relate the two realizations.
\end{proof}

Note that the situation changes in the two-party loops. In the loop \(BC\) (after tracing out \(A\)), the fermionic state shared by \(B\) and \(C\), conditioned on the outcome \((1,1)\), becomes
\[
\ket{\mathrm{GHZ_2}^{-}}=\tfrac1{\sqrt2}\bigl(\ket{\tilde{0}\tilde{0}}-\ket{\tilde{1}\tilde{1}}\bigr)
\]
for the single-mode realization, since exchanging two fermions gives a minus sign, while it remains \(\ket{\mathrm{GHZ_2}^{+}}\) for Q-QIT (and for its F-QIT encoding obtained by replacing \(\ket0\) and \(\ket1\) with \(\ket{00}\) and \(\ket{11}\)). As we show below, this sign difference is what allows the correlations in the two-party loops to distinguish the two classes of \cref{thm:three-party-loop-informal}(ii), leading to \cref{thm:fermionic-self-testing}.

We now state the second result of this letter, which is that the full set of correlations observed in the rewiring experiment uniquely determines the underlying fermionic strategy, up to local parity-preserving isometries.

\begin{theorem}[Fermionic self-testing]
\label{thm:fermionic-self-testing}
In the rewiring scenario, the family of correlations \(\{P_f^l\}_{l\in\mathcal L}\) of the trusted fermionic strategy self-tests a unique fermionic strategy up to local parity-preserving isometries: the trusted single-mode strategy, with the source states of \cref{eq:trusted-source-state} and the measurements of \cref{eq:trusted-measurements} with the coefficients of \cref{eq:trusted-coefficients-A,eq:trusted-coefficients-BC}. In particular, the double-mode realization of \cref{thm:three-party-loop-informal} is excluded.
\end{theorem}

\begin{proof}[Proof's intuition]
Our locality assumption requires the sources to not adapt to the wiring, and a party's measurement to be the same whenever it receives a subsystem from the same party. By \cref{thm:three-party-loop-informal}, the three-party loops leave two candidate realizations, one using single modes, a second using two modes (and which is the encoding of the Q-QIT strategy). As discussed above, in \(BC\) the single-mode strategy (i) gives \(\ket{\mathrm{GHZ_2}^{-}}\) and the double-mode strategy (ii) gives \(\ket{\mathrm{GHZ_2}^{+}}\). For \(x_B=x_C=1\), they therefore give opposite and equal outcome signs, respectively, and only the former matches the reference correlations.

A double-mode strategy could try to recover the missing sign through local phases in its sources or measurements. But each such phase also enters every other wiring involving the same source or local context, and consistency across all five wirings rules this out: the signs produced by exchanging single fermions cannot be mimicked by local phases. The full proof makes this rigorous; see \cref{thm:fermionic-self-testing-formal} in \cref{sec:Main Theorem's Proof} and the proof roadmap in the End Matter.
\end{proof}

Our results show that in network scenarios, beyond centralized Bell scenarios, a self-testing statement is only meaningful once the underlying physical systems are specified: self-testing proofs that assume the subsystems are qudits may no longer be valid when the underlying systems are febits, for which a new notion of device-independent self-testing emerges.

Several questions remain open. First, our fermionic self-test relies on the hidden rewiring of the network. Is this hiding necessary, or can some fixed network already single out the single-mode realization? Second, can all pure entangled fermionic states compatible with the parity superselection rule be self-tested in networks, as is the case in Q-QIT~\cite{vsupic2023quantum}? Third, can a noise-robust version of our result be obtained, which could lead to an experimental realization with fermionic cold atoms? Fourth, can our results be extended beyond febits, to self-testing strategies based on anyons or paraparticles~\cite{Wang2025ParticleExchange}?

Finally, our results raise a security concern. Some ``device-independent'' protocols in networks, such as randomness certification in the triangle network~\cite{sekatski2023partial}, promise security without any assumption on the inner workings of the devices. Yet their proofs rely on self-testing results that implicitly assume the eavesdropper is restricted to Q-QIT. Theorem~\ref{thm:three-party-loop-informal} shows that this assumption is not innocuous: correlations that self-test a unique qubit strategy may admit inequivalent fermionic realizations. Could an eavesdropper manipulating fermions break such network ``device-independent'' protocols, which are provably secure \emph{only} against qubit-based attacks?

\begin{acknowledgments}
\textit{Acknowledgements.} We thank Salman Beigi, Ludovic Noirie, Lucas Tendick, Remi Varloot and Xiangling Xu for fruitful discussions.
We acknowledge funding by the ANR for the JCJC grants LINKS (No. ANR-23-CE47-0003), the T-ERC QNET (No. ANR24-ERCS-0008), the project QUANTINT, as well as the European Union's Horizon 2020 Research and Innovation Programme under QuantERA Grant Agreements No. 731473 and No. 101017733. 
The research ideas, the results, the structure of the presentation, and all proofs were developed by the authors. The AI assistance was only used to check the proofs for missing or incomplete arguments and to polish the language of the manuscript. All suggestions were reviewed and verified by the authors,
who take full responsibility for the content of this work.
\end{acknowledgments}
\bibliography{bibliography}

\clearpage
\onecolumngrid
\begin{center}
    \textbf{END MATTER}
\end{center}

\section*{Trusted fermionic strategy}
\label{sec:DetailedScenario}

The source states are given by \cref{eq:trusted-source-state}.
After redistribution through the unknown network, the modes held by
party \(i\) are denoted by \(i_R\) and \(i_L\).
The corresponding PVM elements are
\begin{equation}
\label{eq:trusted-measurements}
\begin{aligned}
\Pi_i^0
&=
\ket{00}\!\bra{00}_{i_Ri_L},
&
\Pi_i^2
&=
\ket{11}\!\bra{11}_{i_Ri_L},
\\
\Pi_{i,x_i}^{1_+}
&=
\ket{m_{i,x_i}^{+}}\!\bra{m_{i,x_i}^{+}},
&
\Pi_{i,x_i}^{1_-}
&=
\ket{m_{i,x_i}^{-}}\!\bra{m_{i,x_i}^{-}},
\end{aligned}
\end{equation}
where
\begin{equation}
\label{eq:trusted-one-particle-states}
\ket{m_{i,x_i}^{\pm}}
=
u_{i,x_i}^{\pm}\ket{01}_{i_Ri_L}
+
v_{i,x_i}^{\pm}\ket{10}_{i_Ri_L},
\end{equation}
or, equivalently,
\begin{equation}
\label{eq:trusted-one-particle-creation}
\ket{m_{i,x_i}^{\pm}}
=
\left(
u_{i,x_i}^{\pm}f_{i_L}^{\dagger}
+
v_{i,x_i}^{\pm}f_{i_R}^{\dagger}
\right)
\ket{\Omega}_{i_Ri_L}.
\end{equation}
The coefficients satisfy
\begin{equation}
\label{eq:trusted-coefficient-conditions}
\begin{aligned}
&\left|u_{i,x_i}^{\pm}\right|^2
+
\left|v_{i,x_i}^{\pm}\right|^2
=
1,
\\
&u_{i,x_i}^{-}
=
\left(v_{i,x_i}^{+}\right)^*,
\qquad
v_{i,x_i}^{-}
=
-\left(u_{i,x_i}^{+}\right)^*.
\end{aligned}
\end{equation}

For party \(A\), the measurement coefficients are
\begin{equation}
\label{eq:trusted-coefficients-A}
\begin{aligned}
u_{A,0}^{+}=-v_{A,0}^{-}=v_{A,1}^{+}=u_{A,1}^{-}
&=\frac{1}{\sqrt{2(2-\sqrt{2})}},
\\
v_{A,0}^{+}=u_{A,0}^{-}=u_{A,1}^{+}=-v_{A,1}^{-}
&=\frac{\sqrt{2}-1}{\sqrt{2(2-\sqrt{2})}}.
\end{aligned}
\end{equation}
For \(i\in\{B,C\}\), the coefficients are
\begin{equation}
\label{eq:trusted-coefficients-BC}
\begin{aligned}
u_{i,0}^{+}=-v_{i,0}^{-}&=1,
&
v_{i,0}^{+}=u_{i,0}^{-}&=0,
\\
u_{i,1}^{+}=-v_{i,1}^{-}&=\frac{1}{\sqrt{2}},
&
v_{i,1}^{+}=u_{i,1}^{-}&=\frac{1}{\sqrt{2}}.
\end{aligned}
\end{equation}
In the logical basis $\ket{\tilde0}=\ket{01}$, $\ket{\tilde1}=\ket{10}$ of the one-particle sector, these coefficients correspond to party $A$ measuring $(\sigma_X+(-1)^{x_A}\sigma_Z)/\sqrt2$, and parties $B$ and $C$ measuring $\sigma_Z$ and $\sigma_X$ for inputs $0$ and $1$, respectively.

\section*{Proof roadmap}
\label{sec:proof-roadmap}

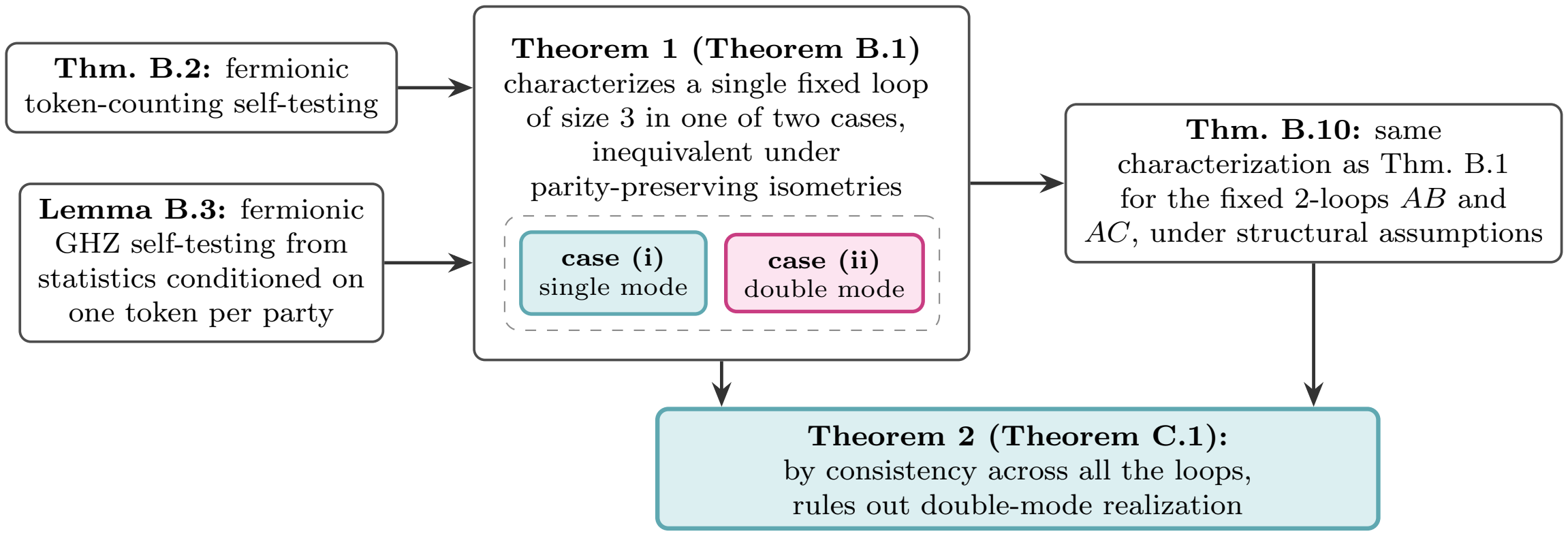
\begin{figure}[!ht]
\centering
\input{RoadmapTikZ.tex}
\caption{\label{fig:roadmap}%
Proof roadmap for \cref{thm:fermionic-self-testing-formal} (stated informally as \cref{thm:fermionic-self-testing}).
Token-counting rigidity and GHZ self-testing classify strategies on a fixed
three-party loop into two classes (\cref{thm:fixed-loop-characterization},
stated informally as \cref{thm:three-party-loop-informal}).
Locality transfers part of this characterization to two-party loops, and
\cref{thm:fixed-loop-characterization-N2} completes it for \(AB\) and \(AC\).
Finally, the proof of \cref{thm:fermionic-self-testing-formal} combines these characterizations
to predict the statistics in \(BC\), and hence rules out the double-mode class.}
\end{figure}

\twocolumngrid
The rigorous argument begins by characterizing the strategies
that reproduce the correlations of a single loop. They fall
into two classes that are inequivalent under local physical
isometries.
We refer to these as the single-mode and
double-mode classes, corresponding to the two trusted
strategies discussed in the intuitive proof of \cref{thm:three-party-loop-informal}.
The two classes are equivalent under local isometries in Q-QIT, and hence are a complete self-test in Q-QIT.

The formal characterization is given in
\cref{thm:fixed-loop-characterization} of \cref{sec:fixed-loop-self-testing}. Its proof
builds on the token-counting rigidity and partial
self-testing results in Q-QIT
\cite{PhysRevA.105.022408,sekatski2023partial} and extends
them to F-QIT, as established in
\cref{thm:exact-fermionic-TC}. This yields parity-preserving
extraction maps and a partial characterization of the
sources and coarse-grained measurements. Fermionic GHZ-state self-testing of the conditional correlations of all parties obtaining \(1_\pm\),
as established in \cref{lem:fermionic_Bell_relations} of the Supplemental Material,
then yields the additional relations needed to complete the characterization.

We then rule out all strategies in the double-mode class by
imposing consistency across all five wirings. We show that
the obstruction illustrated in the intuitive proof of \cref{thm:fermionic-self-testing} applies
to every realization in the double mode class.

We impose the locality constraints by requiring the source
preparations to remain fixed and the measurements to be the
same whenever the local contexts coincide. To organize the
passage from three-party to two-party loops, we establish
\cref{thm:fixed-loop-characterization-N2}, which gives a
version of the fixed-loop characterization under structural
assumptions inherited, through locality, from the three-party
loops. We use this result to characterize the strategies in
\(AB\) and \(AC\). Finally, the proof of
\cref{thm:fermionic-self-testing-formal} combines these
characterizations to show that the realizations in all local contexts
must belong to the same class. The resulting source and
measurement relations determine the distribution in
\(BC\). Conditioned on \(B\) and \(C\) each obtaining an
outcome in \(\{1_+,1_-\}\), the single-mode class gives
opposite signs when \(x_B=x_C=1\), whereas the double-mode
class gives equal signs. The reference fermionic distribution
in \(BC\) requires opposite signs for these inputs, ruling
out the double-mode class and leaving the single-mode
fermionic realization.

\par
\onecolumngrid
\clearpage

\section*{Supplemental Material}

\setcounter{section}{0}
\setcounter{secnumdepth}{3}
\renewcommand{\thesection}{\Alph{section}}

\crefalias{section}{section}
\crefname{section}{Section}{Sections}
\Crefname{section}{Section}{Sections}

\numberwithin{theorem}{section}
\numberwithin{equation}{section}
\noindent\textbf{Contents}\hfill\textbf{Page}\par\smallskip
\noindent\hyperref[sec:TheoreticalDefs]{\ref*{sec:TheoreticalDefs}.\ Self-testing definitions and CHSH examples}\dotfill\pageref{sec:TheoreticalDefs}\par
\noindent\hyperref[sec:fixed-loop-self-testing]{\ref*{sec:fixed-loop-self-testing}.\ Fixed-loop self-testing}\dotfill\pageref{sec:fixed-loop-self-testing}\par
\noindent\hyperref[sec:Main Theorem's Proof]{\ref*{sec:Main Theorem's Proof}.\ Fermionic network self-testing}\dotfill\pageref{sec:Main Theorem's Proof}\par\medskip

\input{DefinitionsSuppMat}

\input{FixedLoopSuppMat}

\input{NetworkSuppMat}

\end{document}

%% file: template.tex
\usepackage{amsmath,amssymb,amsfonts,mathrsfs}
\usepackage{amsthm}
\usepackage{graphicx}
\usepackage{color}
\usepackage{dsfont}
\usepackage{bbold}
\usepackage{faktor}
\usepackage{bbm}
\usepackage{mathtools}
\usepackage{enumitem}

\usepackage{adjustbox}
\usepackage{arydshln}

\usepackage{tikz}
\usetikzlibrary{calc,decorations.pathmorphing}
\usetikzlibrary{quantikz2}
\usepackage{mathrsfs}
\usepackage{romannum}
\usepackage[hidelinks]{hyperref}
\usepackage[capitalise]{cleveref}
\crefname{equation}{Eq.}{Eqs.}
\Crefname{equation}{Equation}{Equations}
\crefname{figure}{Fig.}{Figs.}
\Crefname{figure}{Figure}{Figures}
\crefname{appendix}{Appendix}{Appendices}
\Crefname{appendix}{Appendix}{Appendices}
\usepackage{placeins}

\theoremstyle{plain}
\newtheorem{theorem}{Theorem}
\newtheorem{lemma}[theorem]{Lemma}

\newtheorem{definition}[theorem]{Definition}

\newtheorem*{result*}{Result}

\theoremstyle{definition}

\newcommand{\ii}{\mathrm{i}}

\def\id{\mathbbm{1}}
\newcommand{\ketbra}[2]{|#1\rangle\langle #2|  }

\hypersetup{
  pdftitle    = {Self-testing in Fermionic Networks},
  pdfauthor   = {Fatemeh Moradi Kalarde; Luis Mu\~{n}ecas Tom\'{a}s; Marc-Olivier Renou},
  pdfsubject  = {Quantum information theory; fermionic information theory; self-testing in quantum networks},
  pdfkeywords = {quantum information theory, fermionic information theory, self-testing, device-independent certification, fermions, febits, quantum networks, quantum correlations, quantum nonlocality, parity superselection, local isometries, rewiring networks},
  pdflang     = {en},
  pdfcreator  = {LaTeX with hyperref},
  pdfdisplaydoctitle = true
}

%% file: Figure1TikZ.tex
\begingroup%
\definecolor{fOneA}{HTML}{2F6FB3}%
\definecolor{fOneB}{HTML}{2CA25F}%
\definecolor{fOneC}{HTML}{D95F5F}%
\definecolor{fOneAFill}{HTML}{EAF2FB}%
\definecolor{fOneBFill}{HTML}{EAF7EF}%
\definecolor{fOneCFill}{HTML}{FBECEC}%
\definecolor{fOneAShare}{HTML}{337BBF}%
\definecolor{fOneBShare}{HTML}{28A56C}%
\definecolor{fOneCShare}{HTML}{DD6165}%
\definecolor{fOneIncoming}{HTML}{706D78}%
\definecolor{fOneIncomingDot}{HTML}{6C6C6C}%
\definecolor{fOneCloud}{HTML}{B4B4B4}%
\definecolor{fOneText}{HTML}{111111}%
\definecolor{fOneSumFill}{HTML}{F1F1F1}\definecolor{fOneSumDraw}{HTML}{BDBDBD}
\definecolor{fOneThmFill}{HTML}{DCEFF1}\definecolor{fOneThmDraw}{HTML}{5FA8B1}
\providecommand{\fOneLoopTheorem}{\ref{thm:three-party-loop-informal}}%
\providecommand{\fOneRewiringTheorem}{\ref{thm:fermionic-self-testing}}%
\newcommand{\fOneSans}[1]{\fontfamily{phv}\fontsize{#1pt}{\dimexpr#1pt+1.5pt\relax}\selectfont}%
\newcommand{\fOneSerif}{\fontfamily{ptm}\fontsize{10bp}{10bp}\bfseries\itshape\selectfont}%
\newcommand{\fOneParty}[7]{%
  \path[fill=fOne#1Fill,draw=fOne#1,line width=0.26399mm,
    opacity=0.862471,rounded corners=2.31819mm]
    (#2-3.1075,#3-3.1095) rectangle (#2+3.1075,#3+3.1095);
  \node[text=fOne#1] at (#2+#4,#3+#5)
    {\resizebox{#6}{#7}{{\fOneSerif #1}}};
}%
\newcommand{\fOneWave}[7]{%
  \pgfmathsetmacro{\fOneDx}{#4-#2}%
  \pgfmathsetmacro{\fOneDy}{#5-#3}%
  \pgfmathsetmacro{\fOneLength}{sqrt(\fOneDx*\fOneDx+\fOneDy*\fOneDy)}%
  \pgfmathtruncatemacro{\fOneSamples}{max(35,ceil(\fOneLength/#7*12))}%
  \draw[draw=#1,line width=\fOneWaveWidth]
    plot[domain=0:1,samples=\fOneSamples,variable=\t]
    ({#2+\fOneDx*\t-\fOneDy/\fOneLength*#6*sin(180*\t)*sin(360*\fOneLength/#7*\t)},
     {#3+\fOneDy*\t+\fOneDx/\fOneLength*#6*sin(180*\t)*sin(360*\fOneLength/#7*\t)});
}%
\def\fOneWaveWidth{0.21738mm}%
\newcommand{\fOneDot}[3]{\fill[#1] (#2,#3) circle[radius=0.5645mm];}%
\newcommand{\fOneLink}[5]{%
  \fOneWave{#1}{#2}{#3}{#4}{#5}{0.376447}{1.880876}%
  \fOneDot{#1}{#2}{#3}\fOneDot{#1}{#4}{#5}%
}%
\def\fOneBoxW{26.4}%
\newcommand{\fOneSummary}[6]{%
  \path[fill=#2Fill,draw=#2Draw,line width=0.4pt,rounded corners=2pt]
    (#1-\fOneBoxW,39.90) rectangle (#1+\fOneBoxW,53.40);
  \node[fOneText] at (#1,46.65) {\textbf{#3}\ #4\\[2.5pt]#5\\#6};
}%
\begin{tikzpicture}[x=1mm,y=-1mm,line cap=round,line join=round,
  every node/.style={inner sep=0pt,outer sep=0pt},
  fOnePanel/.style={fill=black!4,draw=black!18,line width=0.5pt,rounded corners=5pt},
  fOneCard/.style={fill=white,draw=black!22,line width=0.5pt,rounded corners=3.5pt},
  fOneTitle/.style={font=\fOneSans{8.5}\bfseries,text=black},
  fOneLegend/.style={font=\fOneSans{7.5},text=black},
  fOneHeading/.style={font=\fOneSans{7.5}\bfseries,text=black},
  fOneText/.style={font=\fOneSans{7.5},text=fOneText,align=center}]
\path[use as bounding box] (0,0) rectangle (175,61.4);
\path[fOnePanel] (0.1,0.1) rectangle (57.88,56.9);
\path[fOnePanel] (59.33,0.1) rectangle (174.9,56.9);
\path[fOneCard] (1.30,7.20) rectangle (56.68,55.70);
\path[fOneCard] (60.53,7.20) rectangle (115.91,55.70);
\path[fOneCard] (118.31,7.20) rectangle (173.69,55.70);
\node[fOneTitle,anchor=west] at (1.6,3.6) {(a)\enspace Centralized self-testing};
\node[fOneTitle,anchor=west] at (60.93,3.6) {(b)\enspace Network self-testing};
\begin{scope}[shift={(38.76,60.325)}]
  \def\fOneWaveWidth{0.19039mm}
  \foreach \x/\y in {0/-1.70,-1.472/0.85,1.472/0.85}{%
    \fOneWave{black}{0}{0}{\x}{\y}{0.188}{0.88}
    \fill[black] (\x,\y) circle[radius=0.4384mm];}
\end{scope}
\node[fOneLegend,anchor=west] at (41.83,59.9) {tripartite source};
\begin{scope}
  \def\fOneWaveWidth{0.19039mm}
  \fOneWave{black}{67.14}{59.9}{72.97}{59.9}{0.188}{0.88}
  \fill[black] (67.14,59.9) circle[radius=0.4384mm];
  \fill[black] (72.97,59.9) circle[radius=0.4384mm];
  \fOneWave{fOneIncoming}{99.69}{59.9}{105.53}{59.9}{0.188}{0.88}
  \fill[fOneIncomingDot] (105.53,59.9) circle[radius=0.4384mm];
\end{scope}
\node[fOneLegend,anchor=west] at (74.57,59.9) {bipartite source};
\node[fOneLegend,anchor=west] at (107.13,59.9) {share of a bipartite source};
\node[fOneHeading] at (28.99,10.25) {Single source};
\node[fOneHeading] at (88.22,10.25) {Fixed loop};
\node[fOneHeading] at (146.00,10.25) {Rewiring scenario};

\begin{scope}[shift={(28.99,25.28)}]
\fOneParty{C}{0.00000}{-7.58333}{0.00000}{-0.97222}{1.971mm}{2.14998mm}
\fOneParty{A}{-8.55556}{6.41667}{-0.77778}{0.68056}{2.020mm}{2.08891mm}
\fOneParty{B}{8.55556}{6.41667}{0.77778}{0.68056}{1.984mm}{2.04642mm}
\fOneWave{black}{0.00000}{1.45833}{0.00000}{-5.58333}{0.376447}{1.880876}
\fOneDot{black}{0.00000}{-5.58333}
\fOneWave{black}{0.00000}{1.45833}{-6.82515}{5.41382}{0.376447}{1.880876}
\fOneDot{black}{-6.82515}{5.41382}
\fOneWave{black}{0.00000}{1.45833}{6.82515}{5.41382}{0.376447}{1.880876}
\fOneDot{black}{6.82515}{5.41382}
\end{scope}

\begin{scope}[shift={(88.22,25.28)}]
\fOneParty{A}{0.00000}{-7.58333}{0.00000}{-0.97222}{2.020mm}{2.08891mm}
\fOneParty{C}{-10.20833}{6.41667}{-0.87500}{0.77778}{1.971mm}{2.14998mm}
\fOneParty{B}{10.20833}{6.41667}{0.87500}{0.77778}{1.984mm}{2.04642mm}
\fOneLink{fOneA}{1.23726}{-5.88652}{8.97107}{4.71985}
\fOneLink{fOneB}{8.10833}{6.41667}{-8.10833}{6.41667}
\fOneLink{fOneC}{-8.97107}{4.71985}{-1.23726}{-5.88652}
\end{scope}

\begin{scope}[shift={(146.00,25.43)}]
\begin{scope}[shift={(-31.749500,-28.614925)}]
\path[fill=fOneCloud]
  (27.65201,25.78170) .. controls (30.21288,24.80808) and (33.79827,24.95791) .. (35.84699,25.85664) .. controls (38.40797,26.15630) and (39.94458,27.20486) .. (40.71284,28.96502) .. controls (41.99341,31.39910) and (40.20069,33.08432) .. (37.63971,33.83337) .. controls (35.07872,35.14399) and (29.18857,35.14399) .. (26.37144,34.20782) .. controls (23.04216,33.64609) and (21.50559,31.21197) .. (22.78616,29.15215) .. controls (23.29831,27.27980) and (25.09103,26.34347) .. (27.65201,25.78170);

\end{scope}
\node[text=fOneText,font=\fontfamily{phv}\fontsize{6.4378bp}{7.5bp}\selectfont] at (0,1.36111) {Hidden network};
\fOneParty{A}{0.00000}{-9.52778}{0.00000}{-1.06944}{2.020mm}{2.08891mm}
\fOneParty{C}{-13.80556}{9.23611}{-0.93333}{0.53472}{1.971mm}{2.14998mm}
\fOneParty{B}{13.80556}{9.23611}{0.93333}{0.53472}{1.984mm}{2.04642mm}
\fOneWave{fOneIncoming}{-0.97222}{-7.58333}{-0.97222}{-1.91528}{0.376447}{1.880876}
\fOneDot{fOneIncomingDot}{-0.97222}{-7.58333}
\fOneLink{fOneAShare}{0.97222}{-7.58333}{0.97222}{-1.91528}
\fOneWave{fOneIncoming}{-12.60000}{7.42778}{-5.58056}{3.43194}{0.376447}{1.880876}
\fOneDot{fOneIncomingDot}{-12.60000}{7.42778}
\fOneLink{fOneCShare}{-11.63750}{9.11944}{-3.57778}{4.52083}
\fOneWave{fOneIncoming}{12.60000}{7.42778}{5.58056}{3.43194}{0.376447}{1.880876}
\fOneDot{fOneIncomingDot}{12.60000}{7.42778}
\fOneLink{fOneBShare}{11.63750}{9.11944}{3.57778}{4.52083}
\end{scope}

\fOneSummary{28.99}{fOneSum}{No difference}{}
  {Q-QIT and F-QIT always}{self-test the same strategies.}
\fOneSummary{88.22}{fOneThm}{Theorem~\fOneLoopTheorem:}{a self-test in Q-QIT only}
  {A correlation with unique Q-QIT strategy}{but with two inequivalent F-QIT strategies}
\fOneSummary{146.00}{fOneThm}{Theorem~\fOneRewiringTheorem:}{a genuine F-QIT self-test}
  {Correlations with no Q-QIT strategy}{but with unique F-QIT strategy}
\end{tikzpicture}%
\endgroup%

%% file: Figure2TikZ.tex
\begingroup%
\definecolor{fTwoA}{HTML}{2F6FB3}\definecolor{fTwoAFill}{HTML}{EAF2FB}\definecolor{fTwoAShare}{HTML}{337BBF}%
\definecolor{fTwoB}{HTML}{2CA25F}\definecolor{fTwoBFill}{HTML}{EAF7EF}\definecolor{fTwoBShare}{HTML}{28A56C}%
\definecolor{fTwoC}{HTML}{D95F5F}\definecolor{fTwoCFill}{HTML}{FBECEC}\definecolor{fTwoCShare}{HTML}{DD6165}%
\definecolor{fTwoIn}{HTML}{706D78}\definecolor{fTwoInDot}{HTML}{6C6C6C}\definecolor{fTwoStep}{HTML}{6F6F6F}%
\definecolor{fTwoOut}{HTML}{121010}\definecolor{fTwoCloud}{HTML}{B4B4B4}%
\definecolor{fTwoText}{HTML}{111111}\definecolor{fTwoLabel}{HTML}{333333}%
\newcommand{\fTwoSans}[1]{\fontfamily{phv}\fontsize{#1bp}{#1bp}\selectfont}%
\newcommand{\fTwoSerif}{\fontfamily{ptm}\fontsize{10bp}{10bp}\bfseries\itshape\selectfont}%
\newcommand{\fTwoBox}[7]{%
  \path[fill=fTwo#1Fill,draw=fTwo#1,line width=#7,opacity=0.862471,rounded corners=#6]
    (#2,#3) rectangle ++(#4,#5);}%
\newcommand{\fTwoPartyA}[3]{\fTwoBox{#1}{#2}{#3}{7.476}{7.481}{2.78869mm}{0.29787mm}}
\newcommand{\fTwoPartyB}[3]{\fTwoBox{#1}{#2}{#3}{6.62555}{6.63014}{2.31819mm}{0.26399mm}}
\newcommand{\fTwoLetter}[5]{\node[text=fTwo#1] at (#2,#3) {\resizebox{#4mm}{#5mm}{{\fTwoSerif #1}}};}%
\newcommand{\fTwoDot}[4]{\fill[#1] (#2,#3) circle[radius=#4mm];}%
\newcommand{\fTwoWave}[9]{%
  \pgfmathsetmacro{\fTwoDx}{#4-#2}\pgfmathsetmacro{\fTwoDy}{#5-#3}%
  \pgfmathsetmacro{\fTwoL}{sqrt(\fTwoDx*\fTwoDx+\fTwoDy*\fTwoDy)}%
  \pgfmathtruncatemacro{\fTwoS}{max(40,ceil(\fTwoL/#7*14))}%
  \draw[draw=#1,line width=#8] plot[domain=0:1,samples=\fTwoS,variable=\t]
    ({#2+\fTwoDx*\t-\fTwoDy/\fTwoL*#6*(#9*sin(180*\t)+1-#9)*sin(360*\fTwoL/#7*\t)},
     {#3+\fTwoDy*\t+\fTwoDx/\fTwoL*#6*(#9*sin(180*\t)+1-#9)*sin(360*\fTwoL/#7*\t)});}%
\newcommand{\fTwoLink}[5]{%
  \fTwoWave{#1}{#2}{#3}{#4}{#5}{0.4012}{2.0066}{0.21738mm}{1}%
  \fTwoDot{#1}{#2}{#3}{0.60157}\fTwoDot{#1}{#4}{#5}{0.60157}}%
\newcommand{\fTwoPath}[3]{\draw[draw=#1,line width=#2] plot coordinates {#3};}%
\begin{tikzpicture}[x=1mm,y=-1mm,line cap=round,line join=round,
  every node/.style={inner sep=0pt,outer sep=0pt},
  fTwoPanel/.style={fill=black!4,draw=black!18,line width=0.5pt,rounded corners=5pt},
  fTwoCard/.style={fill=white,draw=black!22,line width=0.5pt,rounded corners=3.5pt},
  fTwoTitle/.style={font=\fontfamily{phv}\fontsize{8.5pt}{10pt}\selectfont\bfseries,text=black},
  fTwoLegend/.style={font=\fontfamily{phv}\fontsize{7.5pt}{9pt}\selectfont,text=black}]
\path[use as bounding box] (0,0) rectangle (175,51.4);
\path[fTwoPanel] (0.1,0.1) rectangle (57.5,46.9);
\path[fTwoPanel] (59,0.1) rectangle (174.9,46.9);
\path[fTwoCard] (1.3,7.2) rectangle (56.3,45.7);
\path[fTwoCard] (60.2,7.2) rectangle (173.7,45.7);
\node[fTwoTitle,anchor=west] at (1.6,3.6) {(a)\enspace Rewiring game};
\node[fTwoTitle,anchor=west] at (60.6,3.6) {(b)\enspace Possible networks {\scalebox{0.85}{{\fontsize{10pt}{10pt}\selectfont $l\in\mathcal L$}}}};

\fTwoWave{black}{34.90}{49.90}{40.73}{49.90}{0.188}{0.88}{0.19039mm}{1}
\fTwoDot{black}{34.90}{49.90}{0.4384}\fTwoDot{black}{40.73}{49.90}{0.4384}
\node[fTwoLegend,anchor=west] at (42.33,49.90) {locally prepared bipartite state};
\draw[draw=fTwoOut,line width=0.23119mm] (85.04,49.90) -- (89.60,49.90);
\path[fill=fTwoOut,draw=fTwoOut,line width=0.23119mm] (88.90,49.55) -- (89.60,49.90) -- (88.90,50.25) -- cycle;
\fTwoDot{fTwoOut}{85.04}{49.90}{0.4384}
\node[fTwoLegend,anchor=west] at (91.20,49.90) {outgoing share};
\draw[draw=fTwoIn,line width=0.23119mm] (120.20,49.90) -- (115.64,49.90);
\path[fill=fTwoIn,draw=fTwoIn,line width=0.23119mm] (116.34,49.55) -- (115.64,49.90) -- (116.34,50.25) -- cycle;
\fTwoDot{fTwoInDot}{120.20}{49.90}{0.4384}
\node[fTwoLegend,anchor=west] at (121.80,49.90) {incoming share};

\begin{scope}[shift={(-2.53,-2.30)}]
\path[fill=fTwoCloud]
  (27.65201,25.78170) .. controls (30.21288,24.80808) and (33.79827,24.95791) .. (35.84699,25.85664) .. controls (38.40797,26.15630) and (39.94458,27.20486) .. (40.71284,28.96502) .. controls (41.99341,31.39910) and (40.20069,33.08432) .. (37.63971,33.83337) .. controls (35.07872,35.14399) and (29.18857,35.14399) .. (26.37144,34.20782) .. controls (23.04216,33.64609) and (21.50559,31.21197) .. (22.78616,29.15215) .. controls (23.29831,27.27980) and (25.09103,26.34347) .. (27.65201,25.78170);
\draw[draw=fTwoCShare,line width=0.19993mm] (23.50291,35.13839) -- (26.14907,32.49045);
\path[fill=fTwoCShare,draw=fTwoCShare,line width=0.20167mm] (26.02411,32.28924) -- (26.54717,32.09380) -- (26.39409,32.62237) -- cycle;
\draw[draw=fTwoBShare,line width=0.19993mm] (39.31235,34.61691) -- (36.66619,31.96902);
\path[fill=fTwoBShare,draw=fTwoBShare,line width=0.15560mm] (36.62800,32.31948) -- (36.45034,31.74648) -- (37.02054,31.95007) -- cycle;
\draw[draw=fTwoAShare,line width=0.22102mm] (32.73256,23.09341) -- (32.72633,27.13827);
\path[fill=fTwoAShare,draw=fTwoAShare,line width=0.14760mm] (33.01357,26.76531) -- (32.73686,27.30350) -- (32.46217,26.76426) -- cycle;
\draw[draw=fTwoIn,line width=0.14511mm] (21.75300,34.60193) -- (19.74276,36.63423);
\path[fill=fTwoIn,draw=fTwoIn,line width=0.14326mm] (20.11576,36.57218) -- (19.65408,36.73342) -- (19.80405,36.25948) -- cycle;
\draw[draw=fTwoIn,line width=0.15090mm] (41.42958,34.28284) -- (43.37036,36.38161);
\path[fill=fTwoIn,draw=fTwoIn,line width=0.14682mm] (43.32878,36.01252) -- (43.47003,36.48082) -- (43.00319,36.31086) -- cycle;
\draw[draw=fTwoIn,line width=0.19991mm] (31.24452,22.53566) -- (31.23892,19.67615);
\path[fill=fTwoIn,draw=fTwoIn,line width=0.19912mm] (31.02304,20.09473) -- (31.23700,19.65470) -- (31.46457,20.09666) -- cycle;
\fTwoPartyA{A}{28.20074}{11.81684}
\fTwoPartyA{C}{13.44903}{36.66695}
\fTwoPartyA{B}{41.62711}{36.73787}
\fTwoPath{fTwoCShare}{0.19583mm}{
    (19.37,39.10) (19.59,39.11) (19.71,39.09) (19.77,39.05) (19.82,39.00) (19.85,38.92) (19.86,38.79)
    (19.86,38.32) (19.89,38.25) (19.94,38.19) (20.01,38.16) (20.10,38.14) (20.60,38.15) (20.68,38.13)
    (20.75,38.09) (20.80,38.03) (20.82,37.95) (20.83,37.82) (20.83,37.36) (20.86,37.29) (20.91,37.23)
    (20.99,37.20) (21.08,37.18) (21.54,37.20) (21.63,37.18) (21.71,37.15) (21.76,37.10) (21.79,37.02)
    (21.80,36.93) (21.79,36.49) (21.81,36.37) (21.85,36.31) (21.91,36.26) (21.99,36.23) (22.12,36.23)
    (22.58,36.23) (22.66,36.20) (22.72,36.16) (22.75,36.09) (22.77,36.00) (22.77,35.50) (22.79,35.41)
    (22.83,35.35) (22.89,35.30) (22.97,35.28) (23.10,35.27) (23.25,35.28)}
\fTwoPath{fTwoBShare}{0.19583mm}{
    (43.33,38.61) (43.33,38.40) (43.31,38.28) (43.27,38.21) (43.21,38.16) (43.13,38.14) (43.00,38.13)
    (42.54,38.13) (42.47,38.11) (42.41,38.06) (42.37,37.99) (42.35,37.90) (42.36,37.40) (42.34,37.31)
    (42.30,37.25) (42.24,37.20) (42.16,37.18) (42.02,37.17) (41.56,37.18) (41.49,37.15) (41.43,37.10)
    (41.40,37.03) (41.38,36.94) (41.39,36.47) (41.37,36.38) (41.34,36.31) (41.28,36.26) (41.21,36.23)
    (41.11,36.22) (40.68,36.23) (40.56,36.21) (40.49,36.17) (40.45,36.11) (40.42,36.04) (40.41,35.90)
    (40.41,35.44) (40.38,35.37) (40.33,35.31) (40.26,35.28) (40.17,35.26) (39.67,35.27) (39.58,35.25)
    (39.52,35.21) (39.47,35.15) (39.45,35.07) (39.44,34.94) (39.44,34.79)}
\fTwoPath{fTwoStep}{0.19583mm}{
    (24.52,32.20) (24.05,32.18) (23.96,32.20) (23.89,32.23) (23.84,32.28) (23.81,32.36) (23.80,32.45)
    (23.81,32.89) (23.78,33.01) (23.74,33.07) (23.68,33.12) (23.60,33.15) (23.47,33.15) (23.01,33.15)
    (22.94,33.17) (22.88,33.22) (22.84,33.29) (22.83,33.38) (22.83,33.88) (22.81,33.97) (22.77,34.03)
    (22.71,34.08) (22.63,34.10) (22.49,34.11) (22.34,34.10)}
\fTwoPath{fTwoStep}{0.19583mm}{
    (39.00,31.54) (38.98,32.01) (38.99,32.10) (39.03,32.17) (39.08,32.22) (39.16,32.25) (39.25,32.26)
    (39.69,32.25) (39.81,32.28) (39.87,32.31) (39.92,32.37) (39.94,32.45) (39.95,32.59) (39.94,33.05)
    (39.97,33.12) (40.02,33.18) (40.09,33.22) (40.18,33.23) (40.68,33.23) (40.77,33.25) (40.83,33.29)
    (40.88,33.35) (40.90,33.43) (40.91,33.57) (40.90,33.71)}
\fTwoPath{fTwoStep}{0.13916mm}{
    (31.09,26.28) (31.43,25.96) (31.48,25.89) (31.51,25.81) (31.50,25.74) (31.47,25.66) (31.41,25.59)
    (31.10,25.29) (31.03,25.19) (31.01,25.11) (31.02,25.04) (31.06,24.96) (31.14,24.86) (31.47,24.54)
    (31.51,24.47) (31.51,24.39) (31.49,24.32) (31.44,24.24) (31.08,23.89) (31.03,23.82) (31.01,23.74)
    (31.03,23.67) (31.07,23.59) (31.15,23.49) (31.27,23.40)}
\fTwoWave{fTwoAShare}{32.70470}{18.64047}{32.76100}{22.90003}{0.2509}{1.06489}{0.17685mm}{0}
\foreach \c/\x/\y in {fTwoAShare/32.7047/18.2074, fTwoAShare/32.76097/22.90002,
    fTwoCShare/19.36870/39.10255, fTwoCShare/23.22969/35.40288, fTwoInDot/22.13671/34.31812,
    fTwoBShare/43.32751/38.61371, fTwoBShare/39.76194/35.05568, fTwoInDot/41.11577/33.92281,
    fTwoInDot/31.24315/22.95161}{\fTwoDot{\c}{\x}{\y}{0.62627}}
\fTwoLetter{B}{45.58070}{41.57618}{1.99569}{2.05865}
\fTwoLetter{A}{31.84005}{14.63634}{2.03227}{2.10125}
\fTwoLetter{C}{16.63633}{41.27474}{1.98311}{2.16289}
\foreach \c/\x/\y/\g/\s in {fTwoA/28.99990/17.70960/\alpha/R, fTwoAShare/33.81430/23.30590/\alpha/L,
    fTwoCShare/16.02680/38.85220/\gamma/R, fTwoCShare/24.24030/36.09030/\gamma/L}{%
  \node[anchor=base west,text=\c] at (\x,\y) {\scalebox{0.6}[0.6879]{{\fontsize{10bp}{10bp}\selectfont $\g$}}};
  \node[anchor=base west,text=\c] at (\x+1.3634,\y+0.4855) {\scalebox{0.44715}[0.44715]{{\fontsize{10bp}{10bp}\selectfont $\s$}}};}
\foreach \x/\y/\s in {44.24870/38.99050/R, 36.61430/36.90770/L}{%
  \node[anchor=base west,text=fTwoBShare] at (\x,\y) {\scalebox{0.6}[0.6879]{{\fontsize{10bp}{10bp}\selectfont $\beta$}}};
  \node[anchor=base west,text=fTwoBShare] at (\x+1.23,\y+0.4855) {\scalebox{0.44715}[0.44715]{{\fontsize{10bp}{10bp}\selectfont $\s$}}};}
\node[anchor=base west,text=fTwoText,font={\fTwoSans{6.4378}}] at (23.98518,29.91564) {Hidden network};
\node at (31.36,31.58772) {\resizebox{1.11mm}{1.81864mm}{$l$}};
\end{scope}

\begin{scope}[cm={0.938,0,0,0.938,(9.66,0.10)},transform shape]
\fTwoPartyB{A}{65.55392}{21.03578}\fTwoPartyB{C}{55.92737}{33.37644}\fTwoPartyB{B}{74.74537}{33.51328}
\fTwoLink{fTwoA}{70.53275}{26.33649}{76.92465}{34.95228}
\fTwoLink{fTwoB}{76.08369}{36.90059}{61.37891}{36.90059}
\fTwoLink{fTwoC}{60.53781}{34.95228}{66.93000}{26.33649}
\fTwoPartyB{C}{102.53256}{33.53681}\fTwoPartyB{A}{93.09209}{20.95982}\fTwoPartyB{B}{83.69441}{33.35927}
\fTwoLink{fTwoA}{98.20282}{26.32346}{104.59442}{34.93924}
\fTwoLink{fTwoC}{103.75373}{36.88755}{89.04899}{36.88755}
\fTwoLink{fTwoB}{88.20773}{34.93924}{94.60008}{26.32346}
\fTwoPartyB{C}{117.13571}{32.40695}\fTwoPartyB{B}{123.54674}{23.26587}\fTwoPartyB{A}{111.48129}{23.22190}
\fTwoPath{fTwoA}{0.21738mm}{
    (116.50,25.44) (116.62,25.37) (117.00,25.21) (117.12,25.15) (117.23,25.07) (117.33,24.98)
    (117.60,24.65) (117.69,24.56) (117.80,24.48) (117.92,24.43) (118.06,24.42) (118.20,24.43)
    (118.35,24.46) (118.64,24.55) (118.77,24.58) (118.90,24.57) (119.01,24.54) (119.11,24.47)
    (119.20,24.36) (119.46,23.94) (119.56,23.81) (119.66,23.72) (119.78,23.67) (119.90,23.66)
    (120.03,23.71) (120.16,23.79) (120.30,23.90) (120.55,24.14) (120.66,24.24) (120.78,24.31)
    (120.89,24.33) (121.00,24.31) (121.10,24.25) (121.22,24.14) (121.46,23.89) (121.58,23.77)
    (121.71,23.67) (121.84,23.62) (121.97,23.61) (122.08,23.65) (122.19,23.73) (122.30,23.85)
    (122.57,24.28) (122.66,24.40) (122.75,24.48) (122.86,24.52) (122.98,24.53) (123.11,24.50)
    (123.55,24.33) (123.70,24.30) (123.84,24.29) (123.96,24.32) (124.08,24.38) (124.18,24.48)
    (124.44,24.83) (124.53,24.93) (124.63,25.02) (124.74,25.09) (124.86,25.14) (125.25,25.27)
    (125.37,25.33) (125.49,25.40) (125.60,25.49)}
\fTwoPath{fTwoB}{0.21738mm}{
    (125.52,27.66) (125.41,27.73) (125.03,27.90) (124.92,27.96) (124.81,28.03) (124.71,28.12)
    (124.45,28.45) (124.35,28.55) (124.25,28.63) (124.13,28.68) (124.00,28.71) (123.86,28.70)
    (123.72,28.67) (123.42,28.58) (123.29,28.55) (123.16,28.55) (123.05,28.58) (122.95,28.64)
    (122.86,28.74) (122.61,29.16) (122.52,29.29) (122.42,29.39) (122.31,29.46) (122.19,29.48)
    (122.06,29.45) (121.93,29.39) (121.80,29.29) (121.43,28.94) (121.32,28.86) (121.21,28.81)
    (121.10,28.81) (120.99,28.86) (120.89,28.94) (120.53,29.30) (120.40,29.41) (120.28,29.49)
    (120.15,29.53) (120.03,29.52) (119.92,29.46) (119.81,29.36) (119.72,29.23) (119.45,28.82)
    (119.36,28.71) (119.27,28.64) (119.16,28.61) (119.04,28.61) (118.90,28.65) (118.47,28.80)
    (118.33,28.82) (118.19,28.82) (118.07,28.78) (117.96,28.71) (117.86,28.62) (117.61,28.28)
    (117.52,28.18) (117.41,28.09) (117.30,28.03) (117.18,27.98) (116.80,27.83) (116.68,27.77)
    (116.57,27.69)}
\fTwoPath{fTwoC}{0.21738mm}{
    (122.42,34.42) (122.59,34.34) (123.06,34.17) (123.19,34.09) (123.56,33.81) (123.69,33.74)
    (123.82,33.69) (123.95,33.68) (124.08,33.70) (124.21,33.75) (124.32,33.82) (124.61,34.04)
    (124.69,34.10) (124.77,34.13) (124.86,34.14) (124.95,34.13) (125.27,34.06) (125.40,34.04)
    (125.52,34.04) (125.64,34.05) (125.75,34.08) (125.84,34.13) (125.92,34.20) (125.97,34.29)
    (126.00,34.39) (126.00,34.51) (125.98,34.63) (125.95,34.75) (125.76,35.17) (125.74,35.25)
    (125.74,35.32) (125.76,35.39) (125.80,35.46) (125.86,35.53) (126.11,35.78) (126.20,35.88)
    (126.28,35.98) (126.33,36.09) (126.36,36.20) (126.36,36.30) (126.33,36.40) (126.28,36.49)
    (126.19,36.57) (126.09,36.63) (125.98,36.67) (125.85,36.71) (125.41,36.78) (125.34,36.82)
    (125.27,36.86) (125.23,36.92) (125.18,37.00) (125.06,37.33) (124.99,37.44) (124.92,37.55)
    (124.83,37.63) (124.72,37.69) (124.60,37.72) (124.47,37.72) (124.34,37.68) (124.21,37.62)
    (123.96,37.47) (123.83,37.41) (123.70,37.37) (123.57,37.35) (123.10,37.37) (122.93,37.35)
    (122.77,37.31) (122.60,37.23) (122.44,37.14)}
\fTwoPartyB{C}{144.88814}{22.75493}\fTwoPartyB{B}{138.13375}{32.35264}\fTwoPartyB{A}{132.49577}{22.98709}
\fTwoPath{fTwoA}{0.21738mm}{
    (137.83,25.16) (137.94,25.09) (138.31,24.92) (138.43,24.86) (138.53,24.78) (138.63,24.70)
    (138.97,24.27) (139.07,24.18) (139.19,24.12) (139.31,24.10) (139.45,24.10) (139.59,24.13)
    (140.02,24.24) (140.15,24.26) (140.26,24.24) (140.36,24.19) (140.45,24.10) (140.53,23.98)
    (140.78,23.56) (140.88,23.44) (140.98,23.36) (141.10,23.32) (141.22,23.32) (141.34,23.37)
    (141.47,23.45) (141.60,23.56) (141.84,23.80) (141.96,23.90) (142.06,23.96) (142.17,23.99)
    (142.27,23.98) (142.38,23.92) (142.49,23.83) (142.84,23.47) (142.97,23.37) (143.09,23.31)
    (143.21,23.28) (143.33,23.30) (143.44,23.36) (143.54,23.46) (143.64,23.59) (143.89,24.00)
    (143.98,24.09) (144.08,24.16) (144.18,24.19) (144.30,24.19) (144.43,24.16) (144.86,24.03)
    (145.00,24.01) (145.13,24.02) (145.25,24.05) (145.36,24.12) (145.45,24.21) (145.71,24.53)
    (145.80,24.63) (145.90,24.71) (146.01,24.78) (146.49,24.99) (146.61,25.06)}
\fTwoPath{fTwoC}{0.21738mm}{
    (146.67,27.29) (146.56,27.36) (146.19,27.52) (146.07,27.58) (145.97,27.66) (145.87,27.75)
    (145.61,28.07) (145.52,28.17) (145.42,28.26) (145.31,28.31) (145.18,28.34) (145.04,28.34)
    (144.90,28.31) (144.47,28.19) (144.35,28.18) (144.23,28.20) (144.13,28.25) (144.04,28.34)
    (143.96,28.45) (143.70,28.87) (143.61,28.99) (143.50,29.07) (143.39,29.11) (143.27,29.10)
    (143.14,29.05) (143.01,28.97) (142.88,28.86) (142.64,28.62) (142.53,28.52) (142.42,28.46)
    (142.31,28.43) (142.21,28.45) (142.10,28.51) (141.99,28.60) (141.64,28.96) (141.51,29.06)
    (141.38,29.12) (141.26,29.14) (141.15,29.12) (141.04,29.05) (140.94,28.94) (140.84,28.81)
    (140.67,28.53) (140.59,28.41) (140.50,28.31) (140.40,28.25) (140.29,28.22) (140.17,28.23)
    (140.04,28.26) (139.75,28.36) (139.61,28.40) (139.47,28.41) (139.34,28.40) (139.22,28.36)
    (139.11,28.29) (139.02,28.19) (138.76,27.86) (138.67,27.77) (138.57,27.69) (138.46,27.62)
    (138.09,27.47) (137.97,27.41) (137.86,27.34)}
\fTwoPath{fTwoB}{0.21738mm}{
    (143.42,34.24) (143.58,34.17) (144.06,34.02) (144.19,33.95) (144.57,33.68) (144.70,33.61)
    (144.83,33.57) (144.96,33.56) (145.09,33.58) (145.21,33.63) (145.32,33.70) (145.68,34.00)
    (145.76,34.04) (145.84,34.06) (145.93,34.06) (146.37,33.99) (146.50,33.99) (146.62,34.00)
    (146.73,34.03) (146.82,34.08) (146.90,34.15) (146.96,34.24) (146.99,34.34) (146.99,34.45)
    (146.97,34.57) (146.94,34.69) (146.89,34.81) (146.74,35.11) (146.71,35.19) (146.69,35.27)
    (146.71,35.34) (146.74,35.41) (146.79,35.48) (147.04,35.73) (147.12,35.83) (147.20,35.93)
    (147.25,36.04) (147.29,36.15) (147.29,36.26) (147.27,36.36) (147.22,36.45) (147.14,36.52)
    (147.05,36.59) (146.93,36.63) (146.81,36.67) (146.36,36.73) (146.28,36.76) (146.22,36.79)
    (146.16,36.85) (146.12,36.92) (145.98,37.21) (145.92,37.33) (145.85,37.43) (145.76,37.52)
    (145.66,37.59) (145.54,37.63) (145.42,37.63) (145.30,37.60) (145.17,37.55) (144.80,37.33)
    (144.68,37.28) (144.55,37.24) (144.41,37.23) (144.10,37.23) (143.93,37.21) (143.77,37.17)
    (143.61,37.09) (143.46,37.00)}
\fTwoPartyB{C}{166.67724}{22.79150}\fTwoPartyB{B}{153.83701}{22.93717}\fTwoPartyB{A}{159.92616}{32.41571}
\fTwoPath{fTwoB}{0.21738mm}{
    (159.12,25.06) (159.24,24.99) (159.62,24.82) (159.74,24.76) (159.85,24.69) (159.95,24.59)
    (160.22,24.26) (160.32,24.17) (160.43,24.09) (160.56,24.05) (160.69,24.04) (160.84,24.05)
    (161.28,24.17) (161.42,24.20) (161.54,24.19) (161.65,24.15) (161.75,24.07) (161.84,23.95)
    (162.10,23.53) (162.20,23.41) (162.31,23.32) (162.43,23.28) (162.56,23.29) (162.69,23.35)
    (162.82,23.44) (163.21,23.80) (163.33,23.89) (163.44,23.94) (163.55,23.95) (163.66,23.91)
    (163.77,23.82) (164.13,23.45) (164.26,23.33) (164.39,23.25) (164.52,23.22) (164.64,23.23)
    (164.76,23.29) (164.86,23.40) (164.96,23.53) (165.15,23.83) (165.24,23.96) (165.33,24.06)
    (165.43,24.12) (165.55,24.14) (165.68,24.12) (165.82,24.07) (166.11,23.95) (166.27,23.90)
    (166.41,23.88) (166.54,23.89) (166.66,23.94) (166.77,24.02) (166.86,24.13) (167.12,24.49)
    (167.22,24.59) (167.32,24.66) (167.44,24.72) (167.84,24.84) (167.97,24.89) (168.09,24.95)
    (168.21,25.03) (168.32,25.12)}
\fTwoPath{fTwoC}{0.21738mm}{
    (168.24,27.26) (168.12,27.33) (167.75,27.50) (167.63,27.56) (167.52,27.64) (167.42,27.73)
    (167.15,28.06) (167.06,28.16) (166.95,28.23) (166.83,28.28) (166.69,28.30) (166.55,28.29)
    (166.40,28.25) (166.11,28.17) (165.97,28.14) (165.85,28.15) (165.74,28.18) (165.64,28.26)
    (165.55,28.37) (165.29,28.79) (165.20,28.92) (165.09,29.01) (164.98,29.06) (164.85,29.06)
    (164.72,29.02) (164.59,28.94) (164.46,28.83) (164.20,28.59) (164.09,28.49) (163.97,28.42)
    (163.86,28.40) (163.76,28.42) (163.65,28.49) (163.54,28.59) (163.30,28.85) (163.17,28.97)
    (163.04,29.07) (162.92,29.12) (162.79,29.13) (162.67,29.09) (162.56,29.01) (162.46,28.89)
    (162.19,28.46) (162.10,28.35) (162.00,28.26) (161.89,28.22) (161.78,28.22) (161.65,28.25)
    (161.21,28.41) (161.06,28.45) (160.92,28.46) (160.80,28.43) (160.68,28.37) (160.58,28.28)
    (160.49,28.17) (160.32,27.93) (160.23,27.82) (160.13,27.73) (160.02,27.66) (159.90,27.61)
    (159.51,27.48) (159.39,27.42) (159.27,27.35) (159.16,27.27)}
\fTwoPath{fTwoA}{0.21738mm}{
    (165.26,34.32) (165.43,34.24) (165.75,34.14) (165.90,34.08) (166.04,34.01) (166.42,33.73)
    (166.55,33.66) (166.68,33.62) (166.81,33.62) (166.95,33.64) (167.07,33.70) (167.19,33.77)
    (167.47,34.00) (167.55,34.05) (167.64,34.08) (167.72,34.09) (167.82,34.08) (168.15,34.01)
    (168.28,34.00) (168.40,33.99) (168.52,34.01) (168.63,34.05) (168.72,34.10) (168.79,34.18)
    (168.83,34.28) (168.85,34.39) (168.84,34.51) (168.82,34.63) (168.78,34.75) (168.60,35.16)
    (168.59,35.24) (168.60,35.31) (168.63,35.38) (168.68,35.45) (168.93,35.69) (169.10,35.88)
    (169.17,35.99) (169.21,36.09) (169.23,36.20) (169.22,36.30) (169.18,36.40) (169.11,36.48)
    (169.02,36.55) (168.91,36.61) (168.79,36.65) (168.33,36.74) (168.25,36.76) (168.18,36.80)
    (168.13,36.85) (168.09,36.92) (167.97,37.22) (167.92,37.33) (167.85,37.44) (167.77,37.54)
    (167.68,37.62) (167.57,37.67) (167.45,37.69) (167.33,37.68) (167.20,37.63) (167.07,37.57)
    (166.83,37.42) (166.71,37.36) (166.58,37.32) (166.44,37.30) (165.98,37.31) (165.82,37.29)
    (165.65,37.23) (165.50,37.16) (165.34,37.06)}
\foreach \c/\x/\y in {fTwoA/116.50334/25.44235, fTwoA/125.59944/25.48928, fTwoB/125.52450/27.65913,
    fTwoB/116.57323/27.68756, fTwoC/122.42339/34.41659, fTwoC/122.44235/37.14416,
    fTwoA/137.82803/25.15690, fTwoA/146.60785/25.06112, fTwoC/146.67149/27.28851,
    fTwoC/137.85630/27.34093, fTwoB/143.42128/34.23772, fTwoB/143.45652/37.00437,
    fTwoB/159.11786/25.05608, fTwoB/168.31566/25.11972, fTwoC/168.24270/27.26009,
    fTwoC/159.16245/27.26630, fTwoA/165.26060/34.31577, fTwoA/165.33687/37.06464}{%
  \fTwoDot{\c}{\x}{\y}{0.60157}}
\fTwoLetter{A}{68.73137}{24.10872}{2.15352}{2.22698}
\fTwoLetter{C}{58.70533}{36.98417}{2.10170}{2.29209}
\fTwoLetter{B}{78.64859}{37.16821}{2.11505}{2.18168}
\fTwoLetter{A}{96.40139}{24.09568}{2.15366}{2.22698}
\fTwoLetter{B}{86.37548}{36.97115}{2.11485}{2.18168}
\fTwoLetter{C}{106.42742}{36.97115}{2.10156}{2.29213}
\fTwoLetter{A}{114.23292}{26.63684}{2.09848}{2.16982}
\fTwoLetter{B}{127.88503}{26.65454}{2.06069}{2.12555}
\fTwoLetter{C}{120.16911}{35.52288}{2.04780}{2.23334}
\fTwoLetter{B}{141.26360}{35.54679}{2.06070}{2.12570}
\fTwoLetter{C}{148.89367}{26.29192}{2.04780}{2.23334}
\fTwoLetter{A}{135.48578}{26.18746}{2.09848}{2.16982}
\fTwoLetter{A}{163.01728}{35.46655}{2.09844}{2.16982}
\fTwoLetter{C}{170.60892}{26.26963}{2.04780}{2.23350}
\fTwoLetter{B}{156.82575}{26.20819}{2.06068}{2.12570}
\foreach \cx/\cy/\w/\h/\lab/\lx/\ly in {69.03970/16.41683/7.36125/2.52725/ABC/60.960/17.80962,
    96.79425/16.40380/7.27882/2.52725/ACB/88.460/17.78369,
    121.21908/16.45555/4.65534/2.34110/AB/114.200/17.78369,
    142.13538/16.58414/4.73409/2.40967/AC/135.080/17.78369,
    163.74625/16.60324/4.58692/2.40967/BC/156.760/17.78369}{%
  \node[text=fTwoLabel] at (\cx,\cy) {\resizebox{\w mm}{\h mm}{{\fTwoSerif\mdseries \lab}}};
  \node[anchor=south west] at (\lx,\ly) {\resizebox{1.55mm}{2.73496mm}{$l$}};
  \node[anchor=base west,font=\fontfamily{ptm}\fontsize{8.221bp}{8.221bp}\selectfont] at (\lx+1.973,\ly+0.3355) {=};}
\end{scope}
\end{tikzpicture}%
\endgroup%

%% file: RoadmapTikZ.tex
\usetikzlibrary{arrows.meta,fit,calc}%
\begingroup%
\definecolor{rmOkFill}{HTML}{DCEFF1}%
\definecolor{rmOkDraw}{HTML}{5FA8B1}%
\definecolor{rmAltFill}{HTML}{FCE4F0}%
\definecolor{rmAltDraw}{HTML}{C93D82}%
\begin{tikzpicture}[
  x=1cm,y=1cm,
  box/.style    = {rectangle,rounded corners=3pt,draw=black!70,line width=0.7pt,
                   fill=white,align=center,font=\footnotesize,inner xsep=5pt,inner ysep=4pt},
  okbox/.style  = {box,draw=rmOkDraw,line width=1pt,fill=rmOkFill,font=\scriptsize},
  badbox/.style = {box,draw=rmAltDraw,line width=1pt,fill=rmAltFill,font=\scriptsize},
  proves/.style = {-{Stealth[length=2.4mm,width=1.9mm]},line width=0.9pt,black!80},
]

\node[box] (A2) at (0.4,0.85)
  {\textbf{Thm.~\ref{thm:exact-fermionic-TC}:} fermionic\\ token-counting self-testing};
\node[box] (GHZ) at (0.4,-0.85)
  {\textbf{Lemma~\ref{lem:fermionic_Bell_relations}:} fermionic\\
   GHZ self-testing from\\ statistics conditioned on\\ one token per party};

\node[align=center,font=\footnotesize,inner sep=0pt] (A1head) at (5.4,0.55)
  {\textbf{Theorem~\ref{thm:three-party-loop-informal} (Theorem~\ref{thm:fixed-loop-characterization})}\\
   characterizes a single fixed loop\\
   of size 3 in one of two cases,\\
   inequivalent under\\ parity-preserving isometries};
\node[okbox]  (ci)  at (4.4,-0.95) {\textbf{case (i)}\\ single mode};
\node[badbox] (cii) at (6.45,-0.95) {\textbf{case (ii)}\\ double mode};
\node[draw=black!45,dashed,rounded corners=4pt,inner sep=4pt,fit=(ci)(cii)] (grp) {};
\node[box,fill=none,inner sep=8pt,fit=(A1head)(grp)] (A1) {};

\coordinate (rightcol) at (11.2,0);
\node[box] (A10) at (A1.center -| rightcol)
  {\textbf{Thm.~\ref{thm:fixed-loop-characterization-N2}:} same\\
   characterization as Thm.~\ref{thm:fixed-loop-characterization}\\
   for the fixed 2-loops $AB$ and\\ $AC$, under structural assumptions};

\coordinate (mid) at ($(A1.west)!0.5!(A10.east)$);
\coordinate (lowrow) at (0,-2.85);

\node[box,draw=rmOkDraw,line width=1.2pt,fill=rmOkFill,minimum width=7cm] (T1) at (mid |- lowrow)
  {\textbf{Theorem~\ref{thm:fermionic-self-testing} (Theorem~\ref{thm:fermionic-self-testing-formal}):}\\
   by consistency across all the loops,\\
   rules out double-mode realization};

\draw[proves] (A2.east)  -- (A2.east  -| A1.west);
\draw[proves] (GHZ.east) -- (GHZ.east -| A1.west);
\draw[proves] (A1.east)  -- (A10.west);
\draw[proves] (A1.south)  -- (A1.south  |- T1.north);
\draw[proves] (A10.south) -- (A10.south |- T1.north);
\end{tikzpicture}%
\endgroup%

%% file: DefinitionsSuppMat.tex
\onecolumngrid
\section{Self-testing definitions and CHSH examples}
\label{sec:TheoreticalDefs}

We first recall the definition of self-testing in Q-QIT and illustrate it with the paradigmatic example of CHSH self-testing. We use the pure-state formulation throughout this appendix.
\begin{definition}[Q-QIT self-testing] 
The correlations $p(a,b\,|\, x,y)$ self-test a
reference strategy $(\ket{\psi'},\{M'_{a\mid x}\},\{N'_{b\mid y}\})$
if every realization $(\ket{\psi},\{M_{a\mid x}\},\{N_{b\mid y}\})$,
consisting of a normalized state $\ket\psi\in\mathcal H_A\otimes\mathcal H_B$
and local POVMs satisfying
$p(a,b\mid x,y)=\bra\psi(M_{a\mid x}\otimes N_{b\mid y})\ket\psi$,
admits local isometries
\begin{equation}
\begin{aligned}
  \Phi_A&:\mathcal H_A\longrightarrow\mathcal H_{A'}\otimes\mathcal H_{\bar A},\\
  \Phi_B&:\mathcal H_B\longrightarrow\mathcal H_{B'}\otimes\mathcal H_{\bar B},
\end{aligned}
\end{equation}
such that
\begin{equation}\label{eq:stdef-state}
  (\Phi_A\otimes\Phi_B)\ket{\psi}_{AB}
  =\ket{\psi'}_{A'B'}\otimes\ket{\xi}_{\bar A\bar B},
\end{equation}
where $\ket{\xi}_{\bar A\bar B}$ is a normalized auxiliary
\emph{junk} state, and, for every $a,b,x,y$,
\begin{equation}\label{eq:stdef-meas}
(\Phi_A\otimes\Phi_B)\big(M_{a\mid x}\otimes N_{b\mid y}\big)\ket{\psi}_{AB}=\big(M'_{a\mid x}\otimes N'_{b\mid y}\big)
\ket{\psi'}_{A'B'}\otimes\ket{\xi}_{\bar A\bar B}.
\end{equation}
The same isometries and junk state must satisfy both
\cref{eq:stdef-state,eq:stdef-meas} for all inputs and outputs.
\end{definition}

The correlations generated by the trusted strategy in
\cref{eq:CHSHstrategy} self-test its state and measurements,
as shown below.

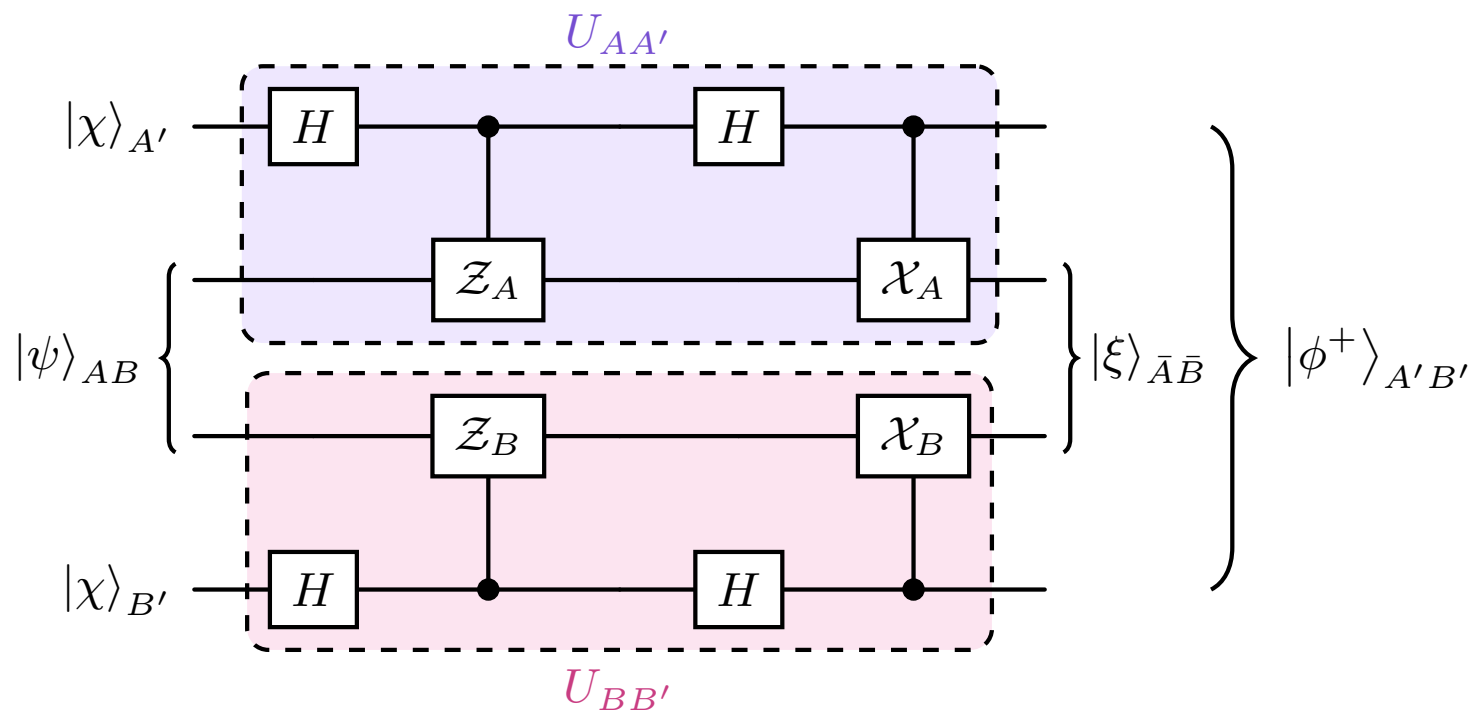
\begin{figure}
    \centering
    \definecolor{swapPurple}{HTML}{8B5CF6}
    \definecolor{swapPink}{HTML}{EC4899}
    \usetikzlibrary{decorations.pathreplacing}
    \begin{adjustbox}{max width=\columnwidth}
    \begin{quantikz}[execute at end picture={
  \draw[decoration={brace,amplitude=8pt},decorate,thick]
    ([xshift=1.10cm]\tikzcdmatrixname-1-7.east) --
    ([xshift=1.10cm]\tikzcdmatrixname-4-7.east);
  \node[right] at
    ([xshift=1.45cm]$(\tikzcdmatrixname-1-7.east)!0.5!(\tikzcdmatrixname-4-7.east)$)
    {$\ket{\phi^{+}}_{A'B'}$};
}]
\lstick{$\ket{\chi}_{A'}$}
    & \gate{H}
      \gategroup[2,steps=5,style={dashed,rounded corners,fill=swapPurple!15,
                 inner xsep=2pt,inner ysep=1pt},background,
                 label style={label position=north,anchor=south,yshift=-0.25cm}]{\textcolor{swapPurple!85!black}{$U_{AA'}$}}
    & \ctrl{1} & & \gate{H} & \ctrl{1} & \\
  \lstick[2]{$\ket{\psi}_{AB}$}
    & & \gate{\mathcal{Z}_A} & & & \gate{\mathcal{X}_A} & \rstick[2]{$\ket{\xi}_{\bar A\bar B}$} \\
    & \gategroup[2,steps=5,style={dashed,rounded corners,fill=swapPink!15,
                 inner xsep=1pt,inner ysep=1pt},background,
                 label style={label position=south,anchor=north,yshift=-0.2cm}]{\textcolor{swapPink!85!black}{$U_{BB'}$}}
    & \gate{\mathcal{Z}_B} & & & \gate{\mathcal{X}_B} & \\
  \lstick{$\ket{\chi}_{B'}$}
    & \gate{H} & \ctrl{-1} & & \gate{H} & \ctrl{-1} &
\end{quantikz}
    \end{adjustbox}
    \caption{The SWAP construction in Q-QIT and F-QIT, using the operators $\mathcal X_i,\mathcal Z_i$ defined in \cref{eq:swap-observables}. In Q-QIT, each party appends one ancillary qubit, initialized in $\ket{\chi}=\ket0$, and the extracted state is $\ket{\phi^+}=\ket{\phi_{\mathrm q}^{+}}=(\ket{0, 0}+\ket{1, 1})/\sqrt2$. In F-QIT, each party appends two febits, initialized in $\ket{\chi}=\ket{00}$, and the extracted state is $\ket{\phi^+}=\ket{\phi_{\mathrm f}^{+}}=(\ket{00,00}+\ket{11,11})/\sqrt2$. The Hadamard gates and the controls then act on the even subspace spanned by $\ket{00}$ and $\ket{11}$. The local unitaries $U_{AA'}$ and $U_{BB'}$, together with the initialized ancillas, define the extraction isometries.}
    \label{fig:swap_qubit_vs_fermion}
\end{figure}

\paragraph*{\normalfont\bfseries Proof of CHSH self-testing in Q-QIT.}
Consider an arbitrary realization with inputs $x,y\in\{0,1\}$
and outcomes $a,b\in\{\pm1\}$, where $M_{a \mid x}$ and $N_{b \mid y}$ are projectors. Define the Hermitian dichotomic observables
$A_x=M_{+1\mid x}-M_{-1\mid x}$ and
$B_y=N_{+1\mid y}-N_{-1\mid y}$.
Let \(\mathcal B=A_0\otimes(B_0+B_1)+A_1\otimes(B_0-B_1)\)
be the CHSH operator, so that
\(\beta=\bra{\psi}\mathcal B\ket{\psi}\).
Writing \(\beta_Q=2\sqrt2\), we start from the sum-of-squares bound
\begin{equation}\label{eq:sos}
  \beta_Q\id-\mathcal B=\frac1{\sqrt2}\Big(A_0-\frac{B_0+B_1}{\sqrt2}\Big)^{\!2}\quad+\frac1{\sqrt2}\Big(A_1-\frac{B_0-B_1}{\sqrt2}\Big)^{\!2}
  \geq 0
\end{equation}

where $A_x$ and $B_y$ stand for $A_x\otimes\id_B$ and
$\id_A\otimes B_y$, respectively; tensor products with identity
operators are left implicit here and below. At $\langle\mathcal B\rangle=\beta_Q$,
both squares have zero expectation. This implies
\begin{equation}\label{eq:sosrel}
  A_0\ket{\psi}=\frac{B_0+B_1}{\sqrt2}\ket{\psi},
  \quad
  A_1\ket{\psi}=\frac{B_0-B_1}{\sqrt2}\ket{\psi}.
\end{equation}
By definition, $A_x^2=B_y^2=\id$ on the local reduced support of the state at each party. Therefore, \cref{eq:sosrel} gives
\begin{equation}\label{eq:anticomm}
  \{A_0,A_1\}\ket{\psi}=0,
  \qquad
  \{B_0,B_1\}\ket{\psi}=0.
\end{equation}
Define
\begin{equation}\label{eq:swap-observables}
\begin{aligned}
\mathcal X_A&=A_0, &\mathcal Z_A&=A_1,\\
\mathcal X_B&=\frac{B_0+B_1}{\sqrt2},
&\mathcal Z_B&=\frac{B_0-B_1}{\sqrt2}.
\end{aligned}
\end{equation}
By \cref{eq:sosrel,eq:anticomm}, the operators $\mathcal X_A,\mathcal Z_A,\mathcal X_B,\mathcal Z_B$ are Hermitian
unitaries on their local reduced supports and preserve these supports.
Moreover,
\begin{equation}\label{eq:swap-relations}
\begin{aligned}
&\{\mathcal X_i,\mathcal Z_i\}\ket{\psi}=0,
\qquad i=A,B,\\
&\mathcal Z_A\ket{\psi}=\mathcal Z_B\ket{\psi},\qquad
\mathcal X_A\ket{\psi}=\mathcal X_B\ket{\psi}.
\end{aligned}
\end{equation}
For each $i=A,B$, we append a qubit with $\mathcal H_{i'}\cong\mathbb C^2$,
initialized in $\ket{\chi}_{i'}=\ket0_{i'}$.
The unitary $U_{ii'}$ in \cref{fig:swap_qubit_vs_fermion} then defines
the isometry $\Phi_i$, where $\bar i$ labels the local output systems other than the extracted register $i'$.
Here $H$ is the Hadamard gate. The controlled gates apply $\mathcal Z_i$
or $\mathcal X_i$ when the ancillary qubit is in $\ket1$, and act as
the identity when it is in $\ket0$. Using \cref{eq:swap-relations},
the joint action of these isometries gives
\begin{equation}
  (\Phi_A\otimes\Phi_B)\ket{\psi}_{AB}
  =\ket{\phi_{\mathrm q}^{+}}_{A'B'}\otimes\ket{\xi}_{\bar A\bar B}.
\end{equation}

On the local reduced supports, the same circuit satisfies
\[
\Phi_i\mathcal X_i
=(\sigma_X\otimes\id_{\bar i})\Phi_i,
\qquad
\Phi_i\mathcal Z_i
=(\sigma_Z\otimes\id_{\bar i})\Phi_i,
\qquad i=A,B,
\]
where the Pauli operators act on the extracted qubit.
The measurement projectors are
\(M_{a\mid x}=(\id+aA_x)/2\) and
\(N_{b\mid y}=(\id+bB_y)/2\).
Define their reference counterparts $M'_{a\mid x}$ and $N'_{b\mid y}$
by the same formulas using
\[
A'_0=\sigma_X,\qquad A'_1=\sigma_Z,\qquad
B'_y=\frac{\sigma_X+(-1)^y\sigma_Z}{\sqrt2}.
\]
It thus follows that, for every \(a,b\in\{\pm1\}\) and
\(x,y\in\{0,1\}\),
\begin{equation}
\label{eq:swap-measurement-self-testing}
(\Phi_A\otimes\Phi_B)
(M_{a\mid x}\otimes N_{b\mid y})\ket{\psi}_{AB}
=
\bigl[(M'_{a\mid x}\otimes N'_{b\mid y})
\ket{\phi_{\mathrm q}^{+}}_{A'B'}\bigr]
\otimes\ket{\xi}_{\bar A\bar B}.
\end{equation}

We now extend the definition of self-testing to F-QIT and revisit the Bell scenario within this framework.
\begin{definition}[F-QIT self-testing]
The correlations $p(a,b\,|\, x,y)$ self-test a
fermionic reference strategy $(\ket{\psi'},\{M'_{a\mid x}\},\{N'_{b\mid y}\})$
if every fermionic realization $(\ket{\psi},\{M_{a\mid x}\},\{N_{b\mid y}\})$,
consisting of a normalized state $\ket\psi\in\mathcal H_A\wedge\mathcal H_B$
of definite total parity and local parity-preserving POVMs satisfying
$p(a,b\mid x,y)=\bra\psi(M_{a\mid x}\wedge N_{b\mid y})\ket\psi$,
admits local parity-preserving isometries
\[
\begin{aligned}
\Phi_A &: \mathcal H_A\longrightarrow
\mathcal H_{A'}\wedge\mathcal H_{\bar A},\\
\Phi_B &: \mathcal H_B\longrightarrow
\mathcal H_{B'}\wedge\mathcal H_{\bar B},
\end{aligned}
\]
such that
\[
(\Phi_A\wedge\Phi_B)\ket{\psi}_{AB}
=\ket{\psi'}_{A'B'}\wedge\ket{\xi}_{\bar A\bar B},
\]
where $\ket{\xi}_{\bar A\bar B}$ is a normalized auxiliary
fermionic \emph{junk} state, and, for every $a,b,x,y$,
\[
\begin{aligned}
&(\Phi_A\wedge\Phi_B)
\bigl[(M_{a\mid x}\wedge N_{b\mid y})\ket{\psi}_{AB}\bigr]=\bigl[(M'_{a\mid x}\wedge N'_{b\mid y})
\ket{\psi'}_{A'B'}\bigr]\wedge\ket{\xi}_{\bar A\bar B}.
\end{aligned}
\]
The same isometries and junk state must satisfy both
relations for all inputs and outputs.
\end{definition}

Under this definition, maximal CHSH violation self-tests the fermionic
reference state and measurements described in the main text, as shown below.

\paragraph*{\normalfont\bfseries Proof of CHSH self-testing in F-QIT.}
Consider an arbitrary fermionic realization attaining $\beta=\beta_Q$,
and define the dichotomic observables $A_x,B_y$ from the corresponding PVMs, as in the Q-QIT case. Hence, $A_x^2=B_y^2=\id$ on the local
reduced supports.
Here, $A_x,B_y$ denote $A_x\wedge\id_B,\id_A\wedge B_y$.
These observables commute because they preserve local parity, so the
same sum-of-squares argument gives \cref{eq:sosrel,eq:anticomm}.
Defining $\mathcal X_i,\mathcal Z_i$ as in \cref{eq:swap-observables}
then gives \cref{eq:swap-relations}.

For the final SWAP construction, we introduce ancillary
spaces $\mathcal H_{A'}\cong\mathbb C^2$ and
$\mathcal H_{B'}\cong\mathbb C^2$, each realized as the even
subspace $\operatorname{span}\{\ket{00},\ket{11}\}$ of two
febits. We initialize them in $\ket{\chi}_{A'}=\ket{00}_{A'}$
and $\ket{\chi}_{B'}=\ket{00}_{B'}$, respectively.
The Hadamard gates act as
\[
\begin{aligned}
H\ket{00}&=\frac{\ket{00}+\ket{11}}{\sqrt2},\\
H\ket{11}&=\frac{\ket{00}-\ket{11}}{\sqrt2}.
\end{aligned}
\]
The controlled gates apply $\mathcal Z_i$ or $\mathcal X_i$
to the physical system when the ancillary state is
$\ket{11}$, and act as the identity when it is $\ket{00}$.
All these gates preserve parity. Appending the ancillas
and applying the local circuits therefore define
physical isometries. Using \cref{eq:swap-relations} as in
the Q-QIT construction gives
\begin{equation}\label{eq:Phiplus}
(\Phi_A\wedge\Phi_B)\ket{\psi}_{AB}
=\ket{\phi_{\mathrm f}^{+}}_{A'B'}\wedge\ket{\xi}_{\bar A\bar B}.
\end{equation}
For every \(a,b,x,y\), the same isometries and junk state also satisfy
\begin{equation}\label{eq:fermionic-swap-measurement-self-testing}
(\Phi_A\wedge\Phi_B)
(M_{a\mid x}\wedge N_{b\mid y})\ket{\psi}_{AB}
=
\bigl[(M'_{a\mid x}\wedge N'_{b\mid y})
\ket{\phi_{\mathrm f}^{+}}_{A'B'}\bigr]
\wedge\ket{\xi}_{\bar A\bar B},
\end{equation}
where \(M'_{a\mid x}=(\id_{A'}+aA'_x)/2\) and
\(N'_{b\mid y}=(\id_{B'}+bB'_y)/2\), with
\[
A'_0=\sigma_X^f,\qquad A'_1=\sigma_Z^f,\qquad
B'_y=\frac{\sigma_X^f+(-1)^y\sigma_Z^f}{\sqrt2}.
\]
Here, \(\sigma_X^f=\ket{00}\!\bra{11}+\ket{11}\!\bra{00}\)
and \(\sigma_Z^f=\ket{00}\!\bra{00}-\ket{11}\!\bra{11}\)
are Pauli operators on the extracted even-parity registers.

%% file: FixedLoopSuppMat.tex
\clearpage
\onecolumngrid
\section{Fixed-loop self-testing}
\label{sec:fixed-loop-self-testing}

We extend the fixed-loop self-testing result in Q-QIT,
Theorem~B.2 of Ref.~\cite{kalarde2026fermions}, to the
fermionic setting. The resulting characterization admits two
classes of strategies that are inequivalent under local
parity-preserving isometries. We also establish a version for
loops of size \(N\geq2\) under structural assumptions that,
in the full game, follow from the self-testing of larger loops.

\begin{theorem}[Fixed-loop self-testing]
\label{thm:fixed-loop-characterization}
Consider the \(N\)-party generalization of the fermionic game
described in the section ``The fermionic self-testing scenario'', with \(N\geq 3\), and fix a loop configuration \(l\)
containing all \(N\) parties. Label the parties
\(A_1,\ldots,A_N\) according to their order along the loop, with
indices understood modulo \(N\). For each
\(i\in\{1,\ldots,N\}\), source \(i\) connects parties \(A_i\) and
\(A_{i+1}\) and distributes a bipartite fermionic state
\[
\ket{\psi_i}_{i_L,(i+1)_R}
\in
\mathcal H_{i_L}\wedge\mathcal H_{(i+1)_R}.
\]
The global source state is thus
\[
\ket{\Psi}
:=
\bigwedge_{i=1}^{N}
\ket{\psi_i}_{i_L,(i+1)_R}.
\]

Party \(A_i\) receives the subsystems \(i_R\) and \(i_L\) from its two
neighbouring sources and, for each input \(x_i\in\{0,1\}\), performs a
parity-preserving four-outcome POVM
\[
\left\{
\Pi_{i,x_i}^{a_i}
\right\}_{a_i\in\{0,1_+,1_-,2\}}
\]
on \(\mathcal H_{i_R}\wedge\mathcal H_{i_L}\). The observed behavior is
\[
P(\mathbf a\mid\mathbf x)
=
\bra{\Psi}
\left(
\bigwedge_{i=1}^{N}
\Pi_{i,x_i}^{a_i}
\right)
\ket{\Psi}.
\]

For each party \(i\), introduce a fermionic ancillary space
\(\mathcal H_{M_i}\), initialized in the normalized even state
\(\ket0_{M_i}\). Let
\[
\left\{\bar\Pi_{i,x_i}^{a_i}\right\}_{a_i\in\{0,1_+,1_-,2\}},
\qquad x_i\in\{0,1\},
\]
be parity-preserving PVMs on
\(\mathcal H_{i_R}\wedge\mathcal H_{i_L}\wedge\mathcal H_{M_i}\)
that provide Naimark dilations of the two POVMs, using the same
ancillary space and initialized state for both inputs.
Grouping \(M_i\) with \(i_L\), define
\[
\begin{aligned}
\ket{\widehat\psi_i}
&:=\ket{\psi_i}_{i_L,(i+1)_R}\wedge\ket0_{M_i},\qquad
\ket{\widehat\Psi}
:=\bigwedge_{i=1}^{N}\ket{\widehat\psi_i}.
\end{aligned}
\]

Let \(P_{ f}^{l}\) denote the behavior generated
by the trusted fermionic strategy for the same loop configuration
\(l\). Suppose that
\[
P(\mathbf a\mid\mathbf x)
=
P_{ f}^{l}(\mathbf a\mid\mathbf x)
\]
for every \(\mathbf a\) and \(\mathbf x\). Then the following two conclusions hold.

\begin{enumerate}

\item \emph{Partial self-testing of the token-counting structure.}
There exist local parity-preserving isometries, independent
of the inputs,
\[
\begin{aligned}
\mathcal U_{i_R}&:\mathcal H_{i_R}
\longrightarrow S_{i_R}\wedge J_{i_R},
\\
\mathcal U_{i_L}:=\mathcal U_{i_LM_i}&:\mathcal H_{i_L}\wedge\mathcal H_{M_i}
\longrightarrow S_{i_L}\wedge J_{i_L},
\end{aligned}
\]
where \(S_{i_R}\simeq S_{i_L}\simeq\mathbb C^2\) are
single fermionic modes and \(J_{i_R},J_{i_L}\) are
auxiliary junk spaces, such that
\begin{equation}
\label{eq:fixed-loop-partial-state}
\begin{aligned}
\ket{\Phi_i}
&:=
\left(\mathcal U_{i_L}\wedge\mathcal U_{(i+1)_R}\right)
\ket{\widehat\psi_i}_{i_LM_i,(i+1)_R}
\\
&=
\frac1{\sqrt2}
\left(
\ket{01}_{S_{i_L}S_{(i+1)_R}}
\wedge\ket{\zeta_i^c}_{J_{i_L}J_{(i+1)_R}}
+
\ket{10}_{S_{i_L}S_{(i+1)_R}}
\wedge\ket{\zeta_i^a}_{J_{i_L}J_{(i+1)_R}}
\right),
\end{aligned}
\end{equation}
with normalized junk states \(\ket{\zeta_i^c}\) and
\(\ket{\zeta_i^a}\).

Define
\[
\begin{aligned}
\mathcal U_i&:=\mathcal U_{i_R}\wedge\mathcal U_{i_L},
&
\mathcal U&:=\bigwedge_{i=1}^{N}\mathcal U_i,
\\
S_i&:=S_{i_R}\wedge S_{i_L},
&
J_i&:=J_{i_R}\wedge J_{i_L},
\end{aligned}
\]
and set
\[
\ket{\Phi}:=\mathcal U\ket{\widehat\Psi}.
\]
Coarse-grain the dilated measurements by setting
\[
\bar\Pi_{i,x_i}^{1}
:=
\bar\Pi_{i,x_i}^{1_+}
+
\bar\Pi_{i,x_i}^{1_-},
\]
while retaining the outcomes \(0\) and \(2\), and define
\[
\begin{aligned}
\Pi_{\mathrm{TC},i}^{0}
&:=
\ket{00}\!\bra{00}_{S_i}\wedge\mathds{1}_{J_i},
\\
\Pi_{\mathrm{TC},i}^{1}
&:=
\left(
\ket{01}\!\bra{01}
+
\ket{10}\!\bra{10}
\right)_{S_i}\wedge\mathds{1}_{J_i},
\\
\Pi_{\mathrm{TC},i}^{2}
&:=
\ket{11}\!\bra{11}_{S_i}\wedge\mathds{1}_{J_i}.
\end{aligned}
\]
For every party \(i\), input \(x_i\in\{0,1\}\), and
coarse-grained outcome \(n\in\{0,1,2\}\),
\begin{equation}
\label{eq:fixed-loop-partial-measurements}
\mathcal U\bar\Pi_{i,x_i}^{n}\ket{\widehat\Psi}
=
\Pi_{\mathrm{TC},i}^{n}\ket{\Phi}.
\end{equation}

\item \emph{Unitary completion of the self-test.}
Fix any choice of isometries and junk states from part~1.
For each source \(i\), let
\(J_{i_L}^{c}\) and \(J_{(i+1)_R}^{c}\) denote the
supports of the two reduced density operators of
\(\ket{\zeta_i^c}\). Similarly, let
\(J_{i_L}^{a}\) and \(J_{(i+1)_R}^{a}\) denote the supports of reduced densities of \(\ket{\zeta_i^a}\).

For \(x_i\in\{0,1\}\), define the signed observable
\[
M_{x_i}^{(i)}
:=
\bar\Pi_{i,x_i}^{1_+}
-
\bar\Pi_{i,x_i}^{1_-},
\]
and set
\[
\mathcal X_1
:=
\frac{M_0^{(1)}+M_1^{(1)}}{\sqrt{2}},
\quad
\mathcal Z_1
:=
\frac{M_0^{(1)}-M_1^{(1)}}{\sqrt{2}},
\]
and
\[
\mathcal X_i:=M_1^{(i)},
\quad
\mathcal Z_i:=M_0^{(i)},
\quad
i\in\{2,\ldots,N\}.
\]
These observables satisfy
\begin{equation}
\label{eq:fixed-loop-partial-XZ}
\begin{aligned}
\mathcal U\mathcal Z_i\ket{\widehat\Psi}
&=
\left[
\left(
\ket{01}\!\bra{01}
-
\ket{10}\!\bra{10}
\right)_{S_i}
\wedge\mathds{1}_{J_i}
\right]\ket{\Phi},
\\
\mathcal U\mathcal X_i\ket{\widehat\Psi}
&=
\left(
\ket{01}\!\bra{10}_{S_i}\wedge T_i
+
\ket{10}\!\bra{01}_{S_i}\wedge T_i^\dagger
\right)\ket{\Phi},
\end{aligned}
\end{equation}
where
\[
T_i=T_i^{(R)}\wedge T_i^{(L)},
\]
with unitary maps
\[
T_i^{(R)}:J_{i_R}^{c}\longrightarrow J_{i_R}^{a},
\qquad
T_i^{(L)}:J_{i_L}^{c}\longrightarrow J_{i_L}^{a}.
\]

All the factors \(T_i^{(R)}\) and \(T_i^{(L)}\), for
all \(i\), have the same definite parity. Accordingly,
the completion of self-testing takes one of the following two forms.

\medskip
\emph{(i) Single-mode systems.}
If all the factors \(T_i^{(R)}\) and \(T_i^{(L)}\)
preserve parity, define
\[
\begin{aligned}
U_{i_R}
&:=
\ket0\!\bra0_{S_{i_R}}\wedge\mathds{1}_{J_{i_R}^{a}}
+
\ket1\!\bra1_{S_{i_R}}\wedge T_i^{(R)},
\\
U_{i_L}
&:=
\ket1\!\bra1_{S_{i_L}}\wedge\mathds{1}_{J_{i_L}^{a}}
+
\ket0\!\bra0_{S_{i_L}}\wedge T_i^{(L)}.
\end{aligned}
\]
These are parity-preserving unitary maps from the
corresponding local reduced supports onto
\(S_{i_R}\wedge J_{i_R}^{a}\) and
\(S_{i_L}\wedge J_{i_L}^{a}\), respectively.

We absorb \(U_{i_R}\) and \(U_{i_L}\) into the initial
isometries, denoting the resulting maps
\[
U_{i_R}\mathcal U_{i_R},
\qquad
U_{i_L}\mathcal U_{i_L}
\]
again by \(\mathcal U_{i_R}\) and \(\mathcal U_{i_L}\),
respectively. Then
\[
\begin{aligned}
&\left(\mathcal U_{i_L}\wedge\mathcal U_{(i+1)_R}\right)
\ket{\widehat\psi_i}_{i_LM_i,(i+1)_R}
=
\frac1{\sqrt2}
\left(e^{\ii\varphi_i}\ket{01}+\ket{10}\right)
_{S_{i_L}S_{(i+1)_R}}
\wedge\ket{\zeta_i^a}_{J_{i_L}^{a}J_{(i+1)_R}^{a}},
\end{aligned}
\]
and
\[
\sum_{i=1}^{N}\varphi_i=0\pmod{2\pi}.
\]

\medskip
\emph{(ii) Double-mode systems.}
If all the factors \(T_i^{(R)}\) and \(T_i^{(L)}\)
change parity, introduce the even subspaces of two
fermionic modes,
\[
\widetilde S_{i_R}\simeq\widetilde S_{i_L}
\simeq\operatorname{span}\{\ket{00},\ket{11}\}.
\]
For \(\sigma\in\{R,L\}\), define the parity-preserving
encoding isometry
\[
\mathcal E_{i_\sigma}
\left(
\ket b_{S_{i_\sigma}}\wedge\ket{\xi}_{J_{i_\sigma}}
\right)
:=
\ket{bb}_{\widetilde S_{i_\sigma}}
\wedge\ket b_{S_{i_\sigma}}
\wedge\ket{\xi}_{J_{i_\sigma}},
\qquad b\in\{0,1\}.
\]
The original single-mode systems are thus retained
as part of the new junk spaces. Define
\[
\begin{aligned}
\widetilde J_{i_R}^{a}
&:=
\operatorname{span}\{\ket0_{S_{i_R}}\}\wedge J_{i_R}^{a},
&
\widetilde J_{i_R}^{c}
&:=
\operatorname{span}\{\ket1_{S_{i_R}}\}\wedge J_{i_R}^{c},
\\
\widetilde J_{i_L}^{a}
&:=
\operatorname{span}\{\ket1_{S_{i_L}}\}\wedge J_{i_L}^{a},
&
\widetilde J_{i_L}^{c}
&:=
\operatorname{span}\{\ket0_{S_{i_L}}\}\wedge J_{i_L}^{c},
\end{aligned}
\]
and the parity-preserving unitary maps
\[
\begin{aligned}
\widetilde T_i^{(R)}
&:=
\ket0\!\bra1_{S_{i_R}}\wedge T_i^{(R)}
:
\widetilde J_{i_R}^{c}\longrightarrow
\widetilde J_{i_R}^{a},
\\
\widetilde T_i^{(L)}
&:=
\ket1\!\bra0_{S_{i_L}}\wedge T_i^{(L)}
:
\widetilde J_{i_L}^{c}\longrightarrow
\widetilde J_{i_L}^{a}.
\end{aligned}
\]
Now define
\[
\begin{aligned}
\widetilde U_{i_R}
&:=
\ket{00}\!\bra{00}_{\widetilde S_{i_R}}
\wedge\mathds{1}_{\widetilde J_{i_R}^{a}}
+
\ket{11}\!\bra{11}_{\widetilde S_{i_R}}
\wedge\widetilde T_i^{(R)},
\\
\widetilde U_{i_L}
&:=
\ket{11}\!\bra{11}_{\widetilde S_{i_L}}
\wedge\mathds{1}_{\widetilde J_{i_L}^{a}}
+
\ket{00}\!\bra{00}_{\widetilde S_{i_L}}
\wedge\widetilde T_i^{(L)}.
\end{aligned}
\]
These are parity-preserving unitary maps from the
corresponding encoded local reduced supports onto
\(\widetilde S_{i_R}\wedge\widetilde J_{i_R}^{a}\)
and
\(\widetilde S_{i_L}\wedge\widetilde J_{i_L}^{a}\),
respectively.

We absorb \(\widetilde U_{i_R}\mathcal E_{i_R}\) and
\(\widetilde U_{i_L}\mathcal E_{i_L}\) into the initial
isometries, denoting the resulting maps
\[
\widetilde U_{i_R}\mathcal E_{i_R}\mathcal U_{i_R},
\qquad
\widetilde U_{i_L}\mathcal E_{i_L}\mathcal U_{i_L}
\]
again by \(\mathcal U_{i_R}\) and \(\mathcal U_{i_L}\),
respectively. Then
\[
\begin{aligned}
&\left(\mathcal U_{i_L}\wedge\mathcal U_{(i+1)_R}\right)
\ket{\widehat\psi_i}_{i_LM_i,(i+1)_R}
=
\frac1{\sqrt2}
\left(
e^{\ii\varphi_i}\ket{00,11}+\ket{11,00}
\right)_{\widetilde S_{i_L}\widetilde S_{(i+1)_R}}
\wedge
\ket{\widetilde{\zeta}_i^a}_{\widetilde J_{i_L}^{a}
\widetilde J_{(i+1)_R}^{a}},
\end{aligned}
\]
where the normalized junk state is
\[
\ket{\widetilde{\zeta}_i^a}
:=\ket{10}_{S_{i_L}S_{(i+1)_R}}
\wedge\ket{\zeta_i^a}_{J_{i_L}J_{(i+1)_R}},
\]
and the phases satisfy
\[
\sum_{i=1}^{N}\varphi_i=(N-1)\pi\pmod{2\pi}.
\]

\medskip
These two strategies are not related by local
parity-preserving isometries and therefore belong
to distinct isometry classes.

Using the final extraction maps in each case, define
\[
\begin{aligned}
\mathcal U_i
&:=\mathcal U_{i_R}\wedge\mathcal U_{i_L},
&
\mathcal U
&:=\bigwedge_{i=1}^{N}\mathcal U_i,
\end{aligned}
\]
and let
\(
\ket{\widehat\Phi}
:=
\mathcal U\ket{\widehat\Psi}
\)
denote the extracted global state. In case~(i), set
\[
S_i:=S_{i_R}\wedge S_{i_L},
\qquad
J_i^a:=J_{i_R}^a\wedge J_{i_L}^a,
\]
and in case~(ii), set
\[
\widetilde S_i:=\widetilde S_{i_R}\wedge\widetilde S_{i_L},
\qquad
\widetilde J_i^a:=\widetilde J_{i_R}^a\wedge\widetilde J_{i_L}^a.
\]

In case~(i), for every party \(i\) and every input \(x_i\in\{0,1\}\),
the dilated projectors \(\bar\Pi_{i,x_i}^{0}\) and
\(\bar\Pi_{i,x_i}^{2}\), together with the signed
observables, satisfy
\begin{equation}
\label{eq:fixed-loop-final-measurement-identities}
\begin{aligned}
\mathcal U\bar\Pi_{i,x_i}^{0}\ket{\widehat\Psi}
&=
\Pi^0_{S_i}
\ket{\widehat\Phi},
\\
\mathcal U\bar\Pi_{i,x_i}^{2}\ket{\widehat\Psi}
&=
\Pi^2_{S_i}
\ket{\widehat\Phi},
\\
\mathcal U\mathcal Z_i\ket{\widehat\Psi}
&=
Z_{S_i}
\ket{\widehat\Phi},
\\
\mathcal U\mathcal X_i\ket{\widehat\Psi}
&=
X_{S_i}
\ket{\widehat\Phi},
\end{aligned}
\end{equation}
with \(S_i\) replaced throughout by \(\widetilde S_i\)
in case~(ii).
The extracted operators are defined as follows.
In case~(i),
\[
\begin{aligned}
\Pi^0_{S_i}
&:=\ket{00}\!\bra{00}_{S_{i_R}S_{i_L}}
\wedge\mathds{1}_{J_i^a},
\\
\Pi^2_{S_i}
&:=\ket{11}\!\bra{11}_{S_{i_R}S_{i_L}}
\wedge\mathds{1}_{J_i^a},
\\
Z_{S_i}
&:=
\bigl(
\ket{01}\!\bra{01}-\ket{10}\!\bra{10}
\bigr)_{S_{i_R}S_{i_L}}\wedge\mathds{1}_{J_i^a},
\\
X_{S_i}
&:=
\bigl(
\ket{01}\!\bra{10}+\ket{10}\!\bra{01}
\bigr)_{S_{i_R}S_{i_L}}\wedge\mathds{1}_{J_i^a}.
\end{aligned}
\]
In case~(ii),
\[
\begin{aligned}
\Pi^0_{\widetilde S_i}
&:=\ket{00,00}\!\bra{00,00}_{\widetilde S_{i_R}\widetilde S_{i_L}}
\wedge\mathds{1}_{\widetilde J_i^a},
\\
\Pi^2_{\widetilde S_i}
&:=\ket{11,11}\!\bra{11,11}_{\widetilde S_{i_R}\widetilde S_{i_L}}
\wedge\mathds{1}_{\widetilde J_i^a},
\\
Z_{\widetilde S_i}
&:=
\bigl(
\ket{00,11}\!\bra{00,11}
-
\ket{11,00}\!\bra{11,00}
\bigr)_{\widetilde S_{i_R}\widetilde S_{i_L}}
\wedge\mathds{1}_{\widetilde J_i^a},
\\
X_{\widetilde S_i}
&:=
\bigl(
\ket{00,11}\!\bra{11,00}
+
\ket{11,00}\!\bra{00,11}
\bigr)_{\widetilde S_{i_R}\widetilde S_{i_L}}
\wedge\mathds{1}_{\widetilde J_i^a}.
\end{aligned}
\]

\end{enumerate}

\end{theorem}

\begin{proof}

We first coarse-grain the distribution $P(\mathbf a\mid\mathbf x)$ by merging the outcomes $1_\pm$ into a single outcome $1$. The resulting distribution is independent of the inputs and coincides with the token-counting distribution of Ref.~\cite{PhysRevA.105.022408}. Our first step is to obtain a partial characterization of the underlying strategy based on this correlation, following Ref.~\cite{sekatski2023partial}. Since that result was established for Q-QIT, we must first extend it to the fermionic setting. The following theorem states this extension.

\begin{theorem}[Partial token-counting self-testing of fermionic strategies]
\label{thm:exact-fermionic-TC}
Consider an \(N\)-cycle, with \(N\geq3\), whose independent sources
distribute states
\[
\ket{\psi_i}_{i_L,(i+1)_R}
\in
\mathcal H_{i_L}\wedge\mathcal H_{(i+1)_R},
\qquad i\in\{1,\ldots,N\}.
\]
The global state is
\[
\ket{\Psi}
:=
\bigwedge_{i=1}^{N}
\ket{\psi_i}_{i_L,(i+1)_R}.
\]
Party \(A_i\) performs a parity-preserving three-outcome POVM
\(\{E_i^{n_i}\}_{n_i=0}^2\) on
\(\mathcal H_{i_R}\wedge\mathcal H_{i_L}\).
Fix a parity preserving Naimark dilation
\(\{\bar\Pi_i^{n_i}\}_{n_i=0}^2\) on
\[
\mathcal H_{i_R}\wedge
\mathcal H_{i_L}\wedge
\mathcal H_{M_i},
\]
where \(M_i\) is initialized in an even state \(\ket{0}_{M_i}\).
Set
\[
\begin{aligned}
\ket{\widehat\psi_i}
&:=
\ket{\psi_i}_{i_L,(i+1)_R}
\wedge\ket{0}_{M_i},\qquad
\ket{\widehat\Psi}
:=
\bigwedge_{i=1}^{N}\ket{\widehat\psi_i},
\end{aligned}
\]
The resulting distribution is
\[
P(\mathbf n)
=
\bra{\widehat\Psi}
\left(
\prod_{i=1}^{N}\bar\Pi_i^{n_i}
\right)
\ket{\widehat\Psi}.
\]

Suppose that
\[
P(\mathbf n)
=
P_{\boldsymbol{1/2}}^{\mathrm{TC}}(\mathbf n)
\]
for every \(\mathbf n\in\{0,1,2\}^N\), where
\(P_{\boldsymbol{1/2}}^{\mathrm{TC}}\) is the fair token-counting
distribution. Then the following statements hold.

\begin{itemize}

\item There exist two-dimensional extracted systems
\(S_{i_R}\simeq S_{i_L}\simeq\mathbb C^2\), auxiliary junk spaces
\(J_{i_R},J_{i_L}\), and local isometries
\[
\begin{aligned}
\mathcal U_{i_R}&:
\mathcal H_{i_R}
\longrightarrow S_{i_R}\wedge J_{i_R},\\
\mathcal U_{i_LM_i}&:
\mathcal H_{i_L}\wedge\mathcal H_{M_i}
\longrightarrow S_{i_L}\wedge J_{i_L}.
\end{aligned}
\]
These isometries preserve local fermionic parity.

\item For every source \(i\), there exist normalized states
\(\ket{\zeta_i^c}\) and \(\ket{\zeta_i^a}\) on
\(J_{i_L}\wedge J_{(i+1)_R}\) such that
\begin{equation}
\label{eq:exact-fermionic-TC-state}
\begin{aligned}
&\left(
\mathcal U_{i_LM_i}
\wedge
\mathcal U_{(i+1)_R}
\right)
\ket{\widehat\psi_i}
 =
\frac{1}{\sqrt2}
\Bigl(
\ket{01}_{S_{i_L}S_{(i+1)_R}}
\wedge\ket{\zeta_i^c}
+
\ket{10}_{S_{i_L}S_{(i+1)_R}}
\wedge\ket{\zeta_i^a}
\Bigr).
\end{aligned}
\end{equation}

\item Define
\[
\begin{aligned}
\mathcal U_i
&:=
\mathcal U_{i_R}\wedge\mathcal U_{i_LM_i},
&
\mathcal U
&:=
\bigwedge_{i=1}^{N}\mathcal U_i,\\
S_i&:=S_{i_R}\wedge S_{i_L},
&
J_i&:=J_{i_R}\wedge J_{i_L},
\end{aligned}
\]
and let \(\ket{\Phi}:=\mathcal U\ket{\widehat\Psi}\).
Define the ideal token-counting projectors
\begin{equation}
\label{eq:exact-fermionic-TC-projectors}
\begin{aligned}
\Pi_{\mathrm{TC},i}^{0}
&:=
\ket{00}\!\bra{00}_{S_i}\wedge\mathds{1}_{J_i},\\
\Pi_{\mathrm{TC},i}^{1}
&:=
\bigl(
\ket{01}\!\bra{01}
+
\ket{10}\!\bra{10}
\bigr)_{S_i}
\wedge\mathds{1}_{J_i},\\
\Pi_{\mathrm{TC},i}^{2}
&:=
\ket{11}\!\bra{11}_{S_i}\wedge\mathds{1}_{J_i}.
\end{aligned}
\end{equation}
Then for every \(i\) and \(n_i\in\{0,1,2\}\),
\begin{equation}
\label{eq:exact-fermionic-TC-measurements}
\mathcal U\bar\Pi_i^{n_i}\ket{\widehat\Psi}
=
\Pi_{\mathrm{TC},i}^{n_i}\ket{\Phi}.
\end{equation}

\end{itemize}
\end{theorem}

\begin{proof}
First, we explain Naimark dilation in fermionic systems.

\noindent\emph{Parity-preserving Naimark dilation.—}
Let
\(\mathcal H=\mathcal H^{\mathrm e}\oplus\mathcal H^{\mathrm o}\)
be a fermionic Hilbert space, and let
\(\{E^t\}_{t=0}^{m-1}\) be an \(m\)-outcome POVM on \(\mathcal H\),
whose elements are all even. Thus,
\[
E^t\geq0,
\qquad
\sum_{t=0}^{m-1}E^t=\mathds{1}_{\mathcal H}.
\]
Let the ancillary system be the graded space
\[
\mathcal H_M
=
\mathcal H_M^{\mathrm e}\oplus\mathcal H_M^{\mathrm o},
\qquad
\dim\mathcal H_M^{\mathrm e}
=
\dim\mathcal H_M^{\mathrm o}
=
m,
\]
with orthonormal bases
\(\{\ket{t_{\mathrm e}}_M\}_{t=0}^{m-1}\) and
\(\{\ket{t_{\mathrm o}}_M\}_{t=0}^{m-1}\) for $\mathcal H_M^{\mathrm e}$ and $\mathcal H_M^{\mathrm o}$ respectively, and fix a normalized
even state \(\ket{0}_M\). Define
\[
\begin{aligned}
V:\mathcal H
&\longrightarrow
\mathcal H\wedge\mathcal H_M,\\
V\ket{\psi}
&:=
\sum_{t=0}^{m-1}
\sqrt{E^t}\ket{\psi}
\wedge\ket{t_{\mathrm e}}_M.
\end{aligned}
\]
Since
\[
V^\dagger V
=
\sum_{t=0}^{m-1}E^t
=
\mathds{1}_{\mathcal H},
\]
\(V\) is an isometry. It is also parity preserving because the operators
\(\sqrt{E^t}\) and the label states
\(\ket{t_{\mathrm e}}_M\) are even. 
Define the unitary $W$ on \(\mathcal H\wedge\mathcal H_M\)
\begin{equation}
\label{eq:fermionic-Naimark-unitary}
\begin{aligned}
W\bigl(\ket{\psi}\wedge\ket{0}_M\bigr)
&:=
V\ket{\psi}
=
\sum_{t=0}^{m-1}
\sqrt{E^t}\ket{\psi}
\wedge\ket{t_{\mathrm e}}_M,
\end{aligned}
\end{equation}
and extend it separately within the even and odd parity sectors of \(\mathcal H\wedge\mathcal H_M\) such that it stays parity preserving.
Define
\[
Q^t
:=
\mathds{1}_{\mathcal H}
\wedge
\left(
\ket{t_{\mathrm e}}\!\bra{t_{\mathrm e}}
+
\ket{t_{\mathrm o}}\!\bra{t_{\mathrm o}}
\right)_M,
\qquad
\bar\Pi^t
:=
W^\dagger Q^tW.
\]
The operators \(Q^t\) form an even PVM. Since \(W\) is unitary,
\[
(\bar\Pi^t)^\dagger=\bar\Pi^t,
\qquad
\bar\Pi^t\bar\Pi^s
=
\delta_{ts}\bar\Pi^t,
\qquad
\sum_{t=0}^{m-1}\bar\Pi^t
=
\mathds{1}_{\mathcal H\wedge\mathcal H_M}.
\]
Therefore, \(\{\bar\Pi^t\}_{t=0}^{m-1}\) is a PVM on
\(\mathcal H\wedge\mathcal H_M\). Moreover, since both \(W\) and
\(Q^t\) preserve parity, so does \(\bar\Pi^t\).

Finally, for arbitrary \(\ket{\varphi},\ket{\psi}\in\mathcal H\),
\[
\begin{aligned}
&\bigl(\bra{\varphi}\wedge\bra{0}_M\bigr)
\bar\Pi^t
\bigl(\ket{\psi}\wedge\ket{0}_M\bigr)
 =
\bra{\varphi}V^\dagger Q^tV\ket{\psi}
=
\bra{\varphi}E^t\ket{\psi}.
\end{aligned}
\]
It follows that the projective measurement \(\{\bar\Pi^t\}_{t=0}^{m-1}\)
reproduces the original POVM, while the original state is merely
enlarged by the fixed even ancillary state \(\ket{0}_M\).

From now on, we work with parity preserving projectors \(\{\bar\Pi^{n_i}_i\}_{n_i=0}^{2}\) for all $i \in \{1, \cdots, N\}$.
We next investigate the token-counting stabilizer. Define the token-counting support
\[
G_N
:=
\left\{
\mathbf n\in\{0,1,2\}^N:
\sum_{i=1}^{N}n_i=N
\right\}.
\]
Define 
\begin{equation}
\label{eq:exact-fermionic-TC-joint-projector}
\bar\Pi_{\mathbf n}
:=
\prod_{i=1}^{N}\bar\Pi_i^{n_i}.
\end{equation}
so \(P(\mathbf n) = \|\bar\Pi_{\mathbf n} \ket{\hat \Psi}\|^2 \). Therefore, exact agreement with
\(P_{\boldsymbol{1/2}}^{\mathrm{TC}}\) implies
\begin{equation}
\label{eq:exact-fermionic-TC-support}
\bar\Pi_{\mathbf n}\ket{\widehat\Psi}=0,
\qquad
\mathbf n\notin G_N,
\end{equation}

For \(n_i\in\{0,1,2\}\), set
\[
\phi_{n_i}
:=
\frac{2\pi(n_i+1/N)}{N+1},
\qquad
\bar U_i
:=
\sum_{n_i=0}^{2}
e^{\ii\phi_{n_i}}\bar\Pi_i^{n_i}.
\]
Each \(\bar U_i\) is an even unitary. Since even operators localized
at distinct parties commute, the global product
\(\bar U:=\prod_i\bar U_i\) is unambiguous and expands as
\[
\begin{aligned}
\bar U
&=
\sum_{\mathbf n\in\{0,1,2\}^N}
\lambda_{\mathbf n}\bar\Pi_{\mathbf n},
\\
\lambda_{\mathbf n}
&=
\exp\!\left[
\frac{2\pi\ii}{N+1}
\left(
\sum_{i=1}^{N}n_i+1
\right)
\right].
\end{aligned}
\]
For every \(\mathbf n\in G_N\), one has
\(\lambda_{\mathbf n}=1\). Together with
\cref{eq:exact-fermionic-TC-support}, and using the fact that $\bar\Pi_{\mathbf n}$ are projectors, this gives a stabilizer of the global state
\begin{equation}
\label{eq:exact-fermionic-TC-stabilizer}
\left(
\prod_{i=1}^{N}\bar U_i
\right)
\ket{\widehat\Psi}
=
\ket{\widehat\Psi}.
\end{equation}
Note that no fermionic reordering sign appears in this relation because all
\(\bar U_i\) are even.

The next step is to show that each unitary $\bar U_i$ factorizes across the bipartition $i_R|i_L$. 

\noindent\emph{Product structure.—}
For the remainder of the argument, we restrict the Hilbert space of each subsystem
to the support of the corresponding reduced state of
\(\ket{\widehat\psi_i}\).

The cyclic product-structure lemma was established and applied in this context
in Ref.~\cite{sekatski2023partial} for finite-dimensional tensor-product systems.
An extension to separable, possibly infinite-dimensional Hilbert spaces is given
in Lemma~K.1 of Ref.~\cite{kalarde2026fermions}.
This result also holds in the fermionic setting, with one important distinction:
the local factors need not commute, because they act on fermionic subsystems
rather than on distinct factors of a tensor-product Hilbert space.

Applying the lemma to \cref{eq:exact-fermionic-TC-stabilizer} yields
\begin{equation}
\label{eq:exact-fermionic-TC-product-unitaries}
\bar U_i \ket{\hat \Psi}=V_{i_R}\wedge W_{i_LM_i}\ket{\hat \Psi},
\end{equation}
where \(V_{i_R}\) and \(W_{i_LM_i}\) are unitaries.
Since \(\bar U_i\) is even and has product form, its two factors
have the same parity. We now exclude the possibility that
both are odd: Let \(P_{i_R}\) denote the parity operator on subsystem \(i_R\). If \(V_{i_R}\) is odd, then
\(P_{i_R}V_{i_R}=-V_{i_R}P_{i_R}\), and hence
\[
\begin{aligned}
(P_{i_R}\wedge\id)\bar U_i
&=(P_{i_R}V_{i_R})\wedge W_{i_LM_i}=-\bar U_i(P_{i_R}\wedge\id).
\end{aligned}
\]
Thus, \(\bar U_i\) anticommutes with the parity operators of \(i_R\)
and \(i_L\) separately, while still commuting with the total parity
operator. This implies that if \(\lambda\) is an eigenvalue of
\(\bar U_i\), then \(-\lambda\) must also be an eigenvalue. However,
\(\bar U_i\) has exactly three distinct eigenvalues \(e^{\ii\phi_n}\),
\(n\in\{0,1,2\}\), which is incompatible with this requirement.
Therefore, both local factors must be even.

\noindent\emph{Spectral projectors and token extraction.—}
Write the spectral decompositions of local unitaries as
\[
\begin{aligned}
V_{i_R}
&=\sum_j e^{\ii v_{i,j}}\Pi_{i_R}^{(j)},\\
W_{i_LM_i}
&=\sum_\ell e^{\ii w_{i,\ell}}
\Pi_{i_LM_i}^{(\ell)},
\end{aligned}
\]
where all these spectral projectors are even. Moreover, the spectrum of
\(V_{i_R} \wedge W_{i_LM_i}\) is contained in the set
\(\{e^{\ii\phi_{n_i}}:n_i=0,1,2\}\).

As in the quantum token-counting proof, define the partition of
spectral index pairs
\[
S_i(n_i):=
\left\{(j,\ell):
e^{\ii(v_{i,j}+w_{i,\ell})}=e^{\ii\phi_{n_i}}
\right\}.
\]
This allows us to write,
\begin{equation}
\label{eq:dilated-projectors-product-form}
\bar\Pi_i^{n_i}
=
\sum_{(j,\ell)\in S_i(n_i)}
\Pi_{i_R}^{(j)}\wedge\Pi_{i_LM_i}^{(\ell)}.
\end{equation}

We now construct a classical model that reproduces the observed
distribution \(P(\mathbf n)\), allowing us to apply the classical
token-counting result. For each source, define
\[
q_i(\ell,j):=
\left\|
\left(
\Pi_{i_LM_i}^{(\ell)}
\wedge\Pi_{(i+1)_R}^{(j)}
\right)\ket{\widehat\psi_i}
\right\|^2.
\]
For every \(i\), \(q_i(\ell,j)\) is a normalized probability
distribution by definition. We define the following classical model:
source \(i\) samples \((\ell_i,j_{i+1})\) according to \(q_i\),
and party \(i\) outputs the unique \(n_i\) such that
\((j_i,\ell_i)\in S_i(n_i)\). This model reproduces the
observed distribution \(P(\mathbf n)\). Indeed,
\cref{eq:exact-fermionic-TC-joint-projector,eq:dilated-projectors-product-form},
together with source independence and the evenness of the spectral
projectors, imply
\[
\begin{aligned}
P(\mathbf n)
&=\sum_{\mathbf j,\boldsymbol\ell}
\left[\prod_{i=1}^{N}q_i(\ell_i,j_{i+1})\right]
\prod_{i=1}^{N}\mathbf1_{(j_i,\ell_i)\in S_i(n_i)},
\end{aligned}
\]
where \(\mathbf{1}[\cdot]\) denotes the indicator function.

Applying classical token-counting rigidity yields binary token functions
$\tau_{i_L}$ and $\tau_{i_R}$, together with a set $\Upsilon$
of spectral-label tuples that has probability one under the
independent source distributions $\{q_i\}_i$.
For every $(\mathbf j,\boldsymbol\ell)\in\Upsilon$ and every $i$,
\begin{align}
\tau_{i_L}(\ell_i)+\tau_{(i+1)_R}(j_{i+1})
&=1,
\label{eq:one-token-per-source}\\
\tau_{i_R}(j_i)+\tau_{i_L}(\ell_i)
&=n_i,
\label{eq:token-counting-response}
\end{align}
where $n_i$ is the unique output such that
$(j_i,\ell_i)\in S_i(n_i)$.
Thus, each source distributes exactly one token, and each party
outputs the total number of tokens it receives.

We now construct local isometries from these token functions.
Let \(S_{i_R}\simeq S_{i_L}\simeq\mathbb C^2\) denote the
extracted system spaces. Choose auxiliary fermionic spaces
with sufficiently many orthogonal states in each parity sector
to encode all spectral labels. For the right subsystem, choose
orthonormal label states \(\ket j\) of parity
\(\tau_{i_R}(j)\); similarly, for the left subsystem, choose
orthonormal label states \(\ket\ell\) of parity
\(\tau_{i_L}(\ell)\). Define the auxiliary spaces
\[
\begin{aligned}
J_{i_R}
&:=\operatorname{span}\{\ket j\}_j\wedge\mathcal H_{i_R},\\
J_{i_L}
&:=\operatorname{span}\{\ket\ell\}_\ell
\wedge(\mathcal H_{i_L}\wedge\mathcal H_{M_i}).
\end{aligned}
\]
For arbitrary vectors \(\ket\varphi\in\mathcal H_{i_R}\) and
\(\ket\chi\in\mathcal H_{i_L}\wedge\mathcal H_{M_i}\), define
\[
\begin{aligned}
\mathcal U_{i_R}\ket\varphi
&:=
\sum_j\ket{\tau_{i_R}(j)}_{S_{i_R}}
\wedge
\bigl(\ket j\wedge\Pi_{i_R}^{(j)}\ket\varphi\bigr)_{J_{i_R}},\\
\mathcal U_{i_LM_i}\ket\chi
&:=
\sum_\ell\ket{\tau_{i_L}(\ell)}_{S_{i_L}}
\wedge
\bigl(\ket\ell\wedge
\Pi_{i_LM_i}^{(\ell)}\ket\chi\bigr)_{J_{i_L}}.
\end{aligned}
\]
As in the Q-QIT proof, each map records the binary token in the
extracted system and retains both the spectral label and the
projected input vector in the junk. The token and label are introduced
jointly. Our choice of encoding ensures that, in each term, they have
the same parity and hence even joint parity. Since also the spectral
projectors are even, the maps preserve the
parity.
Orthogonality and completeness of the spectral PVMs imply
\[
\mathcal U_{i_R}^{\dagger}\mathcal U_{i_R}=\id,
\qquad
\mathcal U_{i_LM_i}^{\dagger}\mathcal U_{i_LM_i}=\id.
\]
Thus the maps above are parity preserving isometries. Their local and global products are therefore well-defined fermionic isometries.

Let \(r_i\) be the probability that source \(i\) sends its token
to party \(i\). The source relation
\cref{eq:one-token-per-source} implies
\[
\begin{aligned}
\bigl(\mathcal U_{i_LM_i}
\wedge\mathcal U_{(i+1)_R}\bigr)
\ket{\widehat\psi_i}\quad=
\sqrt{r_i}\ket{10}\wedge\ket{\zeta_i^a}
+
\sqrt{1-r_i}\ket{01}\wedge\ket{\zeta_i^c}.
\end{aligned}
\]
Here the subsystems are ordered as
\(S_{i_L}\wedge S_{(i+1)_R}\wedge J_{i_L}\wedge J_{(i+1)_R}\).
The junk states \(\ket{\zeta_i^a}\) and \(\ket{\zeta_i^c}\)
are normalized whenever their corresponding coefficients are nonzero.

We now determine \(r_i\).
Party \(i\) receives a token from source \(i\) with probability
\(r_i\), and from source \(i-1\) with probability \(1-r_{i-1}\).
Its expected output is therefore \(r_i+1-r_{i-1}\).
The observed fair token-counting distribution has mean one at every
party, so
\[
\mathbb E[n_i]=r_i+1-r_{i-1}=1.
\]
Thus \(r_i=r_{i-1}\) for every \(i\); denote their common value
by \(r\). Moreover, party \(i\) outputs zero precisely when
neither of its two independent sources sends it a token. Since
the observed probability of this event is \(1/4\),
\[
P(n_i=0)=(1-r_i)r_{i-1}=r(1-r)=\frac14.
\]
so \(r_i=1/2\) for every
\(i\), proving \cref{eq:exact-fermionic-TC-state}.

Finally, \cref{eq:token-counting-response} and the spectral form of
\(\bar\Pi_i^{n_i}\) in \cref{eq:dilated-projectors-product-form} imply
\[
\mathcal U\bar\Pi_i^{n_i}\ket{\widehat\Psi}
=
\Pi_{\mathrm{TC},i}^{n_i}\ket{\Phi},
\qquad n\in\{0,1,2\},
\]
with \(\mathcal U\), \(\ket\Phi\), and
\(\Pi_{\mathrm{TC},i}^{n_i}\) exactly as defined in the statement.
This proves the fermionic partial token-counting self-testing.
\end{proof}

We now apply \cref{thm:exact-fermionic-TC} to the coarse-grained
distribution of the fixed-loop experiment. For each party \(i\),
choose parity-preserving Naimark dilations
\(\{\bar\Pi_{i,x_i}^{a}\}_{a\in\{0,1_+,1_-,2\}}\) of the two
POVMs corresponding to \(x_i=0\) and \(x_i=1\), using a common
ancillary space \(\mathcal H_{M_i}\) initialized in the same
even state \(\ket0_{M_i}\).

Fix the reference input string $\mathbf x = \mathbf 0$ and apply
\cref{thm:exact-fermionic-TC} to the corresponding coarse-grained
distribution.
To distinguish the extracted source states from the physical source
states with their initialized dilation ancillas, define
\[
\ket{\Phi_i}
:=
\bigl(\mathcal U_{i_LM_i}\wedge\mathcal U_{(i+1)_R}\bigr)
\ket{\widehat\psi_i},
\]
where the isometries are those obtained for the reference input string.
The state characterization in \cref{eq:exact-fermionic-TC-state} gives
\begin{equation}
\label{eq:fermionic_fixedloop_partialselftest_state}
\begin{aligned}
\ket{\Phi_i}
=
\frac{1}{\sqrt2}
\Bigl(
\ket{10}_{S_{i_L}S_{(i+1)_R}}
\wedge\ket{\zeta_i^a}_{J_{i_L}J_{(i+1)_R}}
+
\ket{01}_{S_{i_L}S_{(i+1)_R}}
\wedge\ket{\zeta_i^c}_{J_{i_L}J_{(i+1)_R}}
\Bigr).
\end{aligned}
\end{equation}
The extracted global state is
\(\ket\Phi=\mathcal U\ket{\widehat\Psi}
=\bigwedge_{i=1}^N\ket{\Phi_i}\).
Since the extraction isometries preserve parity and the extracted
system states \(\ket{10}\) and \(\ket{01}\) both have odd parity,
the two junk states have the same definite parity:
\[
p(\zeta_i^a)=p(\zeta_i^c)=p(\psi_i)+1\pmod2.
\]

Define the coarse-grained dilated projectors by
\begin{equation}
\bar\Pi_{i,x_i}^{1}
:=\bar\Pi_{i,x_i}^{1_+}+\bar\Pi_{i,x_i}^{1_-},
\end{equation}
while retaining \(\bar\Pi_{i,x_i}^{0}\) and
\(\bar\Pi_{i,x_i}^{2}\) for the zero- and two-token outcomes.

We now show that the coarse-grained projectors are independent
of the input on the support of the local reduced state.
Fix a party \(i\) and set all other parties' inputs to \(0\).
For either choice of \(x_i\in\{0,1\}\), the total token count
is exactly \(N\). Thus, party \(i\) outputs \(n_i\) precisely
when the other parties' outputs sum to \(N-n_i\).

To express this equivalence in terms of the projectors, note that
every joint outcome violating the total-token constraint has zero
probability. Its joint projector therefore annihilates
\(\ket{\widehat\Psi}\). Using completeness of the local PVMs, we
obtain, for each fixed
\(n_i\in\{0,1,2\}\),
\[
\begin{aligned}
\bar\Pi_{i,x_i}^{n_i}\ket{\widehat\Psi}=
\sum_{\substack{\mathbf n_{-i}\\
\sum_{k\ne i}n_k=N-n_i}}
\left(\prod_{k\ne i}\bar\Pi_{k,0}^{n_k}\right)
\ket{\widehat\Psi},
\end{aligned}
\]
where \(\mathbf n_{-i}\) denotes the coarse-grained outcomes
of all parties other than \(i\). The right-hand side contains
only the other parties' projectors and is therefore independent
of \(x_i\). Consequently,
\[
\bigl(\bar\Pi_{i,0}^{n_i}-\bar\Pi_{i,1}^{n_i}\bigr)
\ket{\widehat\Psi}=0.
\]

Since each pure source state has definite total parity, so does
the global state \(\ket{\widehat\Psi}\). To establish equality on
the local support, we choose a Schmidt decomposition across party
\(i\) and the remaining parties in local parity eigenbases
\cite[Theorem~11 and Appendix~D]{vidal2021quantum}:
\[
\ket{\widehat\Psi}
=
\sum_k\sqrt{\lambda_k}\,
\ket{u_k}_i\wedge\ket{v_k}_{-i},
\qquad \lambda_k>0.
\]
Substituting this decomposition into the preceding equality
and using the orthonormality of the vectors
\(\ket{v_k}_{-i}\) gives
\begin{equation}
\label{eq:fermionic_coarse_projectors_input_independent}
\bar\Pi_{i,0}^{n_i}\ket{u_k}_i
=
\bar\Pi_{i,1}^{n_i}\ket{u_k}_i
\qquad\text{for every }k.
\end{equation}
Thus, the two projectors agree on the support of the local reduced
state of $\ket{\widehat \Psi}$. Consequently, the isometries constructed
for the reference input string also apply to every other input string,
giving
\begin{equation}
\label{eq:fermionic_fixedloop_partialselftest_measurement}
\begin{aligned}
\mathcal U \bar{\Pi}_{i,x_i}^{0} \ket{\widehat \Psi}
&=\Pi_{\mathrm{TC},i}^{0} \ket{\Phi} \\
\mathcal U \bar{\Pi}_{i,x_i}^{1} \ket{\widehat \Psi}
&=\Pi_{\mathrm{TC},i}^{1} \ket{\Phi} \\
\mathcal U \bar{\Pi}_{i,x_i}^{2} \ket{\widehat \Psi}
&=\Pi_{\mathrm{TC},i}^{2} \ket{\Phi} ,
\end{aligned}
\end{equation}
where $\Pi_{\mathrm{TC},i}^{n_i}$ are the token-counting projectors
defined in \cref{eq:exact-fermionic-TC-projectors}.

For each party \(i\) and input \(x_i\in\{0,1\}\), define the
dilated signed observable by
\[
M_{x_i}^{(i)}
:=\bar\Pi_{i,x_i}^{1_+}-\bar\Pi_{i,x_i}^{1_-}.
\]
The conditional distribution maximally violates Bell inequality~\Romannum{1} when \(N\) is odd and Bell inequality~\Romannum{2} when \(N\) is even:
\begin{equation}
\begin{aligned}
&\text{Bell inequality \Romannum{1}:}\\[-0.5ex]
&B_1\equiv{}
(N-1)
\bigl(M_0^{(1)}+M_1^{(1)}\bigr)
M_1^{(2)}\cdots M_1^{(N)}
+
\sum_{i=2}^N
\bigl(M_0^{(1)}-M_1^{(1)}\bigr)
M_0^{(i)},
\qquad \langle B_1\rangle\leq\beta_C .
\end{aligned}
\label{eq:Bell_1}
\end{equation}
\begin{equation}
\begin{aligned}
&\text{Bell inequality \Romannum{2}:}\\[-0.5ex]
&B_2\equiv{}
-(N-1)
\bigl(M_0^{(1)}+M_1^{(1)}\bigr)
M_1^{(2)}\cdots M_1^{(N)}
+
\sum_{i=2}^N
\bigl(M_0^{(1)}-M_1^{(1)}\bigr)
M_0^{(i)}, \qquad
\langle B_2\rangle\leq\beta_C .
\end{aligned}
\label{eq:Bell_2}
\end{equation}
where $\beta_C=2 (N-1) $ and $\beta_Q= 2(N-1)\sqrt{2}$. The expectation values above are evaluated on the normalized
dilated conditional state
\[
\ket{\widehat\Psi}^{\mathrm{cond}}
:=
2^{(N-1)/2}
\left(\prod_{i=1}^{N}\bar\Pi_{i,x_i}^{1}\right)
\ket{\widehat\Psi},
\]
where the normalization factor is the inverse square root of
the probability \(2^{1-N}\) that every party obtains one token.

Applying the same conditioning to the extracted state gives
the normalized conditional state
\begin{equation}
\label{eq:fermionic_fixedloop_conditional_state}
\begin{aligned}
\ket\Phi^{\mathrm{cond}}
&=2^{(N-1)/2}
\left(\prod_{i=1}^N\Pi_{\mathrm{TC},i}^{1}\right)\ket\Phi
\\
&=\frac{1}{\sqrt2}\Biggl[
\bigwedge_{i=1}^N
\Bigl(
\ket{10}_{S_{i_L}S_{(i+1)_R}}
\wedge\ket{\zeta_i^a}_{J_{i_L}J_{(i+1)_R}}
\Bigr)
+\bigwedge_{i=1}^N
\Bigl(
\ket{01}_{S_{i_L}S_{(i+1)_R}}
\wedge\ket{\zeta_i^c}_{J_{i_L}J_{(i+1)_R}}
\Bigr)
\Biggr].
\end{aligned}
\end{equation}

The Bell expectation values above are unchanged under the
extraction isometries: the dilated conditional state
\(\ket{\widehat\Psi}^{\mathrm{cond}}\) and signed observables
\(M_{x_i}^{(i)}\) can be replaced by their extracted
counterparts \(\ket\Phi^{\mathrm{cond}}\) and
\(\mathcal U_iM_{x_i}^{(i)}\mathcal U_i^\dagger\),
respectively. We therefore work in the extracted representation
for the remainder of the proof. Let \(\mathcal X^{\prime}_i\) and
\(\mathcal Z^{\prime}_i\) denote the extracted counterparts of the
dilated observables in \cref{thm:fixed-loop-characterization}.
Thus, for the distinguished party \(1\),
\[
\begin{aligned}
\mathcal X^{\prime}_1
&:=\mathcal U_1
\frac{M_0^{(1)}+M_1^{(1)}}{\sqrt2}
\mathcal U_1^\dagger,\\
\mathcal Z^{\prime}_1
&:=\mathcal U_1
\frac{M_0^{(1)}-M_1^{(1)}}{\sqrt2}
\mathcal U_1^\dagger,
\end{aligned}
\]
and, for \(i\in\{2,\ldots,N\}\),
\[
\begin{aligned}
\mathcal X^{\prime}_i&:=\mathcal U_iM_1^{(i)}\mathcal U_i^\dagger,\\
\mathcal Z^{\prime}_i&:=\mathcal U_iM_0^{(i)}\mathcal U_i^\dagger.
\end{aligned}
\]

From maximal violation of the Bell inequalities, we can derive
stabilizer relations for the conditional state. The following
lemma gives these relations and is the fermionic analogue of
Lemmas~F.1 and~F.2 in Supplementary Material~F of
Ref.~\cite{kalarde2026fermions}.

\begin{lemma}
\label{lem:fermionic_Bell_relations}
Under maximal violation of Bell inequality~\Romannum{1} for odd
\(N\), or Bell inequality~\Romannum{2} for even \(N\),
\begin{equation}
\label{eq:stab1}
\mathcal X^{\prime}_1\cdots \mathcal X^{\prime}_N
\ket{\Phi}^{\mathrm{cond}}
=
(-1)^{N-1}
\ket{\Phi}^{\mathrm{cond}} .
\end{equation}
Moreover, for every \(i\in\{1,\dots,N\}\),
\begin{equation}
\label{eq:local_pauli_relations}
\begin{aligned}
&(\mathcal X_i^{\prime})^2
\ket{\Phi}^{\mathrm{cond}}
=
(\mathcal Z_i^{\prime})^2
\ket{\Phi}^{\mathrm{cond}}
=
\ket{\Phi}^{\mathrm{cond}},
\\
&\{\mathcal X^{\prime}_i,\mathcal Z^{\prime}_i\}
\ket{\Phi}^{\mathrm{cond}}
=
0 .
\end{aligned}
\end{equation}
\end{lemma}

\begin{proof}
The proof is the same as that of Lemmas~F.1 and~F.2 in
Supplementary Material~F of Ref.~\cite{kalarde2026fermions}.
\end{proof}

We next introduce the local conditional subspaces. For each party \(i\),
let
\[
\mathcal H_i^{\mathrm{cond}}
:=
\operatorname{supp}
\left(
\operatorname{Tr}_{-i}
\left[
\ket{\Phi}^{\mathrm{cond}}
\bra{\Phi}^{\mathrm{cond}}
\right]
\right),
\]
where \(\operatorname{Tr}_{-i}\) denotes the partial
trace over all parties other than \(i\). Fermionic partial trace is well defined and preserves the
expectation values of all local physical observables~\cite[Sec.~III~A]{vidal2021quantum}.

Define the junk spaces by
\begin{equation}
\label{eq:fermionic_junk_supports}
\begin{aligned}
J_{i_L}^c
&:=
\operatorname{supp}\left(
\operatorname{Tr}_{J_{(i+1)_R}}
\left[\ket{\zeta_i^c}\bra{\zeta_i^c}\right]
\right),\\
J_{i_R}^c
&:=
\operatorname{supp}\left(
\operatorname{Tr}_{J_{(i-1)_L}}
\left[\ket{\zeta_{i-1}^c}\bra{\zeta_{i-1}^c}\right]
\right), \\
J_{i_L}^a
&:=
\operatorname{supp}\left(
\operatorname{Tr}_{J_{(i+1)_R}}
\left[\ket{\zeta_i^a}\bra{\zeta_i^a}\right]
\right),\\
J_{i_R}^a
&:=
\operatorname{supp}\left(
\operatorname{Tr}_{J_{(i-1)_L}}
\left[\ket{\zeta_{i-1}^a}\bra{\zeta_{i-1}^a}\right]
\right),
\end{aligned}
\end{equation}
with source indices understood modulo \(N\).
Since the global junk state in each branch of the conditional state is a product over sources, define the local junk spaces at party \(i\) as
\[
J_i^a:=J_{i_R}^a\wedge J_{i_L}^a,
\qquad
J_i^c:=J_{i_R}^c\wedge J_{i_L}^c.
\]
With the convention of \cref{eq:fermionic_fixedloop_conditional_state},
the anticlockwise term \(a\) corresponds to the system state
\(\ket{01}_{S_{i_R}S_{i_L}}\), while the clockwise term \(c\)
corresponds to the system state \(\ket{10}_{S_{i_R}S_{i_L}}\). Therefore
\[
\mathcal H_i^{\mathrm{cond}}
=
\{\ket{01}_{S_{i_R}S_{i_L}} \}\wedge J_i^a
\;\oplus\;
\{ \ket{10}_{S_{i_R}S_{i_L}} \} \wedge J_i^c.
\]

The following lemma characterizes the operators \(\mathcal Z^{\prime}_i\).
It is the fermionic analogue of Lemma~F.3 in Supplementary
Material~F of Ref.~\cite{kalarde2026fermions}.

\begin{lemma}
\label{lem:fermionic_Z_on_Hcond}
For every party \(i\), on \(\mathcal H_i^{\mathrm{cond}}\),
\begin{equation}
\label{eq:fermionic_Z_on_Hcond}
\mathcal Z^{\prime}_i
=
Z_{S_i}\wedge \mathbb 1_{J_i},
\end{equation}
where
\[
Z_{S_i}
=
\ket{01}\!\bra{01}_{S_{i_R}S_{i_L}}
-
\ket{10}\!\bra{10}_{S_{i_R}S_{i_L}} .
\]
\end{lemma}

\begin{proof}
The proof is the same as that of Lemma~F.3 in Supplementary
Material~F of Ref.~\cite{kalarde2026fermions}.
\end{proof}

Next, we show that the range of \(\mathcal X^{\prime}_i\) lies in the
conditional subspace \(\mathcal H_i^{\mathrm{cond}}\).
The following lemma is the fermionic analogue of Lemma~F.4
in Supplementary Material~F of Ref.~\cite{kalarde2026fermions}.

\begin{lemma}
\label{lem:fermionic_X_preserves_Hcond}
For every party \(i\), \(\mathcal X^{\prime}_i\) preserves the local
conditional subspace:
\begin{equation}
\label{eq:fermionic_X_preserves_Hcond}
\mathcal X^{\prime}_i\mathcal H_i^{\mathrm{cond}}
\subseteq
\mathcal H_i^{\mathrm{cond}} .
\end{equation}
\end{lemma}

\begin{proof}
The proof is the same as that of Lemma~F.4 in Supplementary
Material~F of Ref.~\cite{kalarde2026fermions}.
\end{proof}

We now determine the structure of the operators \(\mathcal X^{\prime}_i\).
The following lemma is the fermionic analogue of Lemma~F.5
in Supplementary Material~F of Ref.~\cite{kalarde2026fermions}.

\begin{lemma}
\label{lem:fermionic_X_block_form}
For every party \(i\), there exists a parity-preserving unitary
\[
T_i:J_i^c\to J_i^a
\]
such that on $\mathcal H_i^{\mathrm{cond}}$,
\begin{equation}
\label{eq:fermionic_X_block_form}
\begin{aligned}
\mathcal X^{\prime}_i
&=
\ket{01}\!\bra{10}_{S_{i_R}S_{i_L}} \wedge T_i
+
\ket{10}\!\bra{01}_{S_{i_R}S_{i_L}} \wedge T_i^\dagger
\end{aligned} .
\end{equation}
\end{lemma}

\begin{proof}
The existence of the unitary \(T_i\) and the block decomposition
follow by the same argument as in the proof of Lemma~F.5 in
Supplementary Material~F of Ref.~\cite{kalarde2026fermions}.
To establish parity preservation, note that the system transition
in each block of \cref{eq:fermionic_X_block_form} preserves
system parity. Moreover, the two terms act on orthogonal system
subspaces.
Therefore, since \(\mathcal X^{\prime}_i\) preserves parity, \(T_i\) and
\(T_i^\dagger\) must both preserve junk parity on
their respective domains \(J_i^c\) and \(J_i^a\).
\end{proof}

The following lemma is the fermionic analogue of Corollary~F.6 in
Supplementary Material~F of Ref.~\cite{kalarde2026fermions}.

\begin{lemma}
\label{lem:fermionic_junk_relations}
The unitaries \(T_i\) satisfy
\begin{equation}
\label{eq:T_relation_c_to_a}
\left(
T_1\wedge\cdots\wedge T_N
\right)
\left(
\bigwedge_{i=1}^{N}\ket{\zeta_i^c}
\right)
=
\bigwedge_{i=1}^{N}\ket{\zeta_i^a},
\end{equation}
and
\begin{equation}
\label{eq:Tdag_relation_a_to_c}
\left(
T_1^\dagger\wedge\cdots\wedge T_N^\dagger
\right)
\left(
\bigwedge_{i=1}^{N}\ket{\zeta_i^a}
\right)
=
\bigwedge_{i=1}^{N}\ket{\zeta_i^c}.
\end{equation}
\end{lemma}

\begin{proof}
The proof follows by substituting the block form
\cref{eq:fermionic_X_block_form} into the stabilizer relation
\cref{eq:stab1} and using the expression for the conditional
state in \cref{eq:fermionic_fixedloop_conditional_state}.
The argument is similar to the proof of Corollary~F.6 in
Supplementary Material~F of Ref.~\cite{kalarde2026fermions},
with the fermionic reordering signs tracked explicitly.

We denote the two components of the conditional state by
\begin{equation}
\label{eq:conditional_a_source_order}
\ket{\Phi}^{\mathrm{cond},a}
=
\bigwedge_{i=1}^{N}
\left(
\ket{10}_{S_{i_L}S_{(i+1)_R}}
\wedge
\ket{\zeta_i^a}
\right),
\end{equation}
and
\begin{equation}
\label{eq:conditional_c_source_order}
\ket{\Phi}^{\mathrm{cond},c}
=
\bigwedge_{i=1}^{N}
\left(
\ket{01}_{S_{i_L}S_{(i+1)_R}}
\wedge
\ket{\zeta_i^c}
\right).
\end{equation} 
Here the wedge products are taken in increasing order of the source
label \(i\). Thus
\begin{equation}
\label{eq:conditional_state_ac_components}
\ket{\Phi}^{\mathrm{cond}}
=
\frac{1}{\sqrt{2}}
\left(
\ket{\Phi}^{\mathrm{cond},a}
+
\ket{\Phi}^{\mathrm{cond},c}
\right).
\end{equation}

We first reorder the subsystems so that all system factors precede the junk factors. In the \(a\)-branch, the resulting order is
\[
S_{1_L},S_{2_L},\dots,S_{N_L},
\]
whereas in the \(c\)-branch it is
\[
S_{2_R},S_{3_R},\dots,S_{N_R},S_{1_R}.
\]
Moving the system factor of source \(j\) past the junk factor of each
earlier source \(i<j\) contributes a sign \((-1)^{p_i}\), where
\(p_i:=p(\zeta_i^a)=p(\zeta_i^c)\) denotes the common parity of the two junk components of source \(i\).
The total sign from this reordering is therefore
\begin{equation}
\label{eq:sigma_definition}
(-1)^\sigma,
\qquad
\sigma
=
\sum_{j=1}^N \sum_{ i<j}p_i
=
\sum_{i=1}^{N}(N-i)p_i .
\end{equation}
Because \(\ket{\zeta_i^a}\) and \(\ket{\zeta_i^c}\) have the same parity \(p_i\), the same sign \((-1)^\sigma\) appears in both
branches. Thus
\begin{equation}
\label{eq:conditional_a_systems_first}
\ket{\Phi}^{\mathrm{cond},a}
=
(-1)^\sigma
\left(
\bigwedge_{i=1}^{N}
\ket{10}_{S_{i_L}S_{(i+1)_R}}
\right)
\wedge
\left(
\bigwedge_{i=1}^{N}
\ket{\zeta_i^a}
\right),
\end{equation}
and
\begin{equation}
\label{eq:conditional_c_systems_first}
\ket{\Phi}^{\mathrm{cond},c}
=
(-1)^\sigma
\left(
\bigwedge_{i=1}^{N}
\ket{01}_{S_{i_L}S_{(i+1)_R}}
\right)
\wedge
\left(
\bigwedge_{i=1}^{N}
\ket{\zeta_i^c}
\right).
\end{equation}

We expand \(\mathcal X^{\prime}_1\cdots\mathcal X^{\prime}_N\) using the block
decomposition in \cref{eq:fermionic_X_block_form}.
We reorder the subsystems so that all system factors precede
the junk factors. Since each \(T_i\) preserves parity, it can
be moved past the system factors without introducing
additional fermionic signs.

We now evaluate the action of \(\mathcal X^{\prime}_1\cdots\mathcal X^{\prime}_N\)
on each branch of the conditional state. On the \(c\)-branch,
only the first block of each \(\mathcal X^{\prime}_i\), containing
\(T_i\), contributes. Using the definition of the wedge product
in \cref{eq:WedgeDefiniton}, we write the system transitions
in terms of creation and annihilation operators as
\[
\ket{01}\!\bra{10}_{S_{i_R}S_{i_L}}
=
f_{S_{i_L}}^\dagger f_{S_{i_R}}.
\]
In the same representation, the system part of
\(\ket{\Phi}^{\mathrm{cond},c}\) is
\[
f_{S_{2_R}}^\dagger
f_{S_{3_R}}^\dagger
\cdots
f_{S_{N_R}}^\dagger
f_{S_{1_R}}^\dagger\ket{\Omega}.
\]
Applying the product of these system transitions therefore gives
\begin{equation}
\label{eq:system_sign_c_to_a}
\begin{aligned}
&
\left(
\prod_{i=1}^{N}
f_{S_{i_L}}^\dagger f_{S_{i_R}}
\right)
\left(
f_{S_{2_R}}^\dagger
f_{S_{3_R}}^\dagger
\cdots
f_{S_{N_R}}^\dagger
f_{S_{1_R}}^\dagger
\right)
\ket{\Omega}=
(-1)^{N-1}
f_{S_{1_L}}^\dagger\cdots f_{S_{N_L}}^\dagger
\ket{\Omega},
\end{aligned}
\end{equation}
where the last equality follows from the fermionic
anticommutation relations. This implies
\begin{equation}
\label{eq:X_on_c_branch}
\begin{aligned}
&
\left(\mathcal X^{\prime}_1\wedge\cdots\wedge\mathcal X^{\prime}_N\right)
\ket{\Phi}^{\mathrm{cond},c}
\\
&\qquad =
(-1)^\sigma(-1)^{N-1}
\left(
\bigwedge_{i=1}^{N}
\ket{10}_{S_{i_L}S_{(i+1)_R}}
\right)
\wedge
\left(
T_1\wedge\cdots\wedge T_N
\right)
\left(
\bigwedge_{i=1}^{N}\ket{\zeta_i^c}
\right).
\end{aligned}
\end{equation}

Similarly, on the \(a\)-branch, only the second block of each
\(\mathcal X^{\prime}_i\) in \cref{eq:fermionic_X_block_form},
containing \(T_i^\dagger\), contributes. Its action on the
system parts of \(\ket{\Phi}^{\mathrm{cond},a}\) gives
\begin{equation}
\label{eq:system_sign_a_to_c}
\begin{aligned}
&
\left(
\prod_{i=1}^{N}
f_{S_{i_R}}^\dagger f_{S_{i_L}}
\right)
\left(
f_{S_{1_L}}^\dagger
f_{S_{2_L}}^\dagger
\cdots
f_{S_{N_L}}^\dagger
\right)
\ket{\Omega} =
(-1)^{N-1}
f_{S_{2_R}}^\dagger
f_{S_{3_R}}^\dagger
\cdots
f_{S_{N_R}}^\dagger
f_{S_{1_R}}^\dagger
\ket{\Omega}.
\end{aligned}
\end{equation}
Including the action on the junk factors and the reordering
sign \((-1)^\sigma\), we obtain
\begin{equation}
\label{eq:X_on_a_branch}
\begin{aligned}
&
\left(\mathcal X^{\prime}_1\wedge\cdots\wedge\mathcal X^{\prime}_N\right)
\ket{\Phi}^{\mathrm{cond},a}
\\
&\qquad =
(-1)^\sigma(-1)^{N-1}
\left(
\bigwedge_{i=1}^{N}
\ket{01}_{S_{i_L}S_{(i+1)_R}}
\right)
\wedge
\left(
T_1^\dagger\wedge\cdots\wedge T_N^\dagger
\right)
\left(
\bigwedge_{i=1}^{N}\ket{\zeta_i^a}
\right).
\end{aligned}
\end{equation}

By \cref{eq:conditional_state_ac_components} and the stabilizer
relation \cref{eq:stab1}, we have
\begin{equation}
\begin{aligned}
&
\left(\mathcal X^{\prime}_1\wedge\cdots\wedge\mathcal X^{\prime}_N\right)
\left(\ket{\Phi}^{\mathrm{cond},c}
+\ket{\Phi}^{\mathrm{cond},a}\right)
=
(-1)^{N-1}
\left(\ket{\Phi}^{\mathrm{cond},c}
+\ket{\Phi}^{\mathrm{cond},a}\right).
\end{aligned}
\end{equation}
Using
\cref{eq:conditional_a_systems_first,eq:conditional_c_systems_first,eq:X_on_c_branch,eq:X_on_a_branch}
and comparing the coefficients of the two orthogonal system
branches on both sides, we obtain
\[
\left(T_1\wedge\cdots\wedge T_N\right)
\left(\bigwedge_{i=1}^{N}\ket{\zeta_i^c}\right)
=
\bigwedge_{i=1}^{N}\ket{\zeta_i^a},
\]
and
\[
\left(T_1^\dagger\wedge\cdots\wedge T_N^\dagger\right)
\left(\bigwedge_{i=1}^{N}\ket{\zeta_i^a}\right)
=
\bigwedge_{i=1}^{N}\ket{\zeta_i^c}.
\]

The fermionic reordering sign therefore exactly cancels the sign
in the stabilizer relation \cref{eq:stab1}. This is one of the main
differences from the standard quantum argument.
\end{proof}

The next lemma shows that each \(T_i\) has a product form
over the right and left junk spaces. It is the fermionic
analogue of Lemma~F.7 in Supplementary Material~F of
Ref.~\cite{kalarde2026fermions}.

\begin{lemma}
\label{lem:fermionic_T_factorization}
For each \(i\), the unitary
\[
T_i:J_{i_R}^c \wedge J_{i_L}^c \to J_{i_R}^a \wedge J_{i_L}^a
\]
factorizes across the junk spaces. More precisely, there exist
unitaries
\[
T_i^{(L)}:J_{i_L}^c\to J_{i_L}^a,
\qquad
T_i^{(R)}:J_{i_R}^c\to J_{i_R}^a,
\]
with equal parities such that,
\begin{equation}
\label{eq:fermionic_T_factorization}
T_i \ket{\zeta^c_{i-1}}\ket{\zeta^c_{i}}
=
T_i^{(R)}\wedge T_i^{(L)} \ket{\zeta^c_{i-1}}\ket{\zeta^c_{i}}.
\end{equation}
\end{lemma}

\begin{proof}
The factorization follows by applying the cyclic product-structure lemma
(Lemma~K.1 of Ref.~\cite{kalarde2026fermions}) to
\cref{eq:T_relation_c_to_a}, using its fermionic extension discussed above. 

Since \(T_i\) preserves parity,
\cref{eq:fermionic_T_factorization} shows that
\(T_i^{(R)}\wedge T_i^{(L)}\) preserves parity when acting
on the two-source junk state. To show that this parity
preservation holds on the entire local junk space, write a
Schmidt decomposition of this state across party \(i\)
and the remaining spaces:
\[
\begin{aligned}
\ket{\zeta_{i-1}^c}\wedge\ket{\zeta_i^c}=
\sum_\lambda s_\lambda\,
\ket{u_\lambda}_i\wedge\ket{v_\lambda}_{-i},
\quad s_\lambda>0.
\end{aligned}
\]
Source independence and the definitions of the junk spaces imply
that the local reduced state at party \(i\) has support
\(J_{i_R}^c\wedge J_{i_L}^c\). Thus, the vectors
\(\ket{u_\lambda}_i\) form an orthonormal basis of this space.
Substituting this Schmidt decomposition into
\cref{eq:fermionic_T_factorization} and using the orthonormality
of the vectors \(\ket{v_\lambda}_{-i}\) gives
\begin{equation}
\label{eq:thisone}
T_i\ket{u_\lambda}_i
=
\bigl(T_i^{(R)}\wedge T_i^{(L)}\bigr)\ket{u_\lambda}_i
\qquad\text{for every }\lambda,
\end{equation}
which extends by linearity to the entire space
\(J_{i_R}^c\wedge J_{i_L}^c\). Since \(T_i\) preserves parity,
so does the product \(T_i^{(R)}\wedge T_i^{(L)}\) on this
entire space. 

We now argue that both $T_i^{(R)}$ and $T_i^{(L)}$ have
definite parity on their entire domains.
Since \(J_i^c=J_{i_R}^c\wedge J_{i_L}^c\) is a product space,
we can choose a product basis
\(\{\ket{r_\mu}\wedge\ket{l_\nu}\}_{\mu,\nu}\).
Here, \(\{\ket{r_\mu}\}_\mu\) and \(\{\ket{l_\nu}\}_\nu\)
are orthonormal bases of \(J_{i_R}^c\) and \(J_{i_L}^c\),
respectively, and each basis element has definite parity.
Written in this product basis, \cref{eq:thisone} shows that both
$T_i^{(R)}\ket{r_\mu}$ and $ T_i^{(L)} \ket{l_\nu}$ have definite parity.

Now fix a right basis vector \(\ket{r_{\mu_0}}\). If
\(T_i^{(R)}\) preserves its parity, pairing this fixed
vector with every left basis vector shows that \(T_i^{(L)}\)
must preserve parity on every \(\ket{l_\nu}\). Similarly,
if \(T_i^{(R)}\) changes its parity, \(T_i^{(L)}\) must
change parity on every \(\ket{l_\nu}\). Thus, by linearity,
\(T_i^{(L)}\) either preserves or reverses parity on all of \(J_{i_L}^c\).

Next, fix a left basis vector \(\ket{l_{\nu_0}}\) and pair it
with every right basis vector. Since \(T_i^{(L)}\) has a
definite parity, \(T_i^{(R)}\) must have the same definite parity on every \(\ket{r_\mu}\). By linearity, this holds
on all of \(J_{i_R}^c\). Hence, the two factors are either
both parity-preserving or both parity-changing on their
entire respective junk domains.

\end{proof}

The following lemma is the fermionic analogue of Corollary~F.8 in
Supplementary Material~F of Ref.~\cite{kalarde2026fermions}.

\begin{lemma}
\label{lem:fermionic_local_phase_relations}
For every source \(i\), the unitaries acting on its two junk
subsystems satisfy
\begin{equation}
\label{eq:fermionic_local_phase_relation}
\left(
T_i^{(L)}\wedge T_{i+1}^{(R)}
\right)
\ket{\zeta_i^c}
=
e^{\ii\widehat\varphi_i}
\ket{\zeta_i^a}.
\end{equation}
Let \(q:=p(T_1^{(L)})=p(T_1^{(R)})\). The phases satisfy
\begin{equation}
\label{eq:fermionic_phase_sum}
\sum_{i=1}^N \widehat\varphi_i =q\pi
\qquad \mathrm{mod}\ 2\pi .
\end{equation}
\end{lemma}

\begin{proof}
Combining the factorization in
\cref{eq:fermionic_T_factorization} with
\cref{eq:T_relation_c_to_a}, as in the proof of Corollary~F.8
in Supplementary Material~F of Ref.~\cite{kalarde2026fermions},
gives the local phase relations
\cref{eq:fermionic_local_phase_relation}.

To determine the sum of these phases, we express the product of the
\(T_i\)'s in source order by moving \(T_1^{(R)}\) to the end of
the ordered product. Its exchange with \(T_1^{(L)}\) contributes
\((-1)^q\).  Moving it past \(T_2,\ldots,T_N\) produces
no sign, since each \(T_i\) preserves parity. Thus,
\[
\begin{aligned}
\bigwedge_{i=1}^N T_i
\quad=(-1)^q
\bigwedge_{i=1}^N
\bigl(T_i^{(L)}\wedge T_{i+1}^{(R)}\bigr),
\end{aligned}
\]
with indices understood modulo \(N\). Substituting this into
\cref{eq:T_relation_c_to_a} therefore gives \cref{eq:fermionic_local_phase_relation} and
\[
(-1)^q=\exp\!\left(\ii\sum_{i=1}^N\widehat\varphi_i\right),
\]
which is equivalent to \cref{eq:fermionic_phase_sum}.
\end{proof}

We now show that \(T_i^{(L)}\) and \(T_{i+1}^{(R)}\)
have the same parity on their entire respective junk domains.
To this end, write a Schmidt decomposition
\[
\ket{\zeta_i^c}
=
\sum_\lambda s_\lambda\,
\ket{u_\lambda}\wedge\ket{v_\lambda},
\qquad s_\lambda>0,
\]
where \(\{\ket{u_\lambda}\}_\lambda\) and
\(\{\ket{v_\lambda}\}_\lambda\) are orthonormal bases of
\(J_{i_L}^c\) and \(J_{(i+1)_R}^c\), respectively, with
each basis vector chosen to have definite parity.
Thus, for every \(\lambda\),
\[
p(u_\lambda)+p(v_\lambda)
=
p(\zeta_i^c)\pmod 2.
\]
Substituting this decomposition into
\cref{eq:fermionic_local_phase_relation} gives
\[
\begin{aligned}
&\sum_\lambda s_\lambda
(-1)^{p(T_{i+1}^{(R)})p(u_\lambda)}
\times
\bigl(T_i^{(L)}\ket{u_\lambda}\bigr)
\wedge
\bigl(T_{i+1}^{(R)}\ket{v_\lambda}\bigr)
=
e^{\ii\widehat\varphi_i}\ket{\zeta_i^a},
\end{aligned}
\]
where the sign comes from the graded fermionic wedge product.

By \cref{lem:fermionic_T_factorization}, each factor has
a definite parity on its entire domain. The transformed
Schmidt vectors therefore also have definite parity.
Moreover, the Schmidt terms remain orthogonal because both
factors are unitary. Since \(\ket{\zeta_i^a}\) and
\(\ket{\zeta_i^c}\) have the same parity, each term on the
left-hand side must have this same parity. Thus, for every
\(\lambda\),
\[
\begin{aligned}
p\bigl(T_i^{(L)}u_\lambda\bigr)
+p\bigl(T_{i+1}^{(R)}v_\lambda\bigr)
=
p(u_\lambda)+p(v_\lambda) \; \pmod 2.
\end{aligned}
\]
The two operators \(T_i^{(L)}\) and \(T_{i+1}^{(R)}\) therefore have the same parity action on each pair of Schmidt basis vectors. These vectors form bases of their respective junk domains, and each operator
has a definite parity throughout its domain by
\cref{lem:fermionic_T_factorization}.
Hence, \(T_i^{(L)}\) and \(T_{i+1}^{(R)}\) are either both parity-preserving or both parity-changing on their entire respective domains.

Combining the fact that \(T_i^{(L)}\) and \(T_i^{(R)}\)
have equal parities with the above result, we conclude that
all factors \(T_i^{(L)}\) and \(T_i^{(R)}\), for all \(i\),
share the same parity \(q\) on their entire respective domains.

We now have two possibilities: either $q=0$ or $q=1$.

\noindent\emph{Case(i): q=0.—}
For each party \(i\), define the following parity-preserving
unitaries, analogously to Eq.~(F.29) of
Ref.~\cite{kalarde2026fermions}:
\begin{equation}
\label{eq:fermionic_local_unitaries}
\begin{aligned}
U_{i_R}
&:=
\ket{0}\!\bra{0}_{S_{i_R}}
\wedge \mathbb 1_{J_{i_R}^a}
+
\ket{1}\!\bra{1}_{S_{i_R}}
\wedge T_i^{(R)},
\\
U_{i_L}
&:=
\ket{1}\!\bra{1}_{S_{i_L}}
\wedge \mathbb 1_{J_{i_L}^a}
+
\ket{0}\!\bra{0}_{S_{i_L}}
\wedge T_i^{(L)} .
\end{aligned}
\end{equation}
Their domains and codomains are
\begin{equation}
\label{eq:fermionic_local_unitary_spaces}
\begin{aligned}
U_{i_R}:{}&
\bigl(\{\ket0_{S_{i_R}}\}\wedge J_{i_R}^a\bigr)
\oplus
\bigl(\{\ket1_{S_{i_R}}\}\wedge J_{i_R}^c\bigr)
\longrightarrow S_{i_R}\wedge J_{i_R}^a,
\\[0.5ex]
U_{i_L}:{}&
\bigl(\{\ket1_{S_{i_L}}\}\wedge J_{i_L}^a\bigr)
\oplus
\bigl(\{\ket0_{S_{i_L}}\}\wedge J_{i_L}^c\bigr)
\longrightarrow S_{i_L}\wedge J_{i_L}^a.
\end{aligned}
\end{equation}
Their action is illustrated in \cref{fig:selftest_unitaries}.

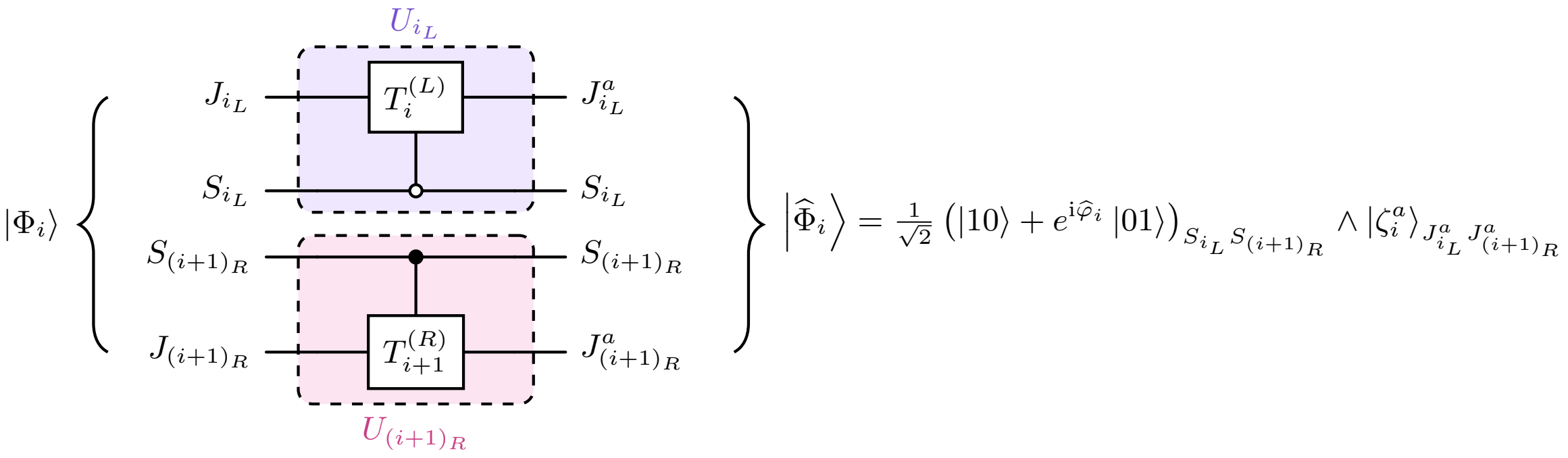
\begin{figure}[!ht]
    \centering
    \definecolor{swapPurple}{HTML}{8B5CF6}
    \definecolor{swapPink}{HTML}{EC4899}
    \begin{adjustbox}{max width=\columnwidth}
    \begin{quantikz}[execute at end picture={
  \draw[decoration={brace,amplitude=8pt},decorate,thick]
    ([xshift=-1.55cm]\tikzcdmatrixname-4-1.west) --
    ([xshift=-1.55cm]\tikzcdmatrixname-1-1.west);
  \node[left] at
    ([xshift=-1.90cm]$(\tikzcdmatrixname-1-1.west)!0.5!(\tikzcdmatrixname-4-1.west)$)
    {$\ket{\Phi_i}$};
  \draw[decoration={brace,amplitude=8pt},decorate,thick]
    ([xshift=1.65cm]\tikzcdmatrixname-1-5.east) --
    ([xshift=1.65cm]\tikzcdmatrixname-4-5.east);
  \node[right,font=\small] at
    ([xshift=2.00cm]$(\tikzcdmatrixname-1-5.east)!0.5!(\tikzcdmatrixname-4-5.east)$)
    {$\ket{\widehat\Phi_i}
      =\frac{1}{\sqrt2}
      \left(\ket{10}+e^{\ii\widehat\varphi_i}\ket{01}\right)_{S_{i_L}S_{(i+1)_R}}
      \wedge\ket{\zeta_i^a}_{J_{i_L}^{a}J_{(i+1)_R}^{a}}$};
}]
\lstick{$J_{i_L}$}
    & \gategroup[2,steps=3,style={dashed,rounded corners,fill=swapPurple!15,
                 inner xsep=2pt,inner ysep=1pt},background,
                 label style={label position=north,anchor=south,yshift=-0.25cm}]{\textcolor{swapPurple!85!black}{$U_{i_L}$}}
    & \gate{T_i^{(L)}} & & \rstick{$J_{i_L}^{a}$} \\
\lstick{$S_{i_L}$}
    & & \ctrl[open]{-1} & & \rstick{$S_{i_L}$} \\
\lstick{$S_{(i+1)_R}$}
    & \gategroup[2,steps=3,style={dashed,rounded corners,fill=swapPink!15,
                 inner xsep=2pt,inner ysep=1pt},background,
                 label style={label position=south,anchor=north,yshift=-0.2cm}]{\textcolor{swapPink!85!black}{$U_{(i+1)_R}$}}
    & \ctrl{1} & & \rstick{$S_{(i+1)_R}$} \\
\lstick{$J_{(i+1)_R}$}
    & & \gate{T_{i+1}^{(R)}} & & \rstick{$J_{(i+1)_R}^{a}$}
\end{quantikz}
    \end{adjustbox}
    \caption{Local unitary completion of the self-test in the single-mode case. The input is the partially extracted source state $\ket{\Phi_i}$ defined in \cref{eq:fixed-loop-partial-state}. The open control applies $T_i^{(L)}$ when $S_{i_L}$ is in $\ket0$, whereas the filled control applies $T_{i+1}^{(R)}$ when $S_{(i+1)_R}$ is in $\ket1$. The output is $\ket{\widehat\Phi_i}$ defined in \cref{eq:fermionic_source_after_U}, with its factorization into system and junk states shown to the right. The output spaces are $S_{i_L}\wedge J_{i_L}^{a}$ and $S_{(i+1)_R}\wedge J_{(i+1)_R}^{a}$; the wire layout is schematic and does not specify a fermionic mode ordering. The same circuit applies to the double-mode case with the replacements specified after \cref{eq:fermionic_even_encoded_unitaries}.}
    \label{fig:selftest_unitaries}
\end{figure}

Applying these unitaries to the partially extracted source state gives
\begin{equation}
\label{eq:fermionic_source_after_U}
\begin{aligned}
&
\ket{\widehat \Phi_i}:=\left(
U_{i_L}\wedge U_{(i+1)_R}
\right)
\ket{\Phi_i}
 =
\frac{1}{\sqrt{2}}
\left(
\ket{10}
+
e^{\ii\widehat\varphi_i}
\ket{01}
\right)_{S_{i_L}S_{(i+1)_R}}
\wedge 
\ket{\zeta_i^a}_{J^a_{i_L}J^a_{(i+1)_R}} .
\end{aligned}
\end{equation}
To obtain the source-state form in case~(i) of
\cref{thm:fixed-loop-characterization}, set
\(\varphi_i:=\widehat\varphi_i\).
Thus, the extracted source consists of the trusted single-fermion
state with relative local phase \(e^{\ii\widehat\varphi_i}\),
together with an uncorrelated junk state. The phases are globally consistent
by \cref{eq:fermionic_phase_sum} with \(q=0\).
Define the global transformed state by
\begin{equation}
\label{eq:globalfinalstate}
    \ket{\widehat \Phi}= \bigwedge_{i=1}^N \ket{\widehat \Phi_i}.
\end{equation}
The corresponding normalized conditional state is
\begin{equation}
\label{eq:fermionic_conditional_state_after_U}
\begin{aligned}
\ket{\widehat \Phi}^{\mathrm{cond}}=
\frac{1}{\sqrt{2}}
\Biggl(
\bigwedge_{i=1}^{N} & \ket{01}_{S_{i_L} S_{{i+1}_R}}+
\bigwedge_{i=1}^{N}\ket{10}_{S_{i_L} S_{{i+1}_R}}
\Biggr)
\wedge\ket{\boldsymbol{\zeta}^a},
\end{aligned}
\end{equation}
where \(\ket{\boldsymbol{\zeta}^a}:= \bigwedge_{i=1}^N \ket{\zeta^a_i}\).

For each party \(i\), set
\[
U_i:=U_{i_R}\wedge U_{i_L}.
\]
The controlled maps in \cref{eq:fermionic_local_unitaries}
leave the extracted occupation numbers unchanged, so
\begin{equation}
\label{eq:fermionic_number_projectors_after_U}
\begin{aligned}
U_i\Pi_{\mathrm{TC},i}^0 U_i^\dagger
&=
\ket{00}\!\bra{00}_{S_{i_R}S_{i_L}}
\wedge\mathbb 1_{J_i^a},
\\
U_i\Pi_{\mathrm{TC},i}^2 U_i^\dagger
&=
\ket{11}\!\bra{11}_{S_{i_R}S_{i_L}}
\wedge\mathbb 1_{J_i^a}
\\U_i\Pi_{\mathrm{TC},i}^1 U_i^\dagger
&=
\ket{01}\!\bra{01}+ \ket{10}\!\bra{10}_{S_{i_R}S_{i_L}}
\wedge\mathbb 1_{J_i^a}.
\end{aligned}
\end{equation}
We absorb $U_i$ into the previously constructed isometry
$\mathcal{U}_i$. Using
\cref{eq:fermionic_fixedloop_partialselftest_measurement,eq:fermionic_source_after_U,eq:globalfinalstate},
we then obtain
\begin{equation}
\label{eq:fermionic_number_projectors_after_U+iso}
\begin{aligned}
\mathcal{U} \bar \Pi_{i}^0 \ket{\widehat \Psi}
&=
\ket{00}\!\bra{00}_{S_{i_R}S_{i_L}}
\wedge\mathbb 1_{J_i^a} \ket{\widehat \Phi},
\\
\mathcal{U} \bar \Pi_{i}^2 \ket{\widehat \Psi}
&=
\ket{11}\!\bra{11}_{S_{i_R}S_{i_L}}
\wedge\mathbb 1_{J_i^a} \ket{\widehat \Phi}
\\
\mathcal{U} \bar \Pi_{i}^1 \ket{\widehat \Psi}
&=
\ket{01}\!\bra{01}+ \ket{10}\!\bra{10}_{S_{i_R}S_{i_L}}
\wedge\mathbb 1_{J_i^a} \ket{\widehat \Phi} .
\end{aligned}
\end{equation}

On \(\mathcal H_i^{\mathrm{cond}}\), $U_i$ acts as
\[
U_i
=
\ket{01}\!\bra{01}_{S_i}\wedge\mathbb 1_{J_i^a}
+
\ket{10}\!\bra{10}_{S_i}\wedge T_i.
\]
Combining this expression with the forms of
\(\mathcal X^{\prime}_i\) and \(\mathcal Z^{\prime}_i\)
on \(\mathcal H_i^{\mathrm{cond}}\), given in
\cref{eq:fermionic_X_block_form,eq:fermionic_Z_on_Hcond}, respectively, gives
\begin{equation}
\label{eq:fermionic_Z_after_U}
U_i\mathcal Z^{\prime}_i U_i^\dagger
=
Z_{S_i}\wedge\mathbb 1_{J_i^a},
\end{equation}
and
\begin{equation}
\label{eq:fermionic_X_after_U}
U_i\mathcal X^{\prime}_i U_i^\dagger
=
X_{S_i}\wedge\mathbb 1_{J_i^a},
\end{equation}
on \( U_i \mathcal H_i^{\mathrm{cond}}\),
where
\[
X_{S_i}
=
\ket{01}\!\bra{10}_{S_{i_R}S_{i_L}}
+
\ket{10}\!\bra{01}_{S_{i_R}S_{i_L}}.
\]
Since \(\mathcal H_i^{\mathrm{cond}}\) is the support of the
reduced state of \(\ket{\Phi}^{\mathrm{cond}}\) at party \(i\),
these identities also hold when applied to
\(\ket{\widehat \Phi}^{\mathrm{cond}}\).

Recall that \(\mathcal X^{\prime}_i\) and
\(\mathcal Z^{\prime}_i\) denote the extracted
representations of the physical dilated observables
\(\mathcal X_i\) and \(\mathcal Z_i\), respectively.
Returning to these physical observables and using the composed
isometries already defined above, we write
\(\mathcal U:=\bigwedge_{k=1}^N\mathcal U_k\) and obtain
\begin{equation}
\begin{aligned}
\label{eq:fermionic_XZ_after_unitary+isometry}
\mathcal U\mathcal X_i\ket{\widehat\Psi}^{\mathrm{cond}}
&=
\left(X_{S_i}\wedge\mathbb 1_{J_i^a}\right)
\ket{\widehat\Phi}^{\mathrm{cond}},\\
\mathcal U\mathcal Z_i\ket{\widehat\Psi}^{\mathrm{cond}}
&=
\left(Z_{S_i}\wedge\mathbb 1_{J_i^a}\right)
\ket{\widehat\Phi}^{\mathrm{cond}}.
\end{aligned}
\end{equation}

To extend these identities to the full state, we use the
coarse-grained projectors \(\bar\Pi_i^{n_i}\). These are
independent of the input \(x_i\) on the local reduced support
of \(\ket{\widehat\Psi}\), as established in
\cref{eq:fermionic_coarse_projectors_input_independent}.
Since they sum to the identity,
\[
\ket{\widehat\Psi}
=
\bar\Pi_i^1\ket{\widehat\Psi}
+
(\bar\Pi_i^0+\bar\Pi_i^2)\ket{\widehat\Psi}.
\]
By definition, the observables \(\mathcal X_i\) and
\(\mathcal Z_i\) are linear combinations of
\(\bar\Pi_{i,x_i}^{1_+}\) and \(\bar\Pi_{i,x_i}^{1_-}\),
whose ranges are contained in that of \(\bar\Pi_{i}^{1}\).
This coarse-grained projector is orthogonal to
\(\bar\Pi_{i}^{0}\) and \(\bar\Pi_{i}^{2}\).
Consequently, \(\mathcal X_i\) and \(\mathcal Z_i\) annihilate
the second term in the decomposition above, giving
\begin{equation}
\label{eq:fermionic_XZ_one_token_component}
\begin{aligned}
\mathcal X_i\ket{\widehat\Psi}
&=\mathcal X_i\bar\Pi_i^1\ket{\widehat\Psi},\\
\mathcal Z_i\ket{\widehat\Psi}
&=\mathcal Z_i\bar\Pi_i^1\ket{\widehat\Psi}.
\end{aligned}
\end{equation}
It remains to determine their action on
\(\bar\Pi_i^1\ket{\widehat\Psi}\).

For this purpose, we return to
\(\ket{\Phi}^{\mathrm{cond}}\) and
\(\Pi_{\mathrm{TC},i}^1\ket{\Phi}\), obtained with the initial
extraction isometries before applying the controlled unitaries
\(U_i\). We show that their local reduced supports at party
\(i\) coincide:
\begin{equation}
\label{eq:fermionic_one_token_reduced_states}
\begin{aligned}
&\operatorname{supp}\operatorname{Tr}_{-i}
\left[
\Pi_{\mathrm{TC},i}^1\ket{\Phi}\bra{\Phi}
\Pi_{\mathrm{TC},i}^1
\right]=
\operatorname{supp}\operatorname{Tr}_{-i}
\left[
\ket{\Phi}^{\mathrm{cond}}\bra{\Phi}^{\mathrm{cond}}
\right].
\end{aligned}
\end{equation}
By \cref{eq:fermionic_fixedloop_partialselftest_state,eq:exact-fermionic-TC-projectors},
party \(i\) has one token precisely when its two adjacent
sources are both in their \(a\)-branches or both in their
\(c\)-branches. These two possibilities give the local system
states \(\ket{01}_{S_i}\) and \(\ket{10}_{S_i}\), with junk
supports \(J_i^a\) and \(J_i^c\), respectively. Therefore,
\[
\operatorname{supp}\operatorname{Tr}_{-i}
\left[
\Pi_{\mathrm{TC},i}^1\ket{\Phi}\bra{\Phi}
\Pi_{\mathrm{TC},i}^1
\right] = \mathcal H_i^{\rm cond}.
\]
In \(\ket{\Phi}^{\mathrm{cond}}\), all sources are in their
\(a\)-branches or all are in their \(c\)-branches, as shown in
\cref{eq:fermionic_fixedloop_conditional_state}. The same two
local components occur, with the same reduced junk states, giving
\[
\operatorname{supp}\operatorname{Tr}_{-i}
\left[
\ket{\Phi}^{\mathrm{cond}}\bra{\Phi}^{\mathrm{cond}}
\right]= \mathcal H_i^{\rm cond}.
\]
This proves \cref{eq:fermionic_one_token_reduced_states}.

The initial extraction isometry maps
\(\ket{\widehat\Psi}^{\mathrm{cond}}\) to
\(\ket{\Phi}^{\mathrm{cond}}\) and
\(\bar\Pi_i^1\ket{\widehat\Psi}\) to
\(\Pi_{\mathrm{TC},i}^1\ket{\Phi}\).
Since the extraction is a product of local isometries, this
equality of local reduced supports also holds for the physical
states:
\begin{equation}
\label{eq:fermionic_physical_one_token_support}
\begin{aligned}
&\operatorname{supp}\operatorname{Tr}_{-i}
\left[
\bar\Pi_i^1\ket{\widehat\Psi}\bra{\widehat\Psi}\bar\Pi_i^1
\right]=
\operatorname{supp}\operatorname{Tr}_{-i}
\left[
\ket{\widehat\Psi}^{\mathrm{cond}}
\bra{\widehat\Psi}^{\mathrm{cond}}
\right].
\end{aligned}
\end{equation}
A Schmidt decomposition of the conditional state across party
\(i\) and the remaining parties shows that
\cref{eq:fermionic_XZ_after_unitary+isometry} implies the following
identities on the local reduced support of
\(\ket{\widehat\Psi}^{\mathrm{cond}}\):
\begin{equation}
\label{eq:fermionic_XZ_on_physical_conditional_support}
\begin{aligned}
\mathcal U_i\mathcal X_i
&=
\left(X_{S_i}\wedge\mathbb 1_{J_i^a}\right)\mathcal U_i,\\
\mathcal U_i\mathcal Z_i
&=
\left(Z_{S_i}\wedge\mathbb 1_{J_i^a}\right)\mathcal U_i.
\end{aligned}
\end{equation}
By \cref{eq:fermionic_physical_one_token_support}, these local
identities also hold on the local reduced support of
\(\bar\Pi_i^1\ket{\widehat\Psi}\), and therefore, on state \(\bar\Pi_i^1\ket{\widehat\Psi}\).
We combine this with the zero action on the local zero- and
two-token components expressed in
\cref{eq:fermionic_XZ_one_token_component}. Since the reference
operators \(X_{S_i}\) and \(Z_{S_i}\) also vanish on these two
sectors, we obtain
\begin{equation}
\label{eq:fermionic_XZ_after_unitary+isometry+totalstate}
\begin{aligned}
\mathcal U\mathcal X_i\ket{\widehat\Psi}
&=
\left(X_{S_i}\wedge\mathbb 1_{J_i^a}\right)
\ket{\widehat\Phi},\\
\mathcal U\mathcal Z_i\ket{\widehat\Psi}
&=
\left(Z_{S_i}\wedge\mathbb 1_{J_i^a}\right)
\ket{\widehat\Phi}.
\end{aligned}
\end{equation}
This completes the result for \(q=0\).

\noindent\emph{Case(ii): q=1.—}
The controlled maps in \cref{eq:fermionic_local_unitaries}
contain an even term and an odd term involving
\(T_i^{(R)}\) or \(T_i^{(L)}\), and therefore do not preserve
parity. To construct parity-preserving isometries for
self-testing, we encode the system registers \(S_{i_R}\) and
\(S_{i_L}\) in the even sectors of new two-mode registers
\(\widetilde{S}_{i_R}\) and \(\widetilde{S}_{i_L}\), respectively,
while retaining the original systems as part of the new junk.

Concretely, for the right subsystem with junk support
\(J_{i_R}^c\), define the isometry
\begin{equation}
\label{eq:fermionic_even_encoding}
\begin{aligned}
&\mathcal E_{i_R}:{}
S_{i_R}\wedge J_{i_R}^c
\longrightarrow
\widetilde{S}_{i_R}\wedge S_{i_R}\wedge J_{i_R}^c,
\\[0.5ex]
&\mathcal E_{i_R}
\bigl(\ket b_{S_{i_R}}\wedge\ket\phi_{J_{i_R}^c}\bigr)=
\ket{bb}_{\widetilde{S}_{i_R}}
\wedge\ket b_{S_{i_R}}
\wedge\ket\phi_{J_{i_R}^c},
\end{aligned}
\end{equation}
for \(b\in\{0,1\}\) and \(\ket\phi\in J_{i_R}^c\).
The same definition applies on \(S_{i_R}\wedge J_{i_R}^a\)
and defines \(\mathcal E_{i_L}\) on
\(S_{i_L}\wedge J_{i_L}^a\) and
\(S_{i_L}\wedge J_{i_L}^c\).
The map preserves inner products. It also preserves parity:
the newly appended state \(\ket{bb}\) is even, while the original
system and junk states are retained.

Applying these isometries to the partially self-tested source
state in \cref{eq:fermionic_fixedloop_partialselftest_state} gives
\begin{equation}
\label{eq:fermionic_even_encoded_source}
\begin{aligned}
&\bigl(\mathcal E_{i_L}\wedge\mathcal E_{(i+1)_R}\bigr)
\ket{\Phi_i}
=\frac1{\sqrt2}\Bigl[
\ket{11,00}_{\widetilde{S}_{i_L}\widetilde{S}_{(i+1)_R}}
\wedge\ket{\widetilde{\zeta}_i^a}
+
\ket{00,11}_{\widetilde{S}_{i_L}\widetilde{S}_{(i+1)_R}}
\wedge\ket{\widetilde{\zeta}_i^c}
\Bigr],
\end{aligned}
\end{equation}
where 
\begin{equation}
\label{eq:fermionic_even_encoded_junk_states}
\begin{aligned}
\ket{\widetilde{\zeta}_i^a}
&:=\ket{10}_{S_{i_L}S_{(i+1)_R}}
\wedge\ket{\zeta_i^a}_{J_{i_L}J_{(i+1)_R}},
\\
\ket{\widetilde{\zeta}_i^c}
&:=\ket{01}_{S_{i_L}S_{(i+1)_R}}
\wedge\ket{\zeta_i^c}_{J_{i_L}J_{(i+1)_R}},
\end{aligned}
\end{equation}

Analogously to \cref{eq:fermionic_junk_supports}, we define the
local supports of the new clockwise and anticlockwise junk
states by
\begin{equation}
\label{eq:fermionic_even_encoded_junk_supports}
\begin{aligned}
\widetilde{J}_{i_R}^a
&:=\{\ket0_{S_{i_R}}\}\wedge J_{i_R}^a,
&\widetilde{J}_{i_L}^a
&:=\{\ket1_{S_{i_L}}\}\wedge J_{i_L}^a,
\\
\widetilde{J}_{i_R}^c
&:=\{\ket1_{S_{i_R}}\}\wedge J_{i_R}^c,
&\widetilde{J}_{i_L}^c
&:=\{\ket0_{S_{i_L}}\}\wedge J_{i_L}^c,
\end{aligned}
\end{equation}
and set \(\widetilde{J}_i^a
:=\widetilde{J}_{i_R}^a\wedge\widetilde{J}_{i_L}^a\) and \(\widetilde{J}_i^c
:=\widetilde{J}_{i_R}^c\wedge\widetilde{J}_{i_L}^c\).

Define the maps \(\widetilde{T}_i^{(R)}:
\widetilde{J}_{i_R}^c
\longrightarrow
\widetilde{J}_{i_R}^a \) and \(\widetilde{T}_i^{(L)}:
\widetilde{J}_{i_L}^c
\longrightarrow
\widetilde{J}_{i_L}^a \) by
\begin{equation}
\label{eq:fermionic_even_encoded_T}
\begin{aligned}
\widetilde{T}_i^{(R)}
&:=\ket0\!\bra1_{S_{i_R}}\wedge T_i^{(R)},
\\
\widetilde{T}_i^{(L)}
&:=\ket1\!\bra0_{S_{i_L}}\wedge T_i^{(L)}.
\end{aligned}
\end{equation}
The operators \(T_i^{(R)}\) and \(T_i^{(L)}\) are both unitary,
and each accompanying system transition maps one normalized
basis state to the other.
Hence, \(\widetilde{T}_i^{(R)}\) and \(\widetilde{T}_i^{(L)}\)
are unitary between the stated domains and codomains.
Moreover, both factors in each wedge product in
\cref{eq:fermionic_even_encoded_T} are odd, so both
\(\widetilde{T}_i^{(R)}\) and \(\widetilde{T}_i^{(L)}\) preserve
parity. Set \(\widetilde{T}_i:= \widetilde{T}_i^{(R)} \wedge \widetilde{T}_i^{(L)}\).

We now show how the conditional operators
\( \mathcal X^{\prime}_i\) and \(\mathcal Z^{\prime}_i\)
transform under
\(\mathcal E_i:=\mathcal E_{i_R}\wedge\mathcal E_{i_L}\). Set
\(\widetilde{S}_i:=\widetilde{S}_{i_R}\wedge\widetilde{S}_{i_L}\).
Using
\cref{eq:fermionic_X_block_form,eq:fermionic_T_factorization,eq:fermionic_Z_on_Hcond}
and the fermionic composition rule gives
\begin{equation}
\label{eq:fermionic_even_encoded_X_block}
\begin{aligned}
\mathcal E_i\mathcal X^{\prime}_i\mathcal E_i^\dagger
={}\ket{00,11}\!\bra{11,00}_{\widetilde{S}_i}
\wedge
 (\widetilde{T}_i)_{\widetilde{J}_i^c} + \ket{11,00}\!\bra{00,11}_{\widetilde{S}_i}
\wedge (\widetilde{T}_i^\dagger)_{\widetilde{J}_i^a}
\end{aligned}
\end{equation}
and
\begin{equation}
\label{eq:fermionic_even_encoded_Z_block}
\begin{aligned}
\mathcal E_i\mathcal Z^{\prime}_i\mathcal E_i^\dagger
={}& \ket{00,11}\!\bra{00,11}_{\widetilde{S}_i} \wedge\mathbb 1_{\widetilde{J}_i^a}
-\ket{11,00}\!\bra{11,00}_{\widetilde{S}_i}
\wedge\mathbb 1_{\widetilde{J}_i^c},
\end{aligned}
\end{equation}
on \(\mathcal E_i\mathcal H_i^{\mathrm{cond}}\).
We give the derivation explicitly for \(\mathcal X^{\prime}_i\).
Write \(\ket{01}\!\bra{10}_{S_i}=f_{S_{i_L}}^\dagger f_{S_{i_R}}\).
Using the fermionic anticommutation relations and the odd parity of
\(T_i^{(R)}\), we obtain
\[
\begin{aligned}
f_{S_{i_L}}^\dagger f_{S_{i_R}} T_i^{(R)}T_i^{(L)}
&=f_{S_{i_R}} T_i^{(R)}f_{S_{i_L}}^\dagger T_i^{(L)}
=\widetilde{T}_i.
\end{aligned}
\]
Grouping the original system modes with their respective junk spaces
and including the adjoint block therefore gives
\(\mathcal X^{\prime}_i=\widetilde{T}_i+\widetilde{T}_i^\dagger\)
on \(\mathcal H_i^{\mathrm{cond}}\), with \(\widetilde{T}_i\) vanishing on \(\widetilde{J}_i^a\) and \(\widetilde{T}_i^\dagger\) vanishing on \(\widetilde{J}_i^c\). By definition, the isometry satisfies
\[
\begin{aligned}
\mathcal E_i\ket{\xi_a}
&=\ket{00,11}_{\widetilde{S}_i}\wedge\ket{\xi_a},\\
\mathcal E_i\ket{\xi_c}
&=\ket{11,00}_{\widetilde{S}_i}\wedge\ket{\xi_c},
\end{aligned}
\]
for \(\ket{\xi_a}\in\widetilde{J}_i^a\) and
\(\ket{\xi_c}\in\widetilde{J}_i^c\).
Note that the new system states are even, so regrouping introduces no sign.
Consider \(\ket{11,00}_{\widetilde{S}_i}\wedge\ket{\xi_c}\). Since \(\widetilde{T}_i\ket{\xi_c}\in\widetilde{J}_i^a\),
\[
\begin{aligned}
&\mathcal E_i\mathcal X^{\prime}_i\mathcal E_i^\dagger
\bigl(\ket{11,00}_{\widetilde{S}_i}\wedge\ket{\xi_c}\bigr)\\
&\qquad=\mathcal E_i
\bigl(\widetilde{T}_i+\widetilde{T}_i^\dagger\bigr)\ket{\xi_c}\\
&\qquad=\mathcal E_i\widetilde{T}_i\ket{\xi_c}\\
&\qquad=\ket{00,11}_{\widetilde{S}_i}
\wedge\widetilde{T}_i\ket{\xi_c}.
\end{aligned}
\]
Starting with \(\ket{00,11}_{\widetilde{S}_i}\wedge\ket{\xi_a}\) gives the other block, proving
\cref{eq:fermionic_even_encoded_X_block}.

We now define the unitaries on the image of
\(\mathcal E\), in analogy with
\cref{eq:fermionic_local_unitaries}, by
\begin{equation}
\label{eq:fermionic_even_encoded_unitaries}
\begin{aligned}
\widetilde{U}_{i_R}
&:=\ket{00}\!\bra{00}_{\widetilde{S}_{i_R}}
\wedge\id_{\widetilde{J}_{i_R}^a} +
\ket{11}\!\bra{11}_{\widetilde{S}_{i_R}}
\wedge\widetilde{T}_i^{(R)},
\\
\widetilde{U}_{i_L}
&:=\ket{11}\!\bra{11}_{\widetilde{S}_{i_L}}
\wedge\id_{\widetilde{J}_{i_L}^a}+
\ket{00}\!\bra{00}_{\widetilde{S}_{i_L}}
\wedge\widetilde{T}_i^{(L)}.
\end{aligned}
\end{equation}
These unitaries are represented by the circuit in \cref{fig:selftest_unitaries}
after replacing \(U,T,S,J\) by their tilded counterparts and
\(\ket0,\ket1\) by \(\ket{00},\ket{11}\), respectively.
The input is now the encoded source state in
\cref{eq:fermionic_even_encoded_source}, and the output is
\(\ket{\widehat\Phi_i}\) in \cref{eq:fermionic_even_encoded_source_after_U},
with junk state \(\ket{\widetilde\zeta_i^a}\) and phase
\(\widetilde\varphi_i\).

Their domains and codomains are
\begin{equation}
\label{eq:fermionic_even_encoded_unitary_spaces}
\begin{aligned}
\widetilde{U}_{i_R}:{}&
\bigl(\{\ket{00}\}_{\widetilde{S}_{i_R}}\wedge\widetilde{J}_{i_R}^a\bigr)
{}\oplus
\bigl(\{\ket{11}\}_{\widetilde{S}_{i_R}}\wedge\widetilde{J}_{i_R}^c\bigr)
\longrightarrow
\{\ket{00},\ket{11}\}_{\widetilde{S}_{i_R}}
\wedge\widetilde{J}_{i_R}^a,
\\[0.5ex]
\widetilde{U}_{i_L}:{}&
\bigl(\{\ket{11}\}_{\widetilde{S}_{i_L}}\wedge\widetilde{J}_{i_L}^a\bigr)
{}\oplus
\bigl(\{\ket{00}\}_{\widetilde{S}_{i_L}}\wedge\widetilde{J}_{i_L}^c\bigr)
\longrightarrow
\{\ket{00},\ket{11}\}_{\widetilde{S}_{i_L}}
\wedge\widetilde{J}_{i_L}^a.
\end{aligned}
\end{equation}
These maps are unitary and parity preserving. Set \(\widetilde{U}_i:= \widetilde{U}_{i_R} \wedge \widetilde{U}_{i_L}\).
Consequently, the composed isometries
\begin{equation}
\label{eq:fermionic_even_encoded_extraction}
\begin{aligned}
&\mathcal V_{i_R}
:=\widetilde{U}_{i_R}\mathcal E_{i_R},
\\
&\mathcal V_{i_L}
:=\widetilde{U}_{i_L}\mathcal E_{i_L},
\\
&\mathcal V_i
:=\mathcal V_{i_R}\wedge\mathcal V_{i_L}
=\widetilde{U}_i\mathcal E_i
\end{aligned}
\end{equation}
preserve fermionic parity.

To determine the final extracted source states, we first compute
\begin{equation}
\label{eq:fermionic_even_encoded_phase_relation}
\begin{aligned}
&\bigl(\widetilde{T}_i^{(L)}\wedge
\widetilde{T}_{i+1}^{(R)}\bigr)
\ket{\widetilde{\zeta}_i^c}
=-\ket{10}_{S_{i_L}S_{(i+1)_R}}
\wedge\bigl(T_i^{(L)}\wedge T_{i+1}^{(R)}\bigr)
\ket{\zeta_i^c}
=-e^{\ii\widehat{\varphi}_i}
\ket{\widetilde{\zeta}_i^a}.
\end{aligned}
\end{equation}
The minus sign above arises when the odd operator \(T_i^{(L)}\)
is moved past the odd transition on \(S_{(i+1)_R}\) in
\cref{eq:fermionic_even_encoded_T}. The last equality follows from
\cref{eq:fermionic_local_phase_relation}.
It follows from \cref{eq:fermionic_even_encoded_source} that
\begin{equation}
\label{eq:fermionic_even_encoded_source_after_U}
\begin{aligned}
& \ket{\widehat{\Phi}_i}:= \bigl(\mathcal V_{i_L}\wedge\mathcal V_{(i+1)_R}\bigr)
\ket{\Phi_i}
=\frac1{\sqrt2}
\Bigl(\ket{11,00}
+e^{\ii\widetilde{\varphi}_i}\ket{00,11}\Bigr)
_{\widetilde{S}_{i_L}\widetilde{S}_{(i+1)_R}}
\wedge
\ket{\widetilde{\zeta}_i^a}_{\widetilde{J}^a_{i_L}\widetilde{J}^a_{(i+1)_R}},
\end{aligned}
\end{equation}
where \(\widetilde{\varphi}_i:=\widehat{\varphi}_i+\pi\).
To obtain the source-state form in case~(ii) of
\cref{thm:fixed-loop-characterization}, set
\(\varphi_i:=\widetilde{\varphi}_i\).
Since \(q=1\), \cref{eq:fermionic_phase_sum} gives
\begin{equation}
\label{eq:fermionic_even_encoded_phase_sum}
\begin{aligned}
\sum_{i=1}^N\widetilde{\varphi}_i
&=\sum_{i=1}^N\widehat{\varphi}_i+N\pi
=(N-1)\pi\pmod{2\pi}.
\end{aligned}
\end{equation}
Thus, the recovered phases sum to zero for odd \(N\) and to
\(\pi\) for even \(N\).

We next characterize the measurements. The self-test of measurements is
analogous to that in case~(i). We absorb each isometry
$\mathcal{V}_i$ into the initial isometry
\(\mathcal{U}_i\), defined before case~(i).
With \(\mathcal U:=\bigwedge_{i=1}^N\mathcal U_i\), the global
extracted state is
\[
\ket{\widehat\Phi}:=\bigwedge_{i=1}^N\ket{\widehat\Phi_i}
=\mathcal U\ket{\widehat\Psi}.
\]
We obtain
\begin{equation}
\label{eq:fermionic_even_encoded_measurements}
\begin{aligned}
\mathcal U \bar \Pi_{i}^0 \ket{\widehat{\Psi}}
&=
\ket{00,00}\!\bra{00,00}_{\widetilde{S}_{i_R}\widetilde{S}_{i_L}}
\wedge\mathbb 1_{\widetilde{J}_i^a} \ket{\widehat{\Phi}},
\\
\mathcal U \bar \Pi_{i}^2 \ket{\widehat{\Psi}}
&=
\ket{11, 11}\!\bra{11,11}_{\widetilde{S}_{i_R} \widetilde{S}_{i_L}}
\wedge\mathbb 1_{\widetilde{J}_i^a} \ket{\widehat{\Phi}}
\\
\mathcal U\mathcal X_i\ket{\widehat{\Psi}}
&=
\left(X_{\widetilde{S}_i}\wedge\mathbb 1_{\widetilde{J}_i^a}\right)
\ket{\widehat{\Phi}},\\
\mathcal U\mathcal Z_i\ket{\widehat{\Psi}}
&=
\left(Z_{\widetilde{S}_i}\wedge\mathbb 1_{\widetilde{J}_i^a}\right)
\ket{\widehat{\Phi}},
\end{aligned}
\end{equation}
where
\begin{equation}
\label{eq:fermionic_even_encoded_XZ}
\begin{aligned}
X_{\widetilde{S}_i}
&:=
\ket{00,11}\!\bra{11,00}_{\widetilde{S}_i}
+
\ket{11,00}\!\bra{00,11}_{\widetilde{S}_i},
\\
Z_{\widetilde{S}_i}
&:=
\ket{00,11}\!\bra{00,11}_{\widetilde{S}_i}
-
\ket{11,00}\!\bra{11,00}_{\widetilde{S}_i}.
\end{aligned}
\end{equation}
This completes the proof for q=1.

Finally, we show that self-testing results of cases (i) and (ii) cannot be related by
local parity-preserving isometries. Suppose, for contradiction, that such isometries
\[
\begin{aligned}
\mathcal W_{i_R}&:S_{i_R}\wedge J_{i_R}^a
\longrightarrow\widetilde S_{i_R}\wedge\widetilde J_{i_R}^a,\\
\mathcal W_{i_L}&:S_{i_L}\wedge J_{i_L}^a
\longrightarrow\widetilde S_{i_L}\wedge\widetilde J_{i_L}^a
\end{aligned}
\]
exist, and set
$\mathcal W_i:=\mathcal W_{i_R}\wedge\mathcal W_{i_L}$.
Let $P_{i_R}$ and $\widetilde P_{i_R}$ denote the parity operators
of the right subsystems of party \(i\), including its junk, in cases (i) and (ii),
respectively. Then we have,
\[
\mathcal W_i X_{S_i}
=
X_{\widetilde S_i}\mathcal W_i
\]
on the local reduced support of the self-tested state in case (i). By the definitions of $X_{S_i}$ and $X_{\widetilde S_i}$ in the
theorem statement, we have
\[
P_{i_R}X_{S_i}P_{i_R}=-X_{S_i},
\qquad
\widetilde P_{i_R}X_{\widetilde S_i}\widetilde P_{i_R}
=X_{\widetilde S_i}.
\]
Since $\mathcal W_{i_R}$ preserves parity, we also have
$\widetilde P_{i_R}\mathcal W_i=\mathcal W_iP_{i_R}$. Note that $P_{i_R}$ preserves the local reduced
support, since each source state has definite parity, so \(\mathcal W_i\)
also applies after acting with $P_{i_R}$.
Combining these identities gives
\[
\begin{aligned}
\mathcal W_iX_{S_i}
&=X_{\widetilde S_i}\mathcal W_i\\
&=\widetilde P_{i_R}X_{\widetilde S_i}
  \widetilde P_{i_R}\mathcal W_i\\
&=\widetilde P_{i_R}\mathcal W_iX_{S_i}P_{i_R}\\
&=\mathcal W_iP_{i_R}X_{S_i}P_{i_R}\\
&=-\mathcal W_iX_{S_i}.
\end{aligned}
\]
In the third equality, we again use that $P_{i_R}$ preserves the local reduced
support.
Thus $\mathcal W_iX_{S_i}=0$ on the local reduced support.
This is impossible, since $\mathcal W_i$ is an isometry and
$X_{S_i}$ acts nontrivially on the one-token sector of that support.
Therefore, the two cases cannot be related by such isometries.

\end{proof}

The following theorem is the fermionic counterpart of
Theorem~F.9 in Supplementary Material~F of
Ref.~\cite{kalarde2026fermions}.

\begin{theorem}[Fixed-loop self-testing under structural assumptions]
\label{thm:fixed-loop-characterization-N2}
Consider the \(N\)-party generalization of the fermionic game
described in the section ``The fermionic self-testing scenario'',
with \(N\geq2\), and fix a loop configuration \(l\) containing
all \(N\) parties.
Label the parties \(A_1,\ldots,A_N\) according to their order
along the loop, with indices understood modulo~\(N\).
Denote the independent normalized source states by
\[
\ket{\psi_i}_{i_L,(i+1)_R}
\in\mathcal H_{i_L}\wedge\mathcal H_{(i+1)_R},
\qquad i\in\{1,\ldots,N\}.
\]
For each input \(x_i\in\{0,1\}\), party \(A_i\) performs a
parity-preserving POVM on \(\mathcal H_{i_R}\wedge\mathcal H_{i_L}\),
with outcomes \(a_i\in\{0,1_+,1_-,2\}\).
For each party \(i\), choose parity-preserving Naimark dilations
of the two POVMs using the same ancillary space \(\mathcal H_{M_i}\)
and the same initial even state \(\ket0_{M_i}\) for both inputs.
Denote the resulting PVM elements by \(\bar\Pi_{i,x_i}^{a_i}\).
Group \(M_i\) with \(i_L\) and set
\[
\begin{aligned}
\ket{\widehat\psi_i}
&:=\ket{\psi_i}_{i_L,(i+1)_R}\wedge\ket0_{M_i},
&
\ket{\widehat\Psi}
&:=\bigwedge_{i=1}^{N}\ket{\widehat\psi_i},\\
\bar\Pi_{i,x_i}^{1}
&:=\bar\Pi_{i,x_i}^{1_+}+\bar\Pi_{i,x_i}^{1_-},
&
M_{x_i}^{(i)}
&:=\bar\Pi_{i,x_i}^{1_+}-\bar\Pi_{i,x_i}^{1_-}.
\end{aligned}
\]
As in \cref{thm:fixed-loop-characterization}, define
\[
\begin{aligned}
\mathcal X_1&:=\frac{M_0^{(1)}+M_1^{(1)}}{\sqrt2},
&
\mathcal Z_1&:=\frac{M_0^{(1)}-M_1^{(1)}}{\sqrt2},\\
\mathcal X_i&:=M_1^{(i)},
&
\mathcal Z_i&:=M_0^{(i)},\qquad i\in\{2,\ldots,N\}.
\end{aligned}
\]

Assume that there exist input-independent local parity-preserving
isometries \(\mathcal U_{i_R}\) and
\(\mathcal U_{i_L}:=\mathcal U_{i_LM_i}\), acting on
\(\mathcal H_{i_R}\) and
\(\mathcal H_{i_L}\wedge\mathcal H_{M_i}\), respectively,
for every \(i\in\{1,\ldots,N\}\), giving the following
single-mode partial characterization. Define
\[
\mathcal U_i:=\mathcal U_{i_R}\wedge\mathcal U_{i_L},
\qquad
\mathcal U:=\bigwedge_{i=1}^{N}\mathcal U_i,
\]
and write
\[
\ket{\Phi_i}:=
\bigl(\mathcal U_{i_L}\wedge\mathcal U_{(i+1)_R}\bigr)
\ket{\widehat\psi_i},
\qquad
\ket\Phi:=\mathcal U\ket{\widehat\Psi}
=\bigwedge_{i=1}^{N}\ket{\Phi_i}.
\]

The codomains of \(\mathcal U_{i_R}\) and \(\mathcal U_{i_L}\)
are \(S_{i_R}\wedge J_{i_R}\) and
\(S_{i_L}\wedge J_{i_L}\), respectively, where
\(S_{i_R}\simeq S_{i_L}\simeq\mathbb C^2\) are single-mode
fermionic spaces, each spanned by \(\ket0\) and \(\ket1\),
and \(J_{i_R}\) and \(J_{i_L}\) are junk spaces. Write
\(S_i=S_{i_R}\wedge S_{i_L}\) and
\(J_i=J_{i_R}\wedge J_{i_L}\).
The extracted source states have the form
\begin{equation}
\label{eq:two_party_single_mode_assumed_state}
\begin{aligned}
\ket{\Phi_i}
&=\frac1{\sqrt2}\Bigl(
\ket{10}_{S_{i_L}S_{(i+1)_R}}
\wedge\ket{\zeta_i^a}_{J_{i_L}J_{(i+1)_R}}
+
\ket{01}_{S_{i_L}S_{(i+1)_R}}
\wedge\ket{\zeta_i^c}_{J_{i_L}J_{(i+1)_R}}
\Bigr).
\end{aligned}
\end{equation}
Here, \(\ket{\zeta_i^a}\) and \(\ket{\zeta_i^c}\) are normalized
junk states with the same definite fermionic parity. For each source
\(i\), let \(J_{i_L}^{a,c}\) and \(J_{(i+1)_R}^{a,c}\) denote
the supports of their reduced density operators on \(J_{i_L}\)
and \(J_{(i+1)_R}\), respectively.

For both inputs and every \(n_i\in\{0,1,2\}\), the
coarse-grained measurements satisfy
\begin{equation}
\label{eq:two_party_single_mode_assumed_coarse_graining}
\mathcal U\bar\Pi_{i,x_i}^{n_i}\ket{\widehat\Psi}
=\Pi_{S_i}^{n_i}\ket\Phi,
\end{equation}
where
\[
\begin{aligned}
\Pi_{S_i}^{0}&:=\ket{00}\!\bra{00}_{S_i}\wedge\id_{J_i},\\
\Pi_{S_i}^{1}&:=\bigl(\ket{01}\!\bra{01}
+\ket{10}\!\bra{10}\bigr)_{S_i}\wedge\id_{J_i},\\
\Pi_{S_i}^{2}&:=\ket{11}\!\bra{11}_{S_i}\wedge\id_{J_i}.
\end{aligned}
\]
The signed observables satisfy
\begin{equation}
\label{eq:two_party_single_mode_assumed_XZ}
\begin{aligned}
&\mathcal U\mathcal Z_i\ket{\widehat\Psi}=\Bigl[\bigl(\ket{01}\!\bra{01}
-\ket{10}\!\bra{10}\bigr)_{S_i}
\wedge\id_{J_i}\Bigr]\ket\Phi,\\
&\mathcal U\mathcal X_i\ket{\widehat\Psi}=\Bigl(\ket{01}\!\bra{10}_{S_i}\wedge T_i
+\ket{10}\!\bra{01}_{S_i}\wedge T_i^\dagger\Bigr)\ket\Phi,
\end{aligned}
\end{equation}
where
\begin{equation}
\label{eq:two_party_T_factorization}
\begin{aligned}
T_i=T_i^{(R)}\wedge T_i^{(L)},
\end{aligned}
\end{equation}
with unitary maps
\(T_i^{(R)}:J_{i_R}^c\longrightarrow J_{i_R}^a\) and
\(T_i^{(L)}:J_{i_L}^c\longrightarrow J_{i_L}^a\).
For each party \(i\), both factors have the same definite parity
\(q_i\in\{0,1\}\), where \(q_i=0\) means parity preserving
and \(q_i=1\) means parity changing. The parity \(q_i\) may
depend on \(i\).

Let \(\ket{\widehat\Psi}^{\mathrm{cond}}\) denote the normalized
dilated state conditioned on every party obtaining \(1_+\) or \(1_-\),
and write \(\ket\Phi^{\mathrm{cond}}
=\mathcal U\ket{\widehat\Psi}^{\mathrm{cond}}\).
Suppose, in addition, that the resulting conditional correlations
maximally violate Bell inequality~\Romannum{1} in \cref{eq:Bell_1}
when \(N\) is odd, and Bell inequality~\Romannum{2} in
\cref{eq:Bell_2} when \(N\) is even:
\[
\begin{aligned}
\langle B_1\rangle_{\mathrm{cond}}
&=2(N-1)\sqrt2, &&N\text{ odd},\\
\langle B_2\rangle_{\mathrm{cond}}
&=2(N-1)\sqrt2, &&N\text{ even}.
\end{aligned}
\]

Then all endpoint factors have the same parity:
\begin{equation}
\label{eq:two_party_common_parity}
q_1=\cdots=q_N=:q.
\end{equation}
The unitary completion takes the following form according to
this common parity.

\begin{itemize}
\item \emph{(i) Single-mode systems: \(q=0\).}
Define \(U_{i_R}\) and \(U_{i_L}\) as in
\cref{eq:fermionic_local_unitaries}, using the factors
\(T_i^{(R)}\) and \(T_i^{(L)}\) assumed above.
These are parity-preserving unitary maps with the domains and
codomains specified in \cref{eq:fermionic_local_unitary_spaces}.
The transformed source states satisfy
\begin{equation}
\label{eq:two_party_single_mode_state}
\begin{aligned}
\ket{\widehat\Phi_i}
&:=\bigl(U_{i_L}\wedge U_{(i+1)_R}\bigr)\ket{\Phi_i}=\frac1{\sqrt2}
\bigl(e^{\ii\varphi_i}\ket{01}+\ket{10}\bigr)
_{S_{i_L}S_{(i+1)_R}} \wedge\ket{\zeta^a_i}_{J_{i_L}J_{(i+1)_R}},
\end{aligned}
\end{equation}
where the phases satisfy
\(\sum_{i=1}^{N}\varphi_i=0\pmod{2\pi}\).
Define
\(U_i:=U_{i_R}\wedge U_{i_L}\) and
\(U:=\bigwedge_{i=1}^{N}U_i\), and set
\(\ket{\widehat\Phi}:=U\ket\Phi
=\bigwedge_{i=1}^{N}\ket{\widehat\Phi_i}\).
For every party \(i\), input \(x_i\in\{0,1\}\), and
\(n_i\in\{0,1,2\}\), the measurements and signed observables satisfy
\begin{equation}
\label{eq:two_party_single_mode_XZ}
\begin{aligned}
U\mathcal U\bar\Pi_{i,x_i}^{n_i}\ket{\widehat\Psi}
&=\Pi_{S_i}^{n_i}\ket{\widehat\Phi},\\
U\mathcal U\mathcal X_i\ket{\widehat\Psi}
&=X_{S_i}\ket{\widehat\Phi},\\
U\mathcal U\mathcal Z_i\ket{\widehat\Psi}
&=Z_{S_i}\ket{\widehat\Phi},
\end{aligned}
\end{equation}
where
\[
\begin{aligned}
X_{S_i}&:=\bigl(\ket{01}\!\bra{10}
+\ket{10}\!\bra{01}\bigr)_{S_i}\wedge\id_{J_i^a},\\
Z_{S_i}&:=\bigl(\ket{01}\!\bra{01}
-\ket{10}\!\bra{10}\bigr)_{S_i}\wedge\id_{J_i^a}.
\end{aligned}
\]

\item \emph{(ii) Double-mode systems: \(q=1\).}
Apply the parity-preserving encoding isometries
\(\mathcal E_{i_R},\mathcal E_{i_L}\) of
\cref{eq:fermionic_even_encoding}. Their extracted system
spaces \(\widetilde S_{i_R},\widetilde S_{i_L}\) are the even
subspaces of two-mode registers spanned by \(\ket{00},\ket{11}\).
Define the encoded junk states and their reduced supports as in
\cref{eq:fermionic_even_encoded_junk_states,eq:fermionic_even_encoded_junk_supports},
and set
\[
\begin{aligned}
\mathcal E_i&:=\mathcal E_{i_R}\wedge\mathcal E_{i_L},
&\mathcal E&:=\bigwedge_{i=1}^{N}\mathcal E_i,\\
\ket{\widetilde\Phi_i}
&:=(\mathcal E_{i_L}\wedge\mathcal E_{(i+1)_R})\ket{\Phi_i},
&\ket{\widetilde\Phi}&:=\mathcal E\ket\Phi.
\end{aligned}
\]
Write \(\widetilde S_i=\widetilde S_{i_R}\wedge\widetilde S_{i_L}\)
and \(\widetilde J_i=\widetilde J_{i_R}\wedge\widetilde J_{i_L}\).
The encoded source states have the form
\begin{equation}
\label{eq:two_party_double_mode_assumed_state}
\begin{aligned}
\ket{\widetilde\Phi_i}
&=\frac1{\sqrt2}\Bigl(
\ket{11,00}_{\widetilde{S}_{i_L}\widetilde{S}_{(i+1)_R}}
\wedge\ket{\widetilde{\zeta}_i^a}_{\widetilde{J}_{i_L}\widetilde{J}_{(i+1)_R}}
+
\ket{00,11}_{\widetilde{S}_{i_L}\widetilde{S}_{(i+1)_R}}
\wedge\ket{\widetilde{\zeta}_i^c}_{\widetilde{J}_{i_L}\widetilde{J}_{(i+1)_R}}
\Bigr).
\end{aligned}
\end{equation}
Here, \(\ket{\widetilde\zeta_i^a}\) and
\(\ket{\widetilde\zeta_i^c}\) are normalized junk states with the
same definite fermionic parity. For each source \(i\), let
\(\widetilde J_{i_L}^{a,c}\) and \(\widetilde J_{(i+1)_R}^{a,c}\)
denote the supports of their reduced density operators on
\(\widetilde J_{i_L}\) and \(\widetilde J_{(i+1)_R}\), respectively.

For both inputs and every \(n_i\in\{0,1,2\}\), the
coarse-grained measurements satisfy
\begin{equation}
\label{eq:two_party_double_mode_assumed_coarse_graining}
\mathcal E\mathcal U\bar\Pi_{i,x_i}^{n_i}\ket{\widehat\Psi}
=\Pi_{\widetilde{S}_i}^{n_i}\ket{\widetilde\Phi},
\end{equation}
where
\[
\begin{aligned}
\Pi_{\widetilde{S}_i}^{0}
&:=\ket{00,00}\!\bra{00,00}_{\widetilde{S}_i}\wedge\id_{\widetilde{J}_i},\\
\Pi_{\widetilde{S}_i}^{1}
&:=\bigl(\ket{00,11}\!\bra{00,11}
+\ket{11,00}\!\bra{11,00}\bigr)_{\widetilde{S}_i}\wedge\id_{\widetilde{J}_i},\\
\Pi_{\widetilde{S}_i}^{2}
&:=\ket{11,11}\!\bra{11,11}_{\widetilde{S}_i}\wedge\id_{\widetilde{J}_i}.
\end{aligned}
\]
The signed observables satisfy
\begin{equation}
\label{eq:two_party_double_mode_assumed_XZ}
\begin{aligned}
&\mathcal E\mathcal U\mathcal Z_i\ket{\widehat\Psi}=\Bigl[\Bigl(\ket{00,11}\!\bra{00,11}_{\widetilde{S}_i}-\ket{11,00}\!\bra{11,00}_{\widetilde{S}_i}\Bigr)
\wedge\id_{\widetilde{J}_i}\Bigr]\ket{\widetilde\Phi},\\
&\mathcal E\mathcal U\mathcal X_i\ket{\widehat\Psi}
=\Bigl(\ket{00,11}\!\bra{11,00}_{\widetilde{S}_i}
\wedge\widetilde{T}_i
+\ket{11,00}\!\bra{00,11}_{\widetilde{S}_i}
\wedge\widetilde{T}_i^\dagger\Bigr)\ket{\widetilde\Phi},
\end{aligned}
\end{equation}
where
\begin{equation}
\label{eq:two_party_even_T_factorization}
\begin{aligned}
\widetilde{T}_i
&=\widetilde{T}_i^{(R)}\wedge\widetilde{T}_i^{(L)},
\end{aligned}
\end{equation}
with the parity-preserving unitary maps defined in
\cref{eq:fermionic_even_encoded_T},
\(\widetilde{T}_i^{(R)}:\widetilde{J}_{i_R}^c
\longrightarrow\widetilde{J}_{i_R}^a\) and
\(\widetilde{T}_i^{(L)}:\widetilde{J}_{i_L}^c
\longrightarrow\widetilde{J}_{i_L}^a\).
Define \(\widetilde U_{i_R}\) and \(\widetilde U_{i_L}\) as in
\cref{eq:fermionic_even_encoded_unitaries}, using the factors
\(\widetilde T_i^{(R)}\) and \(\widetilde T_i^{(L)}\) defined above.
These are parity-preserving unitary maps with the domains and
codomains specified in \cref{eq:fermionic_even_encoded_unitary_spaces}.
The transformed source states satisfy
\begin{equation}
\label{eq:two_party_double_mode_state}
\begin{aligned}
\ket{\widehat\Phi_i}
&:=\bigl(\widetilde{U}_{i_L}\wedge
\widetilde{U}_{(i+1)_R}\bigr)\ket{\widetilde\Phi_i}
=\frac1{\sqrt2}
\bigl(e^{\ii\varphi_i}\ket{00,11}+\ket{11,00}\bigr)
_{\widetilde{S}_{i_L}\widetilde{S}_{(i+1)_R}}
\wedge\ket{\widetilde{\zeta}^a_i}
_{\widetilde{J}_{i_L}\widetilde{J}_{(i+1)_R}},
\end{aligned}
\end{equation}
where the phases satisfy
\(\sum_{i=1}^{N}\varphi_i=(N-1)\pi\pmod{2\pi}\).
Define
\(\widetilde{U}_i:=\widetilde{U}_{i_R}\wedge\widetilde{U}_{i_L}\)
and \(\widetilde{U}:=\bigwedge_{i=1}^{N}\widetilde{U}_i\), and set
\(\ket{\widehat\Phi}:=\widetilde{U}\ket{\widetilde\Phi}
=\bigwedge_{i=1}^{N}\ket{\widehat\Phi_i}\).
For every party \(i\), input \(x_i\in\{0,1\}\), and
\(n_i\in\{0,1,2\}\), the measurements and signed observables satisfy
\begin{equation}
\label{eq:two_party_double_mode_XZ}
\begin{aligned}
\widetilde{U}\mathcal E\mathcal U\bar\Pi_{i,x_i}^{n_i}\ket{\widehat\Psi}
&=\Pi_{\widetilde S_i}^{n_i}
\ket{\widehat\Phi},\\
\widetilde{U}\mathcal E\mathcal U\mathcal X_i\ket{\widehat\Psi}
&=X_{\widetilde{S}_i}
\ket{\widehat\Phi},\\
\widetilde{U}\mathcal E\mathcal U\mathcal Z_i\ket{\widehat\Psi}
&=Z_{\widetilde{S}_i}
\ket{\widehat\Phi},
\end{aligned}
\end{equation}
where
\[
\begin{aligned}
X_{\widetilde{S}_i}
&:=\bigl(\ket{00,11}\!\bra{11,00}
+\ket{11,00}\!\bra{00,11}\bigr)_{\widetilde{S}_i}\wedge\id_{\widetilde{J}_i^a},\\
Z_{\widetilde{S}_i}
&:=\bigl(\ket{00,11}\!\bra{00,11}
-\ket{11,00}\!\bra{11,00}\bigr)_{\widetilde{S}_i}\wedge\id_{\widetilde{J}_i^a}.
\end{aligned}
\]
\end{itemize}

\end{theorem}

\begin{proof}
The stabilizer argument of \cref{lem:fermionic_Bell_relations}
also applies to \(N=2\). Hence maximal violation of Bell
inequality~\Romannum{1} for odd \(N\), or Bell
inequality~\Romannum{2} for even \(N\), gives, after applying
\(\mathcal U\),
\begin{equation}
\label{eq:two_party_conditional_stabilizer}
\mathcal U\left(\prod_{j=1}^{N}\mathcal X_j\right)
\ket{\widehat\Psi}^{\mathrm{cond}}
=(-1)^{N-1}\ket\Phi^{\mathrm{cond}}.
\end{equation}
We next show that the identities in
\cref{eq:two_party_single_mode_assumed_XZ}
remain valid after conditioning on one token at each party.

Fix an input string \(\mathbf x=(x_1,\ldots,x_N)\).
By the source-state form in
\cref{eq:two_party_single_mode_assumed_state},
the extracted state contains exactly \(N\) tokens.
Consequently, one token at each of \(N-1\) parties forces one
token at the remaining party. Only the two branches in which
every source sends its token in the same direction survive
this conditioning, so the event has probability \(2^{1-N}\).
The coarse-grained equations
\cref{eq:two_party_single_mode_assumed_coarse_graining},
together with the factorization of \(\mathcal U\) into local
isometries, therefore give
\begin{equation}
\label{eq:two_party_one_token_conditioning}
\begin{aligned}
\left(\prod_{j\ne i}\bar\Pi_{j,x_j}^{1}\right)\ket{\widehat\Psi}
&=
\left(\prod_{j=1}^{N}\bar\Pi_{j,x_j}^{1}\right)\ket{\widehat\Psi}
=2^{(1-N)/2}\ket{\widehat\Psi}^{\mathrm{cond}},
\end{aligned}
\end{equation}
for every \(i\in\{1,\ldots,N\}\).
Applying \cref{eq:two_party_single_mode_assumed_coarse_graining}
at all parties gives
\[
\begin{aligned}
\ket\Phi^{\mathrm{cond}}
&=2^{(N-1)/2}\mathcal U
\left(\prod_{j=1}^{N}\bar\Pi_{j,x_j}^{1}\right)
\ket{\widehat\Psi}=2^{(N-1)/2}
\left[\prod_{j=1}^{N}
\Pi_{S_j}^{1}\right]
\ket\Phi.
\end{aligned}
\]
The last expression does not depend on the inputs.
Since \(\mathcal U^\dagger\mathcal U=\mathds{1}\), we also have
\(\ket{\widehat\Psi}^{\mathrm{cond}}
=\mathcal U^\dagger\ket\Phi^{\mathrm{cond}}\).
The isometry \(\mathcal U\) is input independent, so the dilated
conditional state is input independent as well.

Recall that the global isometry factorizes as
\(\mathcal U=\bigwedge_{j=1}^{N}\mathcal U_j\).
For example, fix a party \(i\) and apply
\(\prod_{j\ne i}
(\mathcal U_j\bar\Pi_{j,x_j}^{1}\mathcal U_j^\dagger)\)
at all other parties to the assumed identity for
\(\mathcal Z_i\) in
\cref{eq:two_party_single_mode_assumed_XZ}.
Since \(\mathcal U_j^\dagger\mathcal U_j=\mathds{1}\)
for every \(j\), its action on the left-hand side is
\[
\begin{aligned}
&\left[
\prod_{j\ne i}
(\mathcal U_j\bar\Pi_{j,x_j}^{1}\mathcal U_j^\dagger)
\right]
\mathcal U\mathcal Z_i\ket{\widehat\Psi}
=
\mathcal U\mathcal Z_i
\left(\prod_{j\ne i}\bar\Pi_{j,x_j}^{1}\right)
\ket{\widehat\Psi}.
\end{aligned}
\]
Here we also use the fact that
\(\bar\Pi_{j,x_j}^{1}\) and \(\mathcal Z_i\) are even
operators acting on different parties for \(j\ne i\).

On the right-hand side, the same product of projectors commutes
with the diagonal extracted operator in the first line of
\cref{eq:two_party_single_mode_assumed_XZ}, giving
\[
\begin{aligned}
&\left[\bigl(\ket{01}\!\bra{01}-\ket{10}\!\bra{10}\bigr)_{S_i}
\wedge\id_{J_i}\right]
\mathcal U\left(\prod_{j\ne i}\bar\Pi_{j,x_j}^{1}\right)
\ket{\widehat\Psi}\\
&\qquad=
\left[\bigl(\ket{01}\!\bra{01}-\ket{10}\!\bra{10}\bigr)_{S_i}
\wedge\id_{J_i}\right]
\left[\prod_{j\ne i}
\Pi_{S_j}^{1}\right]
\ket{\Phi},
\end{aligned}
\]
where the equality follows by applying the assumed
coarse-grained identity in
\cref{eq:two_party_single_mode_assumed_coarse_graining}
successively at the other parties.

By \cref{eq:two_party_one_token_conditioning}, these projections
produce the corresponding normalized conditional states,
each with a factor \(2^{(1-N)/2}\). Cancelling this factor gives
\[
\mathcal U\mathcal Z_i\ket{\widehat\Psi}^{\mathrm{cond}}
=
\left[\bigl(\ket{01}\!\bra{01}-\ket{10}\!\bra{10}\bigr)_{S_i}
\wedge\id_{J_i}\right]
\ket{\Phi}^{\mathrm{cond}}.
\]
The same projection argument applies to the second line of
\cref{eq:two_party_single_mode_assumed_XZ} for \(\mathcal X_i\)
at every party \(i\).

We now prove \cref{eq:two_party_common_parity}. Evaluate
\cref{eq:two_party_conditional_stabilizer} using
\cref{eq:two_party_single_mode_assumed_XZ} on the conditional
state and the factorization in \cref{eq:two_party_T_factorization}.
Comparing the two orthogonal token branches gives a relation
between products of source junk states. Every endpoint factor
has definite parity, so regrouping the factors by source
introduces only scalar signs. Since the sources are independent
and the junk states are normalized, the corresponding source
vectors are proportional:
\begin{equation}
\label{eq:two_party_unencoded_source_relation}
\bigl(T_i^{(L)}\wedge T_{i+1}^{(R)}\bigr)
\ket{\zeta_i^c}\ \propto\ \ket{\zeta_i^a},
\qquad i\in\{1,\ldots,N\}.
\end{equation}
The proportionality factor is nonzero because both endpoint
maps are unitary. The two junk states in
\cref{eq:two_party_unencoded_source_relation} have the same
definite parity, while the operator on the left has parity
\(q_i+q_{i+1}\) modulo~\(2\). Therefore
\(q_i=q_{i+1}\) for every source, proving
\cref{eq:two_party_common_parity}.

For \(q=0\), all endpoint factors are even and we work with
\(\ket\Phi\). For \(q=1\), apply the encoding
\(\mathcal E=\bigwedge_i(\mathcal E_{i_R}\wedge\mathcal E_{i_L})\)
from \cref{eq:fermionic_even_encoding}. The source calculation
in \cref{eq:fermionic_even_encoded_source} gives
\cref{eq:two_party_double_mode_assumed_state}. The encoding
preserves the token labels, giving
\cref{eq:two_party_double_mode_assumed_coarse_graining}.
The transformations in
\cref{eq:fermionic_even_encoded_T,eq:fermionic_even_encoded_X_block,eq:fermionic_even_encoded_Z_block}
give the observable equations
\cref{eq:two_party_double_mode_assumed_XZ} on the one-token
sector, with even endpoint factors
\(\widetilde T_i^{(R)},\widetilde T_i^{(L)}\).
On the zero- and two-token sectors, both signed observables
vanish by \cref{eq:two_party_single_mode_assumed_XZ}.
Thus \cref{eq:two_party_double_mode_assumed_XZ} holds on the
full state.
Applying \(\mathcal E\) to
\cref{eq:two_party_conditional_stabilizer} preserves its sign
and replaces \(\ket\Phi^{\mathrm{cond}}\) by
\(\ket{\widetilde\Phi}^{\mathrm{cond}}
:=\mathcal E\ket\Phi^{\mathrm{cond}}\).
The projection argument above also gives
\cref{eq:two_party_double_mode_assumed_XZ} on this conditional
state.

Using \cref{eq:two_party_single_mode_assumed_XZ} in case~(i)
and \cref{eq:two_party_double_mode_assumed_XZ} in case~(ii)
on the corresponding conditional states,
together with the source forms in
\cref{eq:two_party_single_mode_assumed_state,eq:two_party_double_mode_assumed_state},
we evaluate the left-hand side of
\cref{eq:two_party_conditional_stabilizer}.
Comparing the two orthogonal system branches and using
\cref{eq:two_party_T_factorization,eq:two_party_even_T_factorization},
we obtain relations between products of source junk states.
Since the sources are independent and the junk states are
normalized, the corresponding factors agree up to phases:
\begin{equation}
\label{eq:two_party_local_phase_relations}
\begin{aligned}
&\text{Case (i):}
\quad
\bigl(T_i^{(L)}\wedge T_{i+1}^{(R)}\bigr)
\ket{\zeta_i^c}
=
e^{\ii\widehat{\varphi}_i}\ket{\zeta_i^a},
\\[0.5ex]
&\text{Case (ii):}
\quad
\bigl(\widetilde{T}_i^{(L)}
\wedge\widetilde{T}_{i+1}^{(R)}\bigr)
\ket{\widetilde{\zeta}_i^c}
=
e^{\ii\widetilde{\varphi}_i}
\ket{\widetilde{\zeta}_i^a},
\end{aligned}
\end{equation}
for \(i\in\{1,\ldots,N\}\), with indices understood modulo~\(N\).

In case~(i), mapping the \(c\)-branch to the \(a\)-branch
contributes the sign \((-1)^{N-1}\) from the fermionic ordering
of the single-mode systems, as in \cref{eq:system_sign_c_to_a}.
This cancels the factor \((-1)^{N-1}\) in
\cref{eq:two_party_conditional_stabilizer}.
Since the junk unitaries preserve parity, regrouping them
by source introduces no additional sign. Thus, as in
\cref{lem:fermionic_local_phase_relations} with \(q=0\),
\[
\sum_{i=1}^{N}\widehat{\varphi}_i
=
0\pmod{2\pi}.
\]

In case~(ii), both basis states of each extracted
subsystem are even, so exchanging the system factors produces
no sign. The junk unitaries are also even, so the factor
\((-1)^{N-1}\) in \cref{eq:two_party_conditional_stabilizer}
remains in the junk relation. Consequently,
\[
\sum_{i=1}^{N}\widetilde{\varphi}_i
=
(N-1)\pi\pmod{2\pi}.
\]

We now construct the extraction unitaries, acting on
\(\ket\Phi\) in case~(i) and on
\(\ket{\widetilde\Phi}\) in case~(ii). In case~(i), use the unitaries
\(U_{i_R}\) and \(U_{i_L}\) defined in
\cref{eq:fermionic_local_unitaries}; in case~(ii), use
\(\widetilde{U}_{i_R}\) and \(\widetilde{U}_{i_L}\) defined in
\cref{eq:fermionic_even_encoded_unitaries}.
These unitaries preserve parity because the corresponding factors \(T_i^{(R)},T_i^{(L)}\) and
\(\widetilde{T}_i^{(R)},\widetilde{T}_i^{(L)}\) are even.

Applied to \(\ket{\Phi_i}\) in case~(i), or to
\(\ket{\widetilde\Phi_i}\) in case~(ii), these unitaries leave the
\(a\)-branch unchanged and transform the junk state of the
\(c\)-branch into that of the \(a\)-branch, with a phase
\(e^{\ii\widehat{\varphi}_i}\) in case~(i) and
\(e^{\ii\widetilde{\varphi}_i}\) in case~(ii), by
\cref{eq:two_party_local_phase_relations}.
In case~(i), set \(\varphi_i=\widehat{\varphi}_i\);
in case~(ii), set \(\varphi_i=\widetilde{\varphi}_i\).
This gives the source-state identities in
\cref{eq:two_party_single_mode_state,eq:two_party_double_mode_state},
with the phase sums established above.

Finally, apply \(U\) in case~(i), or \(\widetilde{U}\) in
case~(ii), to the full-state coarse-grained identities in
\cref{eq:two_party_single_mode_assumed_coarse_graining,eq:two_party_double_mode_assumed_coarse_graining}
and the signed-observable identities in
\cref{eq:two_party_single_mode_assumed_XZ,eq:two_party_double_mode_assumed_XZ}.
These unitaries preserve every token sector, so they leave
the coarse-grained projectors unchanged. They also remove the
junk unitaries from the \(\mathcal X_i\) identities while
leaving the diagonal system operators in the \(\mathcal Z_i\)
identities unchanged.
This yields the measurement identities in
\cref{eq:two_party_single_mode_XZ,eq:two_party_double_mode_XZ}.

\end{proof}

%% file: NetworkSuppMat.tex
\clearpage
\onecolumngrid
\section{Fermionic network self-testing}
\label{sec:Main Theorem's Proof}

This section gives the formal statement and proof of
\cref{thm:fermionic-self-testing}.

\begin{theorem}[Fermionic network self-testing]
\label{thm:fermionic-self-testing-formal}
Consider the three-party fermionic game described in the section
``The fermionic self-testing scenario'', with loop configurations
\[
\mathcal L=\{ABC,ACB,AB,AC,BC\}.
\]
Parties \(A,B,C\) prepare sources \(\alpha,\beta,\gamma\), respectively.
For each \(\theta\in\{\alpha,\beta,\gamma\}\), source \(\theta\)
distributes a normalized pure fermionic state of definite parity
\[
\ket{\psi_\theta}_{\theta_R,\theta_L}
\in\mathcal H_{\theta_R}\wedge\mathcal H_{\theta_L},
\]
where \(\theta_R\) is retained by the preparing party and
\(\theta_L\) is transmitted through the network. The sources are
independent, and their states do not depend on \(l\in\mathcal L\).
The global source state is thus
\[
\ket{\Psi}:=\ket{\psi_\alpha}\wedge\ket{\psi_\beta}
\wedge\ket{\psi_\gamma}.
\]

For a party \(i\) preparing source \(\theta\) and receiving source
\(\eta\), write \(i_L=\theta_R\) and \(i_R=\eta_L\).
We use the same relabelling for the associated spaces and extraction maps.
For each input \(x_i\in\{0,1\}\), the party performs a
parity-preserving four-outcome POVM
\[
\left\{\Pi_{i,x_i}^{(\eta),a_i}\right\}_{a_i\in\{0,1_+,1_-,2\}}
\]
on \(\mathcal H_{i_R}\wedge\mathcal H_{i_L}\).
The measurement depends on the wiring only through the incoming-source
label \(\eta\), so identical local contexts use identical POVMs.
Writing \(\eta_i(l)\) for the source received by party \(i\) in
wiring \(l\), the observed distribution is
\[
P^l(\mathbf a\mid\mathbf x)
=\bra{\Psi}
\left(\bigwedge_{i\in\{A,B,C\}}
\Pi_{i,x_i}^{(\eta_i(l)),a_i}\right)
\ket{\Psi}.
\]

At the party preparing source \(\theta\), introduce a fermionic
ancillary space \(\mathcal H_{M_{\theta_R}}\), initialized in the
normalized even state \(\ket0_{M_{\theta_R}}\).
Let
\[
\left\{\bar\Pi_{i,x_i}^{(\eta),a_i}\right\}_{a_i\in\{0,1_+,1_-,2\}}
\]
be parity-preserving PVMs on
\(\mathcal H_{i_R}\wedge\mathcal H_{i_L}
\wedge\mathcal H_{M_{\theta_R}}\)
providing Naimark dilations of the POVMs,
using the same ancillary space and initialized state for both inputs
and all local contexts. Choose the same dilation whenever the input
and local context agree. Grouping \(M_{\theta_R}\) with the retained
subsystem \(\theta_R\), define
\[
\ket{\widehat\psi_\theta}
:=\ket{\psi_\theta}\wedge\ket0_{M_{\theta_R}},
\qquad
\ket{\widehat\Psi}
:=\ket{\widehat\psi_\alpha}\wedge\ket{\widehat\psi_\beta}
\wedge\ket{\widehat\psi_\gamma}.
\]

For each \(l\in\mathcal L\), let \(P_f^l\) denote the distribution
generated by the trusted fermionic strategy for that wiring.
Suppose that
\[
P^l(\mathbf a\mid\mathbf x)=P_f^l(\mathbf a\mid\mathbf x)
\]
for every \(l\in\mathcal L\), every input \(\mathbf x\), and every
output \(\mathbf a\).

Let \(I(l)\) and \(\Theta(l)\) denote the parties and sources
participating in loop \(l\), respectively. For the extraction maps below,
each input space is restricted to the support of the corresponding
reduced state of \(\ket{\widehat\psi_\theta}\).
Then, for each \(l\in\mathcal L\) and \(\theta\in\Theta(l)\),
there exist local parity-preserving isometries
\[
\begin{aligned}
\mathcal U_{\theta_R}^{l}
&:\mathcal H_{\theta_R}\wedge\mathcal H_{M_{\theta_R}}
\longrightarrow S_{\theta_R}\wedge J_{\theta_R}^{a},
\\
\mathcal U_{\theta_L}^{l}
&:\mathcal H_{\theta_L}
\longrightarrow S_{\theta_L}\wedge J_{\theta_L}^{a},
\end{aligned}
\]
where \(S_{\theta_R}\simeq S_{\theta_L}\simeq\mathbb C^2\)
are single fermionic modes and \(J_{\theta_R}^{a},J_{\theta_L}^{a}\)
are auxiliary junk spaces.
The isometries are independent of the inputs and depend on \(l\)
only through the local context at the party holding the corresponding
subsystem; the same maps are used whenever that context recurs.
They satisfy the following two conclusions.

\begin{enumerate}
\item \emph{Self-testing of the source states.}
For every \(l\in\mathcal L\) and \(\theta\in\Theta(l)\),
\begin{equation}
\label{eq:full-game-final-source-states}
\begin{aligned}
\ket{\widehat\Phi_\theta^{l}}
&:=\left(\mathcal U_{\theta_R}^{l}\wedge
\mathcal U_{\theta_L}^{l}\right)\ket{\widehat\psi_\theta}
=\frac1{\sqrt2}
\left(e^{\ii\varphi_\theta^{l}}\ket{01}+\ket{10}\right)
_{S_{\theta_R}S_{\theta_L}}
\wedge\ket{\zeta_\theta^a}_{J_{\theta_R}^{a}J_{\theta_L}^{a}},
\end{aligned}
\end{equation}
where the junk states \(\ket{\zeta_\theta^a}\) are normalized and
independent of the wiring, and the phases satisfy
\begin{equation}
\label{eq:full-game-final-phase-sums}
\sum_{\theta\in\Theta(l)}\varphi_\theta^{l}=0\pmod{2\pi}
\qquad\text{for every }l\in\mathcal L.
\end{equation}

\item \emph{Self-testing of the measurements.}
For each \(l\in\mathcal L\), use the party-based labels
\(i_R,i_L\) defined above and set
\[
\begin{aligned}
\mathcal U_i^{l}
&:=\mathcal U_{i_R}^{l}\wedge\mathcal U_{i_L}^{l},
&
\mathcal U^{l}&:=\bigwedge_{i\in I(l)}\mathcal U_i^{l},
\\
S_i&:=S_{i_R}\wedge S_{i_L},
&
J_i^a&:=J_{i_R}^a\wedge J_{i_L}^a,
\end{aligned}
\]
Define the dilated state on the participating loop by
\[
\ket{\widehat\Psi^{l}}
:=\bigwedge_{\theta\in\Theta(l)}\ket{\widehat\psi_\theta},
\]
and let
\[
\ket{\widehat\Phi^{l}}
:=\mathcal U^{l}\ket{\widehat\Psi^{l}}
\]
denote the extracted global state. Define the signed observables
\[
M_{x_i}^{(i,l)}
:=\bar\Pi_{i,x_i}^{(\eta_i(l)),1_+}
-\bar\Pi_{i,x_i}^{(\eta_i(l)),1_-},
\]
and set
\[
\mathcal X_A^{l}:=\frac{M_0^{(A,l)}+M_1^{(A,l)}}{\sqrt2},
\qquad
\mathcal Z_A^{l}:=\frac{M_0^{(A,l)}-M_1^{(A,l)}}{\sqrt2}
\]
when \(A\in I(l)\), while
\[
\mathcal X_i^{l}:=M_1^{(i,l)},
\qquad
\mathcal Z_i^{l}:=M_0^{(i,l)},
\qquad i\in I(l)\cap\{B,C\}.
\]
For every \(i\in I(l)\) and \(x_i\in\{0,1\}\),
\begin{equation}
\label{eq:full-game-final-measurement-identities}
\begin{aligned}
\mathcal U^{l}\bar\Pi_{i,x_i}^{(\eta_i(l)),0}
\ket{\widehat\Psi^{l}}
&=\Pi^0_{S_i}\ket{\widehat\Phi^{l}},
\\
\mathcal U^{l}\bar\Pi_{i,x_i}^{(\eta_i(l)),2}
\ket{\widehat\Psi^{l}}
&=\Pi^2_{S_i}\ket{\widehat\Phi^{l}},
\\
\mathcal U^{l}\mathcal Z_i^{l}\ket{\widehat\Psi^{l}}
&=Z_{S_i}\ket{\widehat\Phi^{l}},
\\
\mathcal U^{l}\mathcal X_i^{l}\ket{\widehat\Psi^{l}}
&=X_{S_i}\ket{\widehat\Phi^{l}},
\end{aligned}
\end{equation}
where
\[
\begin{aligned}
\Pi^0_{S_i}
&:=\ket{00}\!\bra{00}_{S_{i_R}S_{i_L}}
\wedge\mathds{1}_{J_i^a},
\\
\Pi^2_{S_i}
&:=\ket{11}\!\bra{11}_{S_{i_R}S_{i_L}}
\wedge\mathds{1}_{J_i^a},
\\
Z_{S_i}
&:=\left(\ket{01}\!\bra{01}-\ket{10}\!\bra{10}\right)
_{S_{i_R}S_{i_L}}\wedge\mathds{1}_{J_i^a},
\\
X_{S_i}
&:=\left(\ket{01}\!\bra{10}+\ket{10}\!\bra{01}\right)
_{S_{i_R}S_{i_L}}\wedge\mathds{1}_{J_i^a}.
\end{aligned}
\]
These identities give the trusted measurements in
\cref{eq:trusted-measurements}, with the coefficients in
\cref{eq:trusted-coefficients-A,eq:trusted-coefficients-BC},
on the extracted state.
\end{enumerate}
\end{theorem}

\begin{proof}
We apply \cref{thm:fixed-loop-characterization} to the
three-party loops \(ABC\) and \(ACB\). Each admits two classes
of strategies that are inequivalent under local parity-preserving
isometries. Following the consistency argument in Supplementary
Material~G of Ref.~\cite{kalarde2026fermions}, we construct a
common partial extraction and show that all participating local
contexts belong to the same parity class. The correlations in
\(BC\) then exclude the double-mode class, leaving the
single-mode source and measurement characterizations.

We use the source states, ancillary systems, and Naimark dilations
fixed in the theorem statement. Apply part~1 of
\cref{thm:fixed-loop-characterization} separately to \(ABC\)
and \(ACB\), and denote the initial extraction isometries by
\[
\mathcal U_{\theta_R}^{l}
:=\mathcal U_{\theta_R M_{\theta_R}}^{l},
\qquad
\mathcal U_{\theta_L}^{l},
\qquad l\in\{ABC,ACB\}.
\]
On each physical reduced support, define the token projections
\begin{equation}
\label{eq:token-projectors}
Q_{\theta_\sigma}^{l}
:=(\mathcal U_{\theta_\sigma}^{l})^\dagger
\bigl(\ket1\!\bra1_{S_{\theta_\sigma}}
\wedge\id_{J_{\theta_\sigma}}\bigr)
\mathcal U_{\theta_\sigma}^{l},
\qquad \sigma\in\{R,L\},
\end{equation}
which are even projections, and the source form in
\cref{eq:fixed-loop-partial-state} shows that they preserve
the corresponding reduced supports of the source state
\(\ket{\widehat\psi_\theta}\), each with expectation
\(1/2\) on it. It also gives
\[
(Q_{\theta_R}^{l}+Q_{\theta_L}^{l})
\ket{\widehat\psi_\theta}=\ket{\widehat\psi_\theta},
\qquad l\in\{ABC,ACB\}.
\]

Fix arbitrary inputs. For a party \(i\) preparing source
\(\theta\) and receiving source \(\eta\) in wiring
\(l\in\{ABC,ACB\}\),
\cref{eq:fixed-loop-partial-measurements,eq:token-projectors}
give, on the local reduced support,
\begin{equation}
\label{eq:full-game-coarse-projectors}
\begin{aligned}
\bar\Pi_{i,x_i}^{(\eta),0}
&=(\id-Q_{\theta_R}^{l})(\id-Q_{\eta_L}^{l}),\\
\bar\Pi_{i,x_i}^{(\eta),1}
&=Q_{\theta_R}^{l}(\id-Q_{\eta_L}^{l})
 +(\id-Q_{\theta_R}^{l})Q_{\eta_L}^{l},\\
\bar\Pi_{i,x_i}^{(\eta),2}
&=Q_{\theta_R}^{l}Q_{\eta_L}^{l},
\end{aligned}
\end{equation}
where
\(\bar\Pi_{i,x_i}^{(\eta),1}
:=\bar\Pi_{i,x_i}^{(\eta),1_+}
+\bar\Pi_{i,x_i}^{(\eta),1_-}\).

In \(AB\), the contexts of \(A\) and \(B\) coincide with
their contexts in \(ACB\) and \(ABC\), respectively.
For the fixed inputs, locality therefore gives the same
dilated PVMs in these matching contexts. Each party also
has the same local reduced state: it holds the same pair
of source subsystems from fixed independent sources, together
with the same ancillary state. Hence the operator equalities
in \cref{eq:full-game-coarse-projectors} remain valid on the
local reduced support of the state in \(AB\), with the token
projections from \(ACB\) at \(A\) and those from \(ABC\) at \(B\).
Since the total count in \(AB\) is always two, we obtain
\[
\bigl(Q_{\alpha_R}^{ACB}+Q_{\beta_L}^{ACB}
+Q_{\beta_R}^{ABC}+Q_{\alpha_L}^{ABC}\bigr)
\ket{\widehat\Psi}=2\ket{\widehat\Psi}.
\]
To compare the token projections for source \(\alpha\), contract
with the normalized source states
\(\ket{\widehat\psi_\beta}\) and
\(\ket{\widehat\psi_\gamma}\). Explicitly, the left-hand side is
\[
\begin{aligned}
&\bigl[(Q_{\alpha_R}^{ACB}+Q_{\alpha_L}^{ABC})
\ket{\widehat\psi_\alpha}\bigr]
\wedge\ket{\widehat\psi_\beta}\wedge\ket{\widehat\psi_\gamma}\\
&\quad+\ket{\widehat\psi_\alpha}\wedge
\bigl[(Q_{\beta_L}^{ACB}+Q_{\beta_R}^{ABC})
\ket{\widehat\psi_\beta}\bigr]
\wedge\ket{\widehat\psi_\gamma}.
\end{aligned}
\]
All token projections are even, so moving them past the other
source factors introduces no sign. Contracting the last two
source factors gives
\[
\begin{aligned}
\Bigl[Q_{\alpha_R}^{ACB}+Q_{\alpha_L}^{ABC}
&+\Bigl(
\bra{\widehat\psi_\beta}Q_{\beta_L}^{ACB}
\ket{\widehat\psi_\beta}
+
\bra{\widehat\psi_\beta}Q_{\beta_R}^{ABC}
\ket{\widehat\psi_\beta}
\Bigr)\id\Bigr]\ket{\widehat\psi_\alpha}
=2\ket{\widehat\psi_\alpha}.
\end{aligned}
\]
Each expectation is \(1/2\), hence
\begin{equation}
\label{eq:full-game-alpha-AB-count}
(Q_{\alpha_R}^{ACB}+Q_{\alpha_L}^{ABC})
\ket{\widehat\psi_\alpha}=\ket{\widehat\psi_\alpha}.
\end{equation}
Similarly, in the loops \(ABC\) and \(ACB\), the total count
is three. Contracting the same two source states gives,
respectively,
\begin{align}
(Q_{\alpha_R}^{ABC}+Q_{\alpha_L}^{ABC})
\ket{\widehat\psi_\alpha}&=\ket{\widehat\psi_\alpha},
\label{eq:full-game-alpha-ABC-count}\\
(Q_{\alpha_R}^{ACB}+Q_{\alpha_L}^{ACB})
\ket{\widehat\psi_\alpha}&=\ket{\widehat\psi_\alpha}.
\label{eq:full-game-alpha-ACB-count}
\end{align}
Subtracting \cref{eq:full-game-alpha-ABC-count,eq:full-game-alpha-ACB-count},
respectively, from \cref{eq:full-game-alpha-AB-count} gives
\[
\begin{aligned}
(Q_{\alpha_R}^{ACB}-Q_{\alpha_R}^{ABC})
\ket{\widehat\psi_\alpha}&=0,\\
(Q_{\alpha_L}^{ABC}-Q_{\alpha_L}^{ACB})
\ket{\widehat\psi_\alpha}&=0.
\end{aligned}
\]
Analogous comparisons using \(BC\) and \(AC\) instead
of \(AB\) identify the token projections for sources
\(\beta\) and \(\gamma\), respectively.
Applying a Schmidt decomposition to each source state then gives
\[
Q_{\theta_\sigma}^{ABC}=Q_{\theta_\sigma}^{ACB},
\qquad
\theta\in\{\alpha,\beta,\gamma\},\quad \sigma\in\{R,L\},
\]
on the corresponding reduced supports of the dilated physical
states. Thus both triangles determine the same token projection
on each source endpoint. By \cref{eq:full-game-coarse-projectors},
the coarse-grained measurements in every local context can be
written in terms of these fixed context-independent token
projections.

We therefore choose the \(ABC\) source isometries for all
wirings and denote them by
\[
\mathcal U_{\theta_R}
:=\mathcal U^{ABC}_{\theta_R M_{\theta_R}},
\qquad
\mathcal U_{\theta_L}:=\mathcal U^{ABC}_{\theta_L}.
\]
For a party \(i\) preparing source \(\theta\) and receiving
source \(\eta\), set
\[
\mathcal U_i^{(\eta)}
:=\mathcal U_{\eta_L}\wedge\mathcal U_{\theta_R}.
\]
Here the incoming subsystem \(\eta_L\) and the retained
subsystem \(\theta_R\), together with its ancilla, correspond
to the party labels \(i_R\) and \(i_L\), respectively, in
\cref{thm:fixed-loop-characterization}.
The expressions for the coarse-grained projectors in
\cref{eq:full-game-coarse-projectors} then give, for every party
and both inputs,
\[
\mathcal U_i^{(\eta)}\bar\Pi_{i,x_i}^{(\eta),n}
=\Pi_{\mathrm{TC},i}^{n}\mathcal U_i^{(\eta)},
\qquad n\in\{0,1,2\},
\]
on the corresponding local reduced supports in both triangles
and, by locality, in each two-party loop. The common extraction
also gives the same source states
\[
\begin{aligned}
\ket{\Phi_\theta}
:=(\mathcal U_{\theta_R}\wedge\mathcal U_{\theta_L})
\ket{\widehat\psi_\theta}
=\frac1{\sqrt2}\bigl(
\ket{01}_{S_{\theta_R}S_{\theta_L}}\wedge\ket{\zeta_\theta^c}
+\ket{10}_{S_{\theta_R}S_{\theta_L}}\wedge\ket{\zeta_\theta^a}
\bigr)
\end{aligned}
\]
in every wiring, with fixed normalized junk states.

We now apply part~2 of \cref{thm:fixed-loop-characterization}
to both triangles using this common extraction. To construct
the unitaries which complete the self-testing, denote the two unitary factors of
\(T_i\) in each local context \((i,\eta)\) by
\(T_i^{(\eta,R)}\) and \(T_i^{(\eta,L)}\), and choose
these factors once for that context. Each context occurs in
exactly one triangle. For a fixed context \((i,\eta)\),
\cref{eq:fixed-loop-partial-XZ} and a Schmidt decomposition of
the triangle state across party \(i\) and the other parties
give, on the local reduced support,
\begin{equation}
\label{eq:full-game-context-XZ}
\begin{aligned}
\mathcal U_i^{(\eta)}\mathcal Z_i
&=\left[
\left(\ket{01}\!\bra{01}-\ket{10}\!\bra{10}\right)_{S_i}
\wedge\id_{J_i}\right]\mathcal U_i^{(\eta)},\\
\mathcal U_i^{(\eta)}\mathcal X_i
&=\left(
\ket{01}\!\bra{10}_{S_i}\wedge T_i
+\ket{10}\!\bra{01}_{S_i}\wedge T_i^\dagger
\right)\mathcal U_i^{(\eta)},
\end{aligned}
\end{equation}
where \(T_i=T_i^{(\eta,R)}\wedge T_i^{(\eta,L)}\).
Whenever the context \((i,\eta)\) occurs in a two-party loop,
the local reduced state, the dilated measurements, and the
initial isometry \(\mathcal U_i^{(\eta)}\) are unchanged.
Consequently, \cref{eq:full-game-context-XZ} holds
in the two-party loops with the same factors
\(T_i^{(\eta,R)}\) and \(T_i^{(\eta,L)}\) chosen in the triangle.

By \cref{thm:fixed-loop-characterization}, the factors
\(T_i^{(\eta,R)},T_i^{(\eta,L)}\) within \(ABC\) and \(ACB\)
have common parities \(q_{ABC}\) and \(q_{ACB}\), respectively,
for all parties \(i\).
In \(AB\), the contexts \((A,\beta)\) and \((B,\alpha)\)
are inherited from \(ACB\) and \(ABC\), respectively.
The common extraction gives the source and coarse-grained
equations
\cref{eq:two_party_single_mode_assumed_state,eq:two_party_single_mode_assumed_coarse_graining}
in \(AB\), while \cref{eq:full-game-context-XZ} gives the
signed-observable equations \cref{eq:two_party_single_mode_assumed_XZ}.
The conditional correlations in \(AB\) maximally violate Bell
inequality~\Romannum{2} in \cref{eq:Bell_2}. We can therefore
apply the common-parity conclusion
\cref{eq:two_party_common_parity} of
\cref{thm:fixed-loop-characterization-N2} to obtain
\[
q_{ACB}=q_{ABC}.
\]
Hence all parties fall in the same parity case in both of
their local contexts:
they give case~(i) when the factors preserve parity and
case~(ii) when they change parity.
In case~(ii), we choose the encoding consistently across all
wirings. By \cref{eq:fermionic_even_encoding}, the encoding
appends the even state \(\ket{bb}\) determined by the token
occupation \(b\), while retaining the original token and junk
states. It therefore does not depend on the measurement context
or the factors \(T_i^{(\eta,R)},T_i^{(\eta,L)}\).
The common initial extraction has fixed the token basis and
junk spaces for each source subsystem. Hence, for each source
\(\theta\), we use the same maps \(\mathcal E_{\theta_R}\)
and \(\mathcal E_{\theta_L}\), regardless of which parties
receive these subsystems. This gives the same encoded source
state in every wiring.

For each local context \((i,\eta)\), take the parity-preserving unitaries 
that complete the self-testing \(U_{i_R}^{(\eta)},U_{i_L}^{(\eta)}\)
from \cref{eq:fermionic_local_unitaries} in case~(i), and
\(\widetilde U_{i_R}^{(\eta)},\widetilde U_{i_L}^{(\eta)}\)
from \cref{eq:fermionic_even_encoded_unitaries} in case~(ii),
after the common encoding. These maps are constructed from
the fixed unitary factors of \(T_i\), so the same maps apply
whenever the local context is unchanged.
We now treat the two cases separately, following the argument
in Supplementary Material~G of Ref.~\cite{kalarde2026fermions}
from Eq.~(G.10) onward: we use the full self-tests in \(ABC\),
\(ACB\), \(AB\), and \(AC\) to determine the conditional
distribution in \(BC\).

\medskip

\noindent\emph{Case (i): single-mode systems.}
Apply part~2 of \cref{thm:fixed-loop-characterization} to
\(ABC\) and \(ACB\), with the common initial isometries and
the completing unitaries fixed above. For \(ABC\), we obtain
\begin{equation}
\label{eq:full-game-ABC-sources}
\begin{aligned}
\bigl(U_{A_L}^{(\gamma)}\wedge U_{B_R}^{(\alpha)}\bigr)
\ket{\Phi_\alpha}
&=\frac1{\sqrt2}
\bigl(e^{\ii\varphi_\alpha^{ABC}}\ket{01}+\ket{10}\bigr)
_{S_{\alpha_R}S_{\alpha_L}}\wedge\ket{\zeta_\alpha^a},\\
\bigl(U_{B_L}^{(\alpha)}\wedge U_{C_R}^{(\beta)}\bigr)
\ket{\Phi_\beta}
&=\frac1{\sqrt2}
\bigl(e^{\ii\varphi_\beta^{ABC}}\ket{01}+\ket{10}\bigr)
_{S_{\beta_R}S_{\beta_L}}\wedge\ket{\zeta_\beta^a},\\
\bigl(U_{C_L}^{(\beta)}\wedge U_{A_R}^{(\gamma)}\bigr)
\ket{\Phi_\gamma}
&=\frac1{\sqrt2}
\bigl(e^{\ii\varphi_\gamma^{ABC}}\ket{01}+\ket{10}\bigr)
_{S_{\gamma_R}S_{\gamma_L}}\wedge\ket{\zeta_\gamma^a},
\end{aligned}
\end{equation}
where
\[
\phi_{ABC}:=\varphi_\alpha^{ABC}+\varphi_\beta^{ABC}
+\varphi_\gamma^{ABC}=0\pmod{2\pi}.
\]
Here \(\varphi_\theta^l\) denotes the phase of source
\(\theta\) in wiring \(l\), and \(\phi_l\) denotes the sum
over the sources in that loop. All phase relations below
are understood modulo~\(2\pi\).
For \(ACB\), the same theorem gives
\begin{equation}
\label{eq:full-game-ACB-sources}
\begin{aligned}
\bigl(U_{A_L}^{(\beta)}\wedge U_{C_R}^{(\alpha)}\bigr)
\ket{\Phi_\alpha}
&=\frac1{\sqrt2}
\bigl(e^{\ii\varphi_\alpha^{ACB}}\ket{01}+\ket{10}\bigr)
_{S_{\alpha_R}S_{\alpha_L}}\wedge\ket{\zeta_\alpha^a},\\
\bigl(U_{B_L}^{(\gamma)}\wedge U_{A_R}^{(\beta)}\bigr)
\ket{\Phi_\beta}
&=\frac1{\sqrt2}
\bigl(e^{\ii\varphi_\beta^{ACB}}\ket{01}+\ket{10}\bigr)
_{S_{\beta_R}S_{\beta_L}}\wedge\ket{\zeta_\beta^a},\\
\bigl(U_{C_L}^{(\alpha)}\wedge U_{B_R}^{(\gamma)}\bigr)
\ket{\Phi_\gamma}
&=\frac1{\sqrt2}
\bigl(e^{\ii\varphi_\gamma^{ACB}}\ket{01}+\ket{10}\bigr)
_{S_{\gamma_R}S_{\gamma_L}}\wedge\ket{\zeta_\gamma^a},
\end{aligned}
\end{equation}
with \(\phi_{ACB}=\varphi_\alpha^{ACB}
+\varphi_\beta^{ACB}+\varphi_\gamma^{ACB}=0\).
The junk states in both sets of equations are the same fixed
\(\ket{\zeta_\theta^a}\), since every completing unitary map leaves
the corresponding \(a\)-branch unchanged.

We next apply \cref{thm:fixed-loop-characterization-N2}
to \(AB\) and \(AC\). In \(AB\), the contexts of \(A\)
and \(B\) are their contexts in \(ACB\) and \(ABC\),
respectively; in \(AC\), the contexts of \(A\) and \(C\)
are their contexts in \(ABC\) and \(ACB\).
The common source forms and coarse-grained identities,
together with the signed-observable identities inherited from
the triangles, satisfy the structural assumptions of that theorem with \(q=0\). The conditional distributions attain the
required Bell maximum, with \(A\) the distinguished party.
The theorem therefore applies using the same completing
unitaries already fixed for these contexts. In \(AB\), it gives
\begin{equation}
\label{eq:full-game-AB-sources}
\begin{aligned}
\bigl(U_{A_L}^{(\beta)}\wedge U_{B_R}^{(\alpha)}\bigr)
\ket{\Phi_\alpha}
&=\frac1{\sqrt2}
\bigl(e^{\ii\varphi_\alpha^{AB}}\ket{01}+\ket{10}\bigr)
_{S_{\alpha_R}S_{\alpha_L}}\wedge\ket{\zeta_\alpha^a},\\
\bigl(U_{B_L}^{(\alpha)}\wedge U_{A_R}^{(\beta)}\bigr)
\ket{\Phi_\beta}
&=\frac1{\sqrt2}
\bigl(e^{\ii\varphi_\beta^{AB}}\ket{01}+\ket{10}\bigr)
_{S_{\beta_R}S_{\beta_L}}\wedge\ket{\zeta_\beta^a},
\end{aligned}
\end{equation}
where \(\phi_{AB}=\varphi_\alpha^{AB}+\varphi_\beta^{AB}=0\).
Similarly, in \(AC\), we obtain
\begin{equation}
\label{eq:full-game-AC-sources}
\begin{aligned}
\bigl(U_{A_L}^{(\gamma)}\wedge U_{C_R}^{(\alpha)}\bigr)
\ket{\Phi_\alpha}
&=\frac1{\sqrt2}
\bigl(e^{\ii\varphi_\alpha^{AC}}\ket{01}+\ket{10}\bigr)
_{S_{\alpha_R}S_{\alpha_L}}\wedge\ket{\zeta_\alpha^a},\\
\bigl(U_{C_L}^{(\alpha)}\wedge U_{A_R}^{(\gamma)}\bigr)
\ket{\Phi_\gamma}
&=\frac1{\sqrt2}
\bigl(e^{\ii\varphi_\gamma^{AC}}\ket{01}+\ket{10}\bigr)
_{S_{\gamma_R}S_{\gamma_L}}\wedge\ket{\zeta_\gamma^a},
\end{aligned}
\end{equation}
where \(\phi_{AC}=\varphi_\alpha^{AC}+\varphi_\gamma^{AC}=0\).
Composing these completing unitaries with the common initial isometries
gives \cref{eq:full-game-final-source-states,eq:full-game-final-phase-sums}
for \(l\in\{ABC,ACB,AB,AC\}\).

We now compare the source identities in
\cref{eq:full-game-ABC-sources,eq:full-game-ACB-sources,eq:full-game-AB-sources,eq:full-game-AC-sources},
following Eqs.~(G.20)--(G.40) of Ref.~\cite{kalarde2026fermions}.
Our aim is to determine the self-tested states of sources
\(\beta\) and \(\gamma\) in \(BC\) from their forms in the
other wirings. For each source \(\theta\), the token states
\(\ket{01}\) and \(\ket{10}\) below are written in the order
\(S_{\theta_R}S_{\theta_L}\).
The completing unitaries in \cref{eq:fermionic_local_unitaries}
preserve each token occupation, so the two token branches
can be compared separately. All these unitaries preserve
parity, so composing maps on the two source subsystems
introduces no fermionic signs.

For source \(\beta\), compare the relations in \(ACB\) and
\(AB\) from
\cref{eq:full-game-ACB-sources,eq:full-game-AB-sources}.
Both contain the same map \(U_{A_R}^{(\beta)}\) on the
transmitted subsystem; only the map on the retained subsystem
changes, from \(U_{B_L}^{(\alpha)}\) to \(U_{B_L}^{(\gamma)}\).
Thus, applying
\(U_{B_L}^{(\gamma)}(U_{B_L}^{(\alpha)})^\dagger\)
on the retained subsystem converts the \(AB\) source relation
into the \(ACB\) source relation. Comparing the \(\ket{01}\)
and \(\ket{10}\) branches separately gives
\begin{equation}
\label{eq:full-game-beta-relative-action}
\begin{aligned}
&\bigl(U_{B_L}^{(\gamma)}(U_{B_L}^{(\alpha)})^\dagger
\wedge\id\bigr)
\bigl(\ket{01}\wedge\ket{\zeta_\beta^a}\bigr)=e^{\ii(\varphi_\beta^{ACB}-\varphi_\beta^{AB})}
\ket{01}\wedge\ket{\zeta_\beta^a},\\
&\bigl(U_{B_L}^{(\gamma)}(U_{B_L}^{(\alpha)})^\dagger
\wedge\id\bigr)
\bigl(\ket{10}\wedge\ket{\zeta_\beta^a}\bigr)
=\ket{10}\wedge\ket{\zeta_\beta^a}.
\end{aligned}
\end{equation}
In \(BC\), source \(\beta\) is acted on by
\(U_{B_L}^{(\gamma)}\) and \(U_{C_R}^{(\beta)}\).
Starting from its \(ABC\) relation in
\cref{eq:full-game-ABC-sources}, we therefore replace only
\(U_{B_L}^{(\alpha)}\) by \(U_{B_L}^{(\gamma)}\).
Using \cref{eq:full-game-beta-relative-action}, we obtain
\begin{equation}
\label{eq:full-game-beta-BC}
\begin{aligned}
&\bigl(U_{B_L}^{(\gamma)}\wedge U_{C_R}^{(\beta)}\bigr)
\ket{\Phi_\beta}\\
&\quad=\bigl(U_{B_L}^{(\gamma)}
(U_{B_L}^{(\alpha)})^\dagger\wedge\id\bigr)
\bigl(U_{B_L}^{(\alpha)}\wedge U_{C_R}^{(\beta)}\bigr)
\ket{\Phi_\beta}\\
&\quad=\frac1{\sqrt2}
\bigl(e^{\ii(\varphi_\beta^{ABC}+\varphi_\beta^{ACB}
-\varphi_\beta^{AB})}\ket{01}+\ket{10}\bigr)
_{S_{\beta_R}S_{\beta_L}}\wedge\ket{\zeta_\beta^a}.
\end{aligned}
\end{equation}

For source \(\gamma\), compare its relations in \(ACB\)
and \(AC\) from
\cref{eq:full-game-ACB-sources,eq:full-game-AC-sources}.
Both contain \(U_{C_L}^{(\alpha)}\) on the retained
subsystem, while the map on the transmitted subsystem changes
from \(U_{A_R}^{(\gamma)}\) in \(AC\) to
\(U_{B_R}^{(\gamma)}\) in \(ACB\).
Applying \(U_{B_R}^{(\gamma)}(U_{A_R}^{(\gamma)})^\dagger\)
on the transmitted subsystem and comparing the two token
branches gives
\begin{equation}
\label{eq:full-game-gamma-relative-action}
\begin{aligned}
&\bigl(\id\wedge U_{B_R}^{(\gamma)}
(U_{A_R}^{(\gamma)})^\dagger\bigr)
\bigl(\ket{01}\wedge\ket{\zeta_\gamma^a}\bigr)=e^{\ii(\varphi_\gamma^{ACB}-\varphi_\gamma^{AC})}
\ket{01}\wedge\ket{\zeta_\gamma^a},\\
&\bigl(\id\wedge U_{B_R}^{(\gamma)}
(U_{A_R}^{(\gamma)})^\dagger\bigr)
\bigl(\ket{10}\wedge\ket{\zeta_\gamma^a}\bigr)
=\ket{10}\wedge\ket{\zeta_\gamma^a}.
\end{aligned}
\end{equation}
In \(BC\), the retained map is \(U_{C_L}^{(\beta)}\), as
in \(ABC\), and the transmitted map is \(U_{B_R}^{(\gamma)}\).
Applying \cref{eq:full-game-gamma-relative-action} to the
\(ABC\) source relation in \cref{eq:full-game-ABC-sources}
therefore gives
\begin{equation}
\label{eq:full-game-gamma-BC}
\begin{aligned}
&\bigl(U_{C_L}^{(\beta)}\wedge U_{B_R}^{(\gamma)}\bigr)
\ket{\Phi_\gamma}\\
&\quad=\bigl(\id\wedge U_{B_R}^{(\gamma)}
(U_{A_R}^{(\gamma)})^\dagger\bigr)
\bigl(U_{C_L}^{(\beta)}\wedge U_{A_R}^{(\gamma)}\bigr)
\ket{\Phi_\gamma}\\
&\quad=\frac1{\sqrt2}
\bigl(e^{\ii(\varphi_\gamma^{ABC}+\varphi_\gamma^{ACB}
-\varphi_\gamma^{AC})}\ket{01}+\ket{10}\bigr)
_{S_{\gamma_R}S_{\gamma_L}}\wedge\ket{\zeta_\gamma^a}.
\end{aligned}
\end{equation}
Denote the phases of the \(\ket{01}\) branches in
\cref{eq:full-game-beta-BC,eq:full-game-gamma-BC} by
\(\varphi_\beta^{BC}\) and \(\varphi_\gamma^{BC}\), respectively.
Thus \cref{eq:full-game-final-source-states} also holds for \(BC\).

To determine their sum, we also need the relation between
the four phases of source \(\alpha\).
Its \(ABC\) and \(AB\) relations in
\cref{eq:full-game-ABC-sources,eq:full-game-AB-sources}
contain the same transmitted map \(U_{B_R}^{(\alpha)}\).
Changing the retained map from \(U_{A_L}^{(\beta)}\) to
\(U_{A_L}^{(\gamma)}\), and comparing the two token branches,
gives
\begin{equation}
\label{eq:full-game-alpha-relative-action}
\begin{aligned}
&\bigl(U_{A_L}^{(\gamma)}(U_{A_L}^{(\beta)})^\dagger
\wedge\id\bigr)
\bigl(\ket{01}\wedge\ket{\zeta_\alpha^a}\bigr)=e^{\ii(\varphi_\alpha^{ABC}-\varphi_\alpha^{AB})}
\ket{01}\wedge\ket{\zeta_\alpha^a},\\
&\bigl(U_{A_L}^{(\gamma)}(U_{A_L}^{(\beta)})^\dagger
\wedge\id\bigr)
\bigl(\ket{10}\wedge\ket{\zeta_\alpha^a}\bigr)
=\ket{10}\wedge\ket{\zeta_\alpha^a}.
\end{aligned}
\end{equation}
The same change of retained map converts the \(ACB\)
source relation into the \(AC\) source relation, since both
contain \(U_{C_R}^{(\alpha)}\) on the transmitted subsystem.
Applying \cref{eq:full-game-alpha-relative-action} to the
\(ACB\) relation in \cref{eq:full-game-ACB-sources} yields
\begin{equation}
\label{eq:full-game-alpha-AC-comparison}
\begin{aligned}
&\bigl(U_{A_L}^{(\gamma)}\wedge U_{C_R}^{(\alpha)}\bigr)
\ket{\Phi_\alpha}\\
&\quad=\bigl(U_{A_L}^{(\gamma)}(U_{A_L}^{(\beta)})^\dagger
\wedge\id\bigr)
\bigl(U_{A_L}^{(\beta)}\wedge U_{C_R}^{(\alpha)}\bigr)
\ket{\Phi_\alpha}\\
&\quad=\frac1{\sqrt2}
\bigl(e^{\ii(\varphi_\alpha^{ACB}+\varphi_\alpha^{ABC}
-\varphi_\alpha^{AB})}\ket{01}+\ket{10}\bigr)
_{S_{\alpha_R}S_{\alpha_L}}\wedge\ket{\zeta_\alpha^a}.
\end{aligned}
\end{equation}
Comparing the coefficient of \(\ket{01}\) in
\cref{eq:full-game-alpha-AC-comparison} with the source
\(\alpha\) relation in \cref{eq:full-game-AC-sources}
gives the first equality below. The other two are the phases
obtained in \cref{eq:full-game-beta-BC,eq:full-game-gamma-BC}:
\begin{equation}
\label{eq:full-game-source-phase-relations}
\begin{aligned}
\varphi_\alpha^{ABC}+\varphi_\alpha^{ACB}
&=\varphi_\alpha^{AB}+\varphi_\alpha^{AC},\\
\varphi_\beta^{BC}
&=\varphi_\beta^{ABC}+\varphi_\beta^{ACB}-\varphi_\beta^{AB},\\
\varphi_\gamma^{BC}
&=\varphi_\gamma^{ABC}+\varphi_\gamma^{ACB}-\varphi_\gamma^{AC}.
\end{aligned}
\end{equation}
The total phase in \(BC\) is the sum of the phases of
sources \(\beta\) and \(\gamma\).
Using \cref{eq:full-game-source-phase-relations} and the
loop constraints \(\phi_{ABC}=\phi_{ACB}=\phi_{AB}=\phi_{AC}=0\)
from \cref{eq:full-game-ABC-sources,eq:full-game-ACB-sources,eq:full-game-AB-sources,eq:full-game-AC-sources},
we obtain
\begin{equation}
\label{eq:full-game-BC-phase}
\begin{aligned}
\phi_{BC}
&:=\varphi_\beta^{BC}+\varphi_\gamma^{BC}\\
&=(\varphi_\beta^{ABC}+\varphi_\beta^{ACB}-\varphi_\beta^{AB})
 +(\varphi_\gamma^{ABC}+\varphi_\gamma^{ACB}-\varphi_\gamma^{AC})\\
&=\phi_{ABC}+\phi_{ACB}-\phi_{AB}-\phi_{AC}\\
&\quad-(\varphi_\alpha^{ABC}+\varphi_\alpha^{ACB}
-\varphi_\alpha^{AB}-\varphi_\alpha^{AC})\\
&=\phi_{ABC}+\phi_{ACB}-\phi_{AB}-\phi_{AC}=0.
\end{aligned}
\end{equation}
Together with the phase constraints above, this establishes
\cref{eq:full-game-final-phase-sums} for every \(l\in\mathcal L\).

It remains to compute the conditional probability in \(BC\).
We work with the states and measurements after applying the
common initial isometries and the completing unitaries fixed above.
The local context of \(B\) is its context in \(ACB\), while
that of \(C\) is its context in \(ABC\). The measurement
relations in \cref{eq:fixed-loop-final-measurement-identities}
therefore hold on the corresponding local reduced supports
in \(BC\), since the physical measurements, reduced supports,
and extraction maps are unchanged in each context.
This establishes \cref{eq:full-game-final-measurement-identities}
for \(BC\), completing the measurement identities for all
\(l\in\mathcal L\).
Neither \(B\) nor \(C\) is the distinguished party, so the
signed observable for input~\(1\) is \(\mathcal X_i\).
On the extracted conditional support, the corresponding
fine-grained projectors are thus
\begin{equation}
\label{eq:full-game-BC-projectors}
\frac12\bigl(\id+s_iX_{S_i}\bigr),
\qquad i\in\{B,C\},\quad s_i\in\{+1,-1\},
\end{equation}
where \(s_i\) labels the outcome \(1_{s_i}\) and \(\id\)
is the identity on that support.

Conditioning on both parties obtaining the coarse-grained
outcome~\(1\) selects the source branches
\(\ket{01,01}\) and \(\ket{10,10}\) from the product of
the completed source states in
\cref{eq:full-game-beta-BC,eq:full-game-gamma-BC}.
Their relative phase is
\(e^{\ii(\varphi_\beta^{BC}+\varphi_\gamma^{BC})}=1\)
by \cref{eq:full-game-BC-phase}. Hence the normalized
conditional state is
\begin{equation}
\label{eq:BCState_parity_proof}
\ket{\widehat\Phi_{BC}}^{\mathrm{cond}}
=\frac1{\sqrt2}\bigl(\ket{01,01}+\ket{10,10}\bigr)
_{S_{B_L}S_{C_R}S_{C_L}S_{B_R}}
\wedge\ket{\zeta_\beta^a}\wedge\ket{\zeta_\gamma^a},
\end{equation}
up to an irrelevant common phase. The system modes are
written in source order and placed before the junk modes.
Since each source junk state has definite parity and the two
branches of each source system have the same parity, moving
a junk state past the other source system contributes the
same sign to both branches.
The extracted measurements act trivially on the junk, so
we can compute their expectations using only the system state.

Reordering the system modes from
\(S_{B_L}S_{C_R}S_{C_L}S_{B_R}\) to the party order
\(S_{B_R}S_{B_L}S_{C_R}S_{C_L}\) gives
\begin{equation}
\label{eq:full-game-BC-system-state}
\frac1{\sqrt2}\bigl(
\ket{01}_{S_B}\wedge\ket{01}_{S_C}
-\ket{10}_{S_B}\wedge\ket{10}_{S_C}\bigr).
\end{equation}
Indeed, the source branch \(\ket{01,01}\) becomes
\(-\ket{10}_{S_B}\wedge\ket{10}_{S_C}\): the occupied
mode \(S_{B_R}\) passes the occupied mode \(S_{C_R}\).
The other source branch, \(\ket{10,10}\), becomes
\(\ket{01}_{S_B}\wedge\ket{01}_{S_C}\) without a sign,
since \(S_{B_R}\) is empty.
The state in \cref{eq:full-game-BC-system-state} has
\(\langle X_{S_B}\rangle=\langle X_{S_C}\rangle=0\) and
\(\langle X_{S_B}\wedge X_{S_C}\rangle=-1\).
Using the projectors in \cref{eq:full-game-BC-projectors},
we therefore obtain
\begin{equation}
\label{target_dist_BC}
\begin{aligned}
&P^{BC}\!\left(1_{s_B},1_{s_C}\,\middle|\,
x_B=x_C=1,\mathrm{cond}\right)\\
&\quad=\frac14\left\langle
\bigl(\id+s_BX_{S_B}\bigr)\wedge
\bigl(\id+s_CX_{S_C}\bigr)\right\rangle\\
&\quad=\frac14(1-s_Bs_C)
=P_f^{BC}\!\left(1_{s_B},1_{s_C}\,\middle|\,
x_B=x_C=1,\mathrm{cond}\right).
\end{aligned}
\end{equation}
Case~(i) therefore gives the required conditional probability.

\medskip

\noindent\emph{Case (ii): double-mode systems.}
We apply the double-mode completion in part~2 of
\cref{thm:fixed-loop-characterization} to \(ABC\) and \(ACB\),
using the common encoding maps \(\mathcal E_{\theta_R}\)
and \(\mathcal E_{\theta_L}\) fixed above.
For \(AB\) and \(AC\), we apply
\cref{thm:fixed-loop-characterization-N2} to the common
single-mode partial characterization, before encoding.
The partially extracted source states and coarse-grained
projectors have the forms in
\cref{eq:two_party_single_mode_assumed_state,eq:two_party_single_mode_assumed_coarse_graining},
while \cref{eq:full-game-context-XZ} gives the required
signed-observable relations with \(q=1\).
Since the conditional distributions attain the required
Bell maximum, case~(ii) of that theorem gives the
double-mode completion with the same encoding maps.

The source relations in
\cref{eq:full-game-ABC-sources,eq:full-game-ACB-sources,eq:full-game-AB-sources,eq:full-game-AC-sources}
then hold with \(U\) replaced by \(\widetilde U\), each
\(\ket{\Phi_\theta}\) replaced by
\((\mathcal E_{\theta_R}\wedge\mathcal E_{\theta_L})
\ket{\Phi_\theta}\), and \(S\) and \(\ket{\zeta_\theta^a}\)
replaced by \(\widetilde S\) and
\(\ket{\widetilde\zeta_\theta^a}\), respectively.
The source basis states \(\ket{01}\) and \(\ket{10}\)
are replaced by \(\ket{00,11}\) and \(\ket{11,00}\).
The phase constraints are now
\[
\phi_{ABC}=\phi_{ACB}=0,
\qquad
\phi_{AB}=\phi_{AC}=\pi.
\]
The unitaries \(\widetilde U\) preserve parity and each
encoded token occupation. We can therefore repeat the
comparisons in
\cref{eq:full-game-beta-relative-action,eq:full-game-gamma-relative-action,eq:full-game-alpha-relative-action,eq:full-game-alpha-AC-comparison}
with these replacements. This gives the same source-phase
relations \cref{eq:full-game-source-phase-relations}, and hence
\[
\phi_{BC}
=\phi_{ABC}+\phi_{ACB}-\phi_{AB}-\phi_{AC}
=-2\pi=0\pmod{2\pi}.
\]

We again compute the conditional probability after the
common initial isometries, now followed by the common
encoding and the completing unitaries \(\widetilde U\).
The measurement relations inherited from \(ACB\) at \(B\)
and from \(ABC\) at \(C\) give the projectors in
\cref{eq:full-game-BC-projectors} with \(X_{S_i}\) replaced
by \(X_{\widetilde S_i}\).
Conditioning on the coarse-grained outcome~\(1\) at both
parties selects the two encoded source branches corresponding
to those in \cref{eq:BCState_parity_proof}, again with
relative phase~\(1\).
Since both \(\ket{00}\) and \(\ket{11}\) have even parity,
reordering the encoded subsystems into party order introduces
no sign. After disregarding the junk factors, on which the
measurements act trivially, the system state is
\begin{equation}
\label{eq:full-game-BC-encoded-system-state}
\frac1{\sqrt2}\bigl(
\ket{00,11}_{\widetilde S_B}\wedge
\ket{00,11}_{\widetilde S_C}
+\ket{11,00}_{\widetilde S_B}\wedge
\ket{11,00}_{\widetilde S_C}\bigr).
\end{equation}
For the state in \cref{eq:full-game-BC-encoded-system-state},
\(\langle X_{\widetilde S_B}\rangle
=\langle X_{\widetilde S_C}\rangle=0\), while
\(\langle X_{\widetilde S_B}\wedge X_{\widetilde S_C}\rangle=+1\).
The same probability calculation as in \cref{target_dist_BC}
therefore yields
\begin{equation}
\label{dist_BC}
P^{BC}\!\left(1_{s_B},1_{s_C}\,\middle|\,
x_B=x_C=1,\mathrm{cond}\right)
=\frac14(1+s_Bs_C),
\end{equation}
which contradicts the required fermionic probability in
\cref{target_dist_BC}. Since all participating local contexts belong to
the same case, this excludes case~(ii) for the full game.
The remaining single-mode case satisfies
\cref{eq:full-game-final-source-states,eq:full-game-final-phase-sums,eq:full-game-final-measurement-identities},
completing the proof.
\end{proof}